\documentclass[12pt]{article}
\usepackage{epsfig}

\catcode`\@=11

\newcount\hour
\newcount\minute
\newtoks\amorpm \hour=\time\divide\hour by 60\minute
=\time{\multiply\hour by 60 \global\advance\minute by-\hour}
\edef\standardtime{{\ifnum\hour<12 \global\amorpm={am}%
        \else\global\amorpm={pm}\advance\hour by-12 \fi
        \ifnum\hour=0 \hour=12 \fi
        \number\hour:\ifnum\minute<10
        0\fi\number\minute\the\amorpm}}
\edef\militarytime{\number\hour:\ifnum\minute<10
0\fi\number\minute}

\def\draftlabel#1{{\@bsphack\if@filesw {\let\thepage\relax
   \xdef\@gtempa{\write\@auxout{\string
      \newlabel{#1}{{\@currentlabel}{\thepage}}}}}\@gtempa
   \if@nobreak \ifvmode\nobreak\fi\fi\fi\@esphack}
        \gdef\@eqnlabel{#1}}
\def\@eqnlabel{}
\def\@vacuum{}
\def\marginnote#1{}
\def\draftmarginnote#1{\marginpar{\raggedright\scriptsize\tt#1}}
\def\draft{
        \pagestyle{plain}
        \overfullrule=2pt
        \oddsidemargin -.1truein
        \def\@oddhead{\sl \phantom{\today\quad\militarytime} \hfil
        \smash{\Large\sl DRAFT} \hfil \today\quad\militarytime}
        \let\@evenhead\@oddhead
        \let\label=\draftlabel
        \let\marginnote=\draftmarginnote
        \def\ps@empty{\let\@mkboth\@gobbletwo
        \def\@oddfoot{\hfil \smash{\Large\sl DRAFT} \hfil}
        \let\@evenfoot\@oddhead}
        \def\@eqnnum{(\theequation)\rlap{\kern\marginparsep\tt\@eqnlabel}%
        \global\let\@eqnlabel\@vacuum}  }

\newcommand{\rf}[1]{(\ref{#1})}
\renewcommand{\theequation}{\thesection.\arabic{equation}}
\renewcommand{\thefootnote}{\fnsymbol{footnote}}
\newcommand{\newsection}{    
\setcounter{equation}{0}\section}
\def\appendix#1{\addtocounter{section}{1}\setcounter{equation}{0}
\renewcommand{\thesection}{\Alph{section}}
\section*{Appendix \thesection\protect\indent \parbox[t]{11.15cm}{#1}}
\addcontentsline{toc}{section}{Appendix \thesection\ \ \ #1}}

\def \lc {{light-cone}}
\def \ci{\cite}
\def\f{{\rm f}}

\def \we {\wedge}

\def \foot{\footnote}
\def \bi{\bibitem}
\def \la {\label}
\def \ov {\over}

\def \x {{\rm x}}
\def \tv {{\rm v}}
\def \four {{1 \ov 4}}
\def \b {\beta}
\def \Om {\Omega}

\def \t {\theta}
\def \s{\sigma}
\def \d {\partial}

\def \del{\partial}

\def \four{{\textstyle {1\ov 4}}}

\def\det{\hbox{det}}

\def\be{\begin{equation}}
\def\ee{\end{equation}}

\def \ci {\cite}

\def \ggg{\it {\Im}}

\def \rtimes {\bowtie}

\begin{document}


\date{November 2002}

\begin{titlepage}

\begin{center}
\hfill CERN-TH/2002-348\\
\hfill  Imperial/TP/02 -3/4 \\

\vspace{1.5cm}

{\Large \bf
 Solvable model of strings\\[.2cm]

%
in  a time-dependent plane-wave background
}
\\[.5cm]

\vspace{0.5cm}
{\large  G. Papadopoulos$^a$,
J.G. Russo$^{b,c}$
 and A.A. Tseytlin$^{d,e,}$\footnote{
 Also at Lebedev Physics Institute, Moscow.}}\\
\end{center}

\vskip 0.1cm
\centerline{\it ${}^a$ Department of Mathematics,  King's College London}
 \centerline{\it London WC2R 2LS, U.K. }

\vskip 0.2cm
\centerline{\it ${}^b$
Theory Division, CERN, Geneve, CH 1211,  Switzerland}

\vskip 0.2cm
\centerline{\it ${}^c$
 Departamento de F\'\i sica, Universidad de
Buenos Aires, }
\centerline {\it Ciudad Universitaria and Conicet, Pab. I, 1428
Buenos Aires, Argentina}

\vskip 0.2 cm
\centerline{\it ${}^d$ Theoretical Physics Group, Blackett Laboratory, }
\centerline{\it Imperial College, London SW7 2BZ, U.K.}

\vskip 0.2 cm
\centerline{\it ${}^e$ Smith Laboratory, The Ohio State University}
\centerline{\it Columbus, OH 43210, USA}

\vskip 1.0 cm

\begin{abstract}
We 
investigate a string model defined by  a special  plane-wave metric
$ds^2 = 2dudv - \lambda (u) x^2 du^2 + dx^2 $
with  $ \lambda  = { k \ov u^2}$ and 
$k$=const $> 0$. This  metric is  a Penrose limit
of some cosmological, Dp-brane and fundamental string backgrounds.
Remarkably, in Rosen coordinates the metric has a ``null cosmology''
interpretation with flat spatial sections and scale  
factor which is a power of the light-cone time $u$.
We show that:
\ \
(i)  This spacetime is a Lorentzian homogeneous
space. 
In particular, it  admits
 a boost  isometry
$u'= \ell u, \ v'= \ell^{-1} v$ similar to Minkowski space. 
\ \
(ii) It is an exact solution of string theory when supplemented
by a $u$-dependent dilaton such that 
 the corresponding  effective string coupling
$e^{\phi(u)}$  goes to zero at $u=\infty$ and at
the singularity  $u=0$, reducing back-reaction effects.
\ \ 
(iii) The classical string equations in this background
 become linear in the light-cone  gauge and
can be solved explicitly  in terms of Bessel's functions,  and thus
the string model can be directly quantized.  This allows one
to address the  issue of  singularity at the
string-theory level.
We  examine the  propagation  of first-quantized point-particle and
string modes in this time-dependent background. Using an
analytic continuation  prescription we argue that  the 
string propagation
through the singularity can be  smooth.

\end{abstract}

\end{titlepage}

\newpage
\setcounter{page}{1}
\renewcommand{\thefootnote}{\arabic{footnote}}
\setcounter{footnote}{0}

 \def \lc {light-cone\ }

\def \ae {{\rm E}}
\def \Epsilon {{\cal E}}
\hoffset=35pt
\voffset=-1.5cm
\textwidth=15.8cm
\textheight=23cm

\hoffset=-25pt
\voffset=-2.5cm
\catcode`\@=11

\def \lc {light cone\ }
\def \vv {{\cal X}}
\def \X {{\cal X}}

\def\bea{\begin{eqnarray}}
\def\eea{\end{eqnarray}}
\def\beann{\begin{eqnarray*}}
\def\eeann{\end{eqnarray*}}
\def\beq{\begin{equation}}
\def\eeq{\end{equation}}
\def\ba{\begin{array}}
\def\ea{\end{array}}
\def\ben{\begin{enumerate}}
\def\een{\end{enumerate}}
 \def \l {\lambda}
 \def\m {\mu}
\def\s {\sigma }

 \def \la {\label}
 \def\be{\begin{equation}}
\def\ee{\end{equation}}

\def \ci {\cite}
\def \la {\label}
\def \const {{\rm const}}
\def \haf{{\textstyle { 1 \ov 2}} }
\def \de{{\textstyle { 1 \ov 9}} }
\def \si{{\textstyle { 1 \ov 6}} }
\def \rt {{\tx { \ta \ov 2}}}
\def \r {\rho}
\def \fo{{ 1 \ov 4}}

\def \he {{\rm h}}

\font\mybb=msbm10 at 11pt
\font\mybbb=msbm10 at 17pt
\def\bb#1{\hbox{\mybb#1}}
\def\bbb#1{\hbox{\mybbb#1}}
\def\bZ {\bb{Z}}
\def\bR {\bb{R}}
\def\bE {\bb{E}}
\def\bT {\bb{T}}
\def\bM {\bb{M}}
\def\bH {\bb{H}}
\def\bC {\bb{C}}
\def\bA {\bb{A}}
\def\e  {\epsilon}
\def\bbC {\bbb{C}}
\def \DD {{\rm D}}
\def \foot {\footnote}
\def \k {\kappa}
\def \ov {\over}
\def \ha { { 1\ov 2}}
\def \we { \wedge}
\def \P { \Phi} \def\ep {\epsilon}
\def \go { g_1}\def \gd { g_2}\def \gt { g_3}\def \gc { g_4}\def \gp { g_5}
\def \F {{\cal F}}
\def \del { \partial}
\def \t {\theta}
\def \p {\phi}
\def \ee {\epsilon}
\def \te {\tilde \epsilon}
\def \ps {\psi}
\def \td {\tilde}
\def \g {\gamma}
\def \bi{\bibitem}
\def\a{\alpha }
\def \p {\phi}
\def \ep {\epsilon}
\def \s {\sigma}
\def \gr {\rho}
\def \r {\rho}
\def \d {\delta}
\def \G {\Gamma}
\def \l {\lambda}
\def \m {\mu}
\def \g {\gamma}
\def \n {\nu}
\def \vp {\varphi}
\def \td {\tilde}
\def \x {{\rm x}}
\def \tv {{\rm v}}
\def \four {{1 \ov 4}}
\def \b {\beta}

\def\be{\begin{equation}}
\def\ee{\end{equation}}
\def \ci {\cite}
\def \bi {\bibitem}
\def \la{\label}
\def \f {{\rm f}}

\def \foot {\footnote}

\def \u {u }
\def \v  {v}
\def \t {\tau}
\def\aa{ {\cal A} }
\def \z {{\rm w}}
\def \B {\Sigma}
\def \ww {\td u}
\def \pp {{\rm p}}

\setcounter{section}{0}
\setcounter{subsection}{0}

\newsection{Introduction}

Some major problems  in string cosmology are  the nature
of cosmological singularity, the initial conditions or the possibility of
 a pre-big bang region and the definition observables
(see, e.g., \ci{que,GV,craps,ban,setu,cope,mart} for  recent reviews, discussions
 and references).
Similar  issues  arise in many  {\it time-dependent} backgrounds.
To address them one may start  with examples
which, in contrast to ``realistic'' (FRW or de Sitter) cosmological backgrounds,
are easier to embed to  and  analyze  in string theory.

Much of our understanding of closed string theory is based
on  few examples of  exactly solvable models. Solvability in this context
means that it is possible
to find explicitly the solutions to the classical string equations,
perform a canonical quantization, determine the spectrum of the
Hamiltonian operator and possibly compute some of the simplest
scattering amplitudes.
Many of such  models are constructed by defining
 closed string theory
on a certain background including
non-vanishing p-form
 field strengths and non-trivial dilaton.
 In superstring theory,  such backgrounds
are usually supergravity solutions that
preserve a large number
of space-time supersymmetries.
There are three broad classes of known exactly solvable closed string theories:

(i) Strings on flat space and its various compact and non-compact orbifolds,
as well as models  related to those on
flat-space  by formal
coordinate and T-duality transformations (see, e.g.,  \ci{RT});

(ii) Strings on compact and non-compact (gauged) WZW
backgrounds\  and their  orbifolds. These models have non-vanishing NS-NS
 two-form gauge potential
 and a  dilaton;

(iii) Strings on
 plane-wave  backgrounds
 with non-vanishing NS-NS and/or  R-R form gauge potentials.

A common characteristic of all of the above models is that the
string background does not receive  $\a'$ corrections
in an essential way (for the case  (iii) see  \ci{AK,hs}).
In addition, the  classical string
 equations can 
be solved explicitly and thus  the theory can be
quantized.
The case (iii) also includes the recently found maximally 
supersymmetric  (BFHP) solution \cite{blabla} of IIB string theory  
and the Lorentzian symmetric spaces  \cite{jfogp}.

\subsection{Why plane wave models ?}

Searching for new examples of solvable string theories
in {\it time-dependent} backgrounds
it is natural to study in detail  strings propagating in a generic
plane-wave space-times with the  metric
\be
 ds^2 = 2d\u d\v + A_{ij}(\u) x^i x^j d\u^2  + dx^2\ ,
 \la{gene}
  \ee
where $dx^2$ denotes the standard metric in the Euclidean space  $\bE^d$
and $x\in \bE^d$.
Here the  corresponding \lc   gauge fixed action is quadratic
 and therefore   the
 string equations are
linear in the  transverse coordinates $x^i$ and
 admit at least a formal solution.
 Properties of  strings propagating in
 such plane-wave backgrounds\ have been previously
 investigated, in particular, in \ci{hs,BROOKS,DS}.
 \foot{There is a more general class of pp-wave
models  recently  discussed in  \ci{MM,RUST}
 for which the non-constant component of the   metric
 is no longer  quadratic
 in the $x^i$ coordinates as it was   in the plane-wave case.
 These models may also be solvable in the cases when   they can be  related
 to integrable two-dimensional field theories.
}

The solutions of string theory in the plane-wave background
 for which
the matrix $A=(A_{ij})$ is {\it constant} negative semi-definite
    {\it and } the  dilaton is constant
  was  discussed extensively in the
 literature,  both in the case of   non-vanishing  NS-NS 2-form
   gauge potential \ci{rusts,fork,kirits})
 and in the case of non-vanishing    R-R  p-form gauge potentials \ci{mets,bmn,mt,ruts}.
 Such plane-waves are Lorentzian symmetric spaces \cite{jfogp} and one of them
 is the BFHP solution \ci{blabla}.
 The  \lc gauge fixed string action does not explicitly depend
 on the world-sheet time and describes a collection of free massive bosons
 and fermions with mass matrix $A$.
The most obvious and simplest  generalization
 is to make the ``mass-matrix'' $A$  dependent on $\u$
 and include appropriate non-vanishing
 form field strengths and/or  a  non-constant
 dilaton.
Examples  of such models have already  appeared 
in, e.g.,  \ci{hs,DS,blau,leo,gim,sekk,obers}.

 Some of our  motivations for investigating plane-wave backgrounds
 with non-constant matrix $A$  are the following.


 (1) Strings  in standard cosmological
  backgrounds, like,  for example,
 the metric of  FRW universe with time dependent scale factor
$a(t)$  and dilaton $\p(t)$,
 are difficult  to solve because  (i)   such  backgrounds
  receive $\a'$ corrections  and (ii) classical
string equations are non-linear.
 Plane-wave  backgrounds  also
exhibit effective  time dependence,
but  they are much simpler to study
 than  cosmological models;
 they
may be viewed as  curved-space generalizations of
previously studied
 flat null orbifold  and  null brane  models    \ci{HS,HSS,liu}.
Moreover,  plane wave  backgrounds are related to cosmological backgrounds
 via a Penrose limit procedure in two different ways. As explained in \ci{blau},
taking  Penrose limit along radial direction of
 the FRW universe gives a plane wave metric of the type \rf{gene}.
 Alternatively,  with  any $(d+1)$-dimensional cosmological metric
$ds^2 = - dt^2 + g_{ij}(t) d\x^i d\x^j$  we can associate
 a $(d+2)$-dimensional plane-wave metric of the form (in Rosen coordinates)
$ds^2 = 2 dud\tv  + g_{ij}(u) d\x^i d\x^j$
by performing
a Penrose limit along one  extra dimension. This suggests a
  ``null cosmology'' interpretation for  the  plane-wave  metric.
All this
  may be viewed as an indication
of some  relevance of the study of the
 plane-wave backgrounds for investigations of  string cosmology.
\foot{The well-known differences with standard cosmology
are of course  the presence of supersymmetry and the
absence of  particle creation and  vacuum
polarization for a plane-wave metric. Also, plane-wave metrics have no horizons \ci{hub}
and  the  nature of singularity is
different: all scalar curvature invariants (though not components
of the curvature) vanish for a plane wave
metric while they blow up at a singularity of a cosmological
metric.}


 (2) Plane waves  appear
 as Penrose limits of various p-brane and ``non-conformal''
 generalizations of $AdS_5 \times M^5$  backgrounds\ \ci{blabla,blau}.\foot{Plane-wave backgrounds
may  be viewed   also as Penrose-type limits
of pp-wave models. In fact,  they can be thought of  as  quadratic approximations
originating from  expansions near the
null geodesic  associated with an extremum of the non-trivial component
 of the pp-wave metric.
 For example, in the case with
 $g_{uu}= - m^2 (x_1^2 - x^2_2)^2 $
 considered in \ci{RT}  one is to expand near  the vacuum  line
 $x_1= a, \ x_2 =a$.}
 One particularly simple class of examples,  which will be a special case
 of models discussed below,  is  the Penrose limit \ci{blau,sekk,Rya}
 of the near-horizon geometries of
the  fundamental string  and Dp-brane   backgrounds.
  Study  of these
 examples  may shed light on some aspects of string-string interactions in
 curved space.
 Indeed,
 in the standard  quantization of strings in a given background
 the back-reaction effect of the various string states
on the geometry of space-time is usually ignored.\foot{A justification for
this is that the ten-dimensional Newton constant scales as
$G_{10}\sim g_s^2$ in terms of the string coupling constant $g_s$.
Therefore,  at small string coupling, the effect on the geometry of
a mass $M$ which is measured by $G_{10} M$ is small. This
argument holds\ also  for certain non-perturbative states of
string theory, like D-branes, for which  masses scale as
$M\sim g_s^{-1}$ with the string coupling.} It is obviously
desirable to learn how to take the back-reaction into account.
To model this,  one may
 consider a source string  with  a large mass located at some
point in space-time and  study  quantum strings
propagating in the  background produced by the source.


 (3) Plane-wave models
are also good examples  to study the role
 of non-constant  dilaton in the context of  first-quantized string theory.
  Non-trivial dilaton appears in many  simple
static (e.g., p-brane) and  non-static (e.g., cosmological)  backgrounds.
   Dilaton couples to string
  world sheet  through the 2-d curvature term \ci{FRT},
 and its  role
  is to ensure that  the resulting two-dimensional
  stress tensor is traceless at the quantum level, i.e. that
  the two-dimensional theory is conformal.
 In the \lc gauge this effectively amounts
  to cancelling the anomalous
 contribution of the ``transverse''
  string coordinates to the expectation value of the
\lc Hamiltonian. We shall illustrate this below   on a plane-wave  example.

\subsection{A homogeneous plane wave }
Despite  apparent
simplicity of  generic  plane-wave backgrounds \rf{gene}
the corresponding  string theory model is hard to analyse
explicitly  using analytic methods.
A further important  simplification occurs
if we   consider the following special case of the
 isotropic ($A_{ij} =- \l(u)  \d_{ij}$)
plane-wave  metric  \rf{gene}
\be\la{kk}
ds^2=2dudv- \l(u)  x^2 du^2+ dx^2\ ,
\ \ \ \ \ \ \ \  \l(u) = {k\over u^2} \ , \ \  \ k= \const \ .
\ee
This case  has several remarkable features.
The presence of an apparent  singularity at $u=0$  makes this model an interesting
laboratory for a study  of the issue of
initial singularity in time-dependent backgrounds. This is apparent in 
 Rosen coordinates where the metric \rf{kk} for $k<{1\over4}$
has a simple ``null cosmology'' form 
\be \la{nulk}
ds^2 = 2 dud\tv + a^2(u)  d\x^i d\x^i \ ,\ \ \ \ \ \ \ \ \ \
a(u) = u^\mu\   , \ee 
where $\ \mu= \ha ( 1 - \sqrt{ 1 - 4k})$.
At the same time,  the classical string equations here can be solved explicitly
 in terms of Bessel's functions and thus  the model is under much more 
 analytic control than in  generic plane wave case.
Previous related  discussions were in \ci{hs,BROOKS,DS,sekk}.
In particular, a  different 
class of 4-dimensional plane-wave models with {\it Ricci-flat} metrics 
\rf{gene} with $A_{ij} x^i x^j = {k \ov u^n} ( x^2_1 - x^2_2) $ was 
studied in  
\ci{DS} (for the case of ``sandwich" waves  which are non-trivial in a
finite interval of $u$), where  
the string equations are similarly solved in terms of Bessel's functions. 
In our model the  metric  \rf{kk} has   a negative  definite
 matrix $A_{ij}$ and thus a  non-zero Ricci tensor which is 
``compensated''  by  non-trivial dilaton field.

For different values of the constant $k$
the  plane wave metric  \rf{kk}
is a    Penrose limit
of the FRW metric \ci{blau}, as well
of the near-horizon regions of Dp-brane backgrounds
($k={(7-p)(p-3)\over16}$) and the fundamental string
background ($k = { 3 \ov 16}$)  \ci{blau,sekk,Rya}.
All plane waves
admit a Heisenberg group of isometries. In the isotropic case we have also
invariance under orthogonal rotations in $\bE^d$.
The above plane-wave metric \rf{kk}
is special in that it
admits an additional {\it scaling isometry}
 $
\u \to \ell u  , \ \
 \v \to \ell^{-1} \v$,  which is the same
$SO(1,1)$ Lorentz symmetry
present in the flat-space limit of \rf{gene}.\foot{While this  symmetry
will be formally  broken by  the dilaton,  it will
have important consequences for the solution of string theory
in  which, as we shall see below, the dilaton will play   rather limited role.}
This symmetry implies {\it scale-invariance} of the geometry:
there is  no dimensional parameter like radius, so the components of the curvature
$R_{uiuj}$
get their  canonical dimension 2 from $ 1/u^2$
dependence on coordinates only.
Another consequence of scale invariance is the independence
of \lc Hamiltonian on $p^+=p^u=p_v$
 in contrast to the string model in \ci{mets} based
on the BFHP solution.

It turns out that the space-time associated with
  the plane-wave \rf{kk}
is a conformally flat  {\it homogeneous} Lorentzian space.
The  plane-wave  spacetime with metric
(\ref{kk}) restricted to $u > 0$
 is geodesically incomplete even though
it is homogeneous. This is unlike
the de-Sitter, AdS
and BFHP plane-wave  \ci{blabla}
which are Lorentzian symmetric spaces and smooth.

The presence of the singularity at $u=0$ is in agreement
with the standard argument   \ci{hs} (see also \ci{sing} for a recent review)
that  all the  metrics \rf{gene}
with $A_{ij}(u)$  divergent  at certain value of $u$, say $u=0$,
are singular:
(i)  the geodesic deviation equation is governed  by the curvature components
$R_{uiuj} =  A_{ij}(u)$ and so the tidal forces are infinite at $u=0$;
(ii) time-like geodesics  reach the point $u=0$ in
 finite proper time,
and so the space-time without $u=0$ point is geodesically incomplete.

As we will  see from the  detailed analysis
of the geometrical structure of the space
 associated with \rf{kk}, all time-like geodesics of the region $u>0$
 focus at $u=0$, $x=0$ at  finite
 proper time. Some of these geodesics cannot be extended 
 to the $u<0$ region but some other {\it can}. 
 In Rosen coordinates the line $u=0,$ $x=0$ is
 mapped to a hyperplane located at the origin of the  light-cone ``time''
 $u=0$.  An alternative global construction
 of the homogeneous plane wave space,  based on  group theory, allows
one to identify this hyperplane as a set of
  fixed points  of the scaling symmetry mentioned above.
 In the coordinates where  (\ref{kk}) is conformal to (d+2)-dimensional
Minkowski spacetime, conformal coordinates,
 the homogeneous plane wave space
can be  identified either with the Minkowski space with the singularity
 located at the origin of a light-cone
 coordinate or with  a strip in Minkowski space bounded
 by two hyperplanes located at two opposite values of a light-cone
 coordinate with the singularity  located at the origin  of this
 coordinate.

Our suggestion  is  to   consider the spacetime (\ref{kk}) for the whole
range $(-\infty, + \infty)$
of the light-cone time $u$, and (in the spirit of the ``null cosmology''
interpretation)  view it
as a universe which starts as  flat space at $u\rightarrow -\infty$,
collapses at $u=0$ and then expands again
 to a flat universe at $u\rightarrow \infty$.
The global definition of string theory on this space will then
depend  on choosing appropriate  boundary  conditions
at $u=0$. As already mentioned above,
 the present problem   thus  has a
similarity with  string cosmology set-up, with
the   advantage of the absence of $\a'$-corrections
and a possibility of an exact solution of first-quantized string model.\foot{
In the \lc gauge, the   equations for $\tau$-dependence of
 string transverse coordinates
 $x$  may be interpreted as
 two-dimensional scalar field equations in a  cosmological
  background (with ``scale-invariant'' choice of
 $\tau$-dependent mass term).} 

Indeed, a remarkable feature of this special example \rf{kk}
  is that the corresponding classical string  equations
  can be solved {\it explicitly} (see also \ci{DS})  so that the  theory can be
  canonically  quantized in a straightforward way
(cf. \ci{hs,DS,JN,gim}).
This  allows one to study string dynamics on this
background in much detail.
As in any time-dependent  background, we are dealing with non-stationary
quantum-mechanical problem
and thus the main observables
are not   masses of string states
 but the time-depedent
 expectation values  and transition amplitudes.

\subsection{Structure  of the paper  }

The rest of the paper is organised as follows.

In sections 2 and 3 we explore the geometrical structure of the space-time
associated with  the plane wave metric \rf{kk}.
In section 2  we find the corresponding  Killing vectors
 and demonstrate that this  spacetime is a
homogeneous space.
Then in  section 3  we give its  global construction
  and, starting with the form of the metric in
 Rosen coordinates,
describe  its conformal compactification.
We find also  the equation for  the conformal boundary.

In section 4 we  embed the metric \rf{kk} into string theory,
i.e.  describe   a  general class  of
exact string backgrounds   (defining 2-d conformal theories)
 with metric \rf{kk} and non-trivial dilaton and/or  5-form fluxes.
We  mention generalizations of these models
 which have no singularity at $u=0$ in both the spacetime metric and  dilaton.
We also discuss   relations  between
plane-wave backgrounds  and  cosmological models.

In sections 5  and 6 we turn to quantum  theory
 of particles and strings
in the background of \rf{kk} supported by  $u$-dependent dilaton.
In section 5 we start with solving   the Klein-Gordon equation
for a scalar field, both in Rosen and Brinkmann coordinates.
 For fixed $p^+ \equiv p^u=p_v$ (in the Fourier representation in $u$)
this equation may be interpreted as time-dependent
($\tau \sim u$)
Schr\"odinger equation
representing \lc gauge  dynamics of a
relativistic particle theory.
The \lc Hamiltonian 
 is that of a collection of oscillators with time-dependent
frequency $ \omega = { \sqrt k \ov \tau}$.
We explain how to diagonalize the Hamiltonian in terms of proper set of
creation/annihilation  operators and compute its  expectation values.
We also demonstrate  that the decrease of the effective string coupling 
$e^{\p(u)}$ reduces back reaction near the singularity.

In section 6
we solve the classical (super)string equations in the plane-wave
background (\ref{kk}) in the \lc gauge and then canonically
quantize the theory. We show that, as in the point-particle case,
 the   string  \lc Hamiltonian operator
which describes an infinite collection of oscillators with time-dependent 
frequencies
can be put into a ``diagonal'' form.
We discuss  the role
of the dilaton in maintaining the conformal invariance of  the quantum theory.
We then study the creation of excitation modes as  
string approaches the singularity  at
 $u=0$. The total number of created modes depends on a choice of 
vacuum in  Fock space;
choosing  the usual vacuum of free massless 
particles at infinity leads to a 
divergent result near the singularity, but another choice
gives  no mode creation at all.
We then study string propagation from $u=-\infty $ to $u=+\infty $
 through the singularity and
describe an analytic continuation
prescription that leads to the conclusion that the transition amplitude
from ``in''
 state  at $u=-\infty$ to ``out'' state at $u=\infty$ is trivial,
 i.e. the string can actually pass through the singularity,
 and the final 
string state at $u=\infty $ is the same as the original state at 
$u=-\infty $.

Section 7 contains  some  concluding remarks.

In appendix A we point out that the space  \rf{kk} admits also a
group-manifold structure with \rf{kk} being a left-invariant metric.
In appendix B we write down 
  the homogeneous plane wave metric in
several different  coordinate systems.
In appendix C  we present the solution for geodesics of the 
space  \rf{kk}
and give the expression for the geodesics
that go through the singularity.
In appendix D we discuss  the form of
the Penrose diagram  for  this  plane
wave spacetime.


\newsection{Geometry of a class of  homogeneous\\
 plane-wave  space-times}

In this section we shall first find the isometries  of the
plane-wave metric \rf{kk} and then show that, like the
metric \rf{gene} with $A_{ij}$=const   in \ci{blabla},
 this is a Lorentzian $G/H$ {\it homogeneous  space}.
However,  unlike the BFHP solution  \cite{blabla},
this space is singular despite being homogeneous.
In contrast to  the Euclidean homomogeneous spaces, the
Lorentzian ones can be singular, i.e. geodesically incomplete.
This plane wave space
 is thus a special case in the class of metrics \rf{gene}
with  singular $A_{ij}(u)$ which generically describe
singular spaces
\ci{hs}.

\subsection{Isometries
:\ \  Killing vectors
}
To find the Killing vectors of the plane-wave background
\rf{kk}, we shall begin with more general  plane-wave metric
\begin{equation}
ds^2=2 du dv -\lambda(u)  x^2 du^2+ ds^2(\bE^d)
\label{typ}
\end{equation}
where $x\in \bE^d$ and $\lambda$ is a function of $u$ only. After some
computation, it is easy to see that such a metric (\ref{typ})
admits the following Killing vectors:
\bea T&=&\partial_v
\nonumber
\\
X_i&=& a \partial_i-\partial_u a x_i \partial_v
\nonumber
\\
R_{ij}&=&x_i\partial_j-x_j\partial_i
\eea
where $a(u)$ satisfies
\bea
\partial_u^2a+\lambda  a=0\
\label{kkeq}
\eea
We raise and lower indices $i,j$  in the $\bE^d$ directions with respect
to the
Euclidean metric. The Killing vectors $R_{ij}$ are associated to
orthogonal rotations in $\bE^d$.
Since the  equation for   $a(u)$  is second order,
it is determined up to two arbitrary constants.
Therefore, there are total of
$1 + 2d + \ha d (d-1)  =1 + \ha d (d+3) $ isometries.
For the special choice of
 \be
\la{lam}
\lambda(u)={k\over u^2}  \ee
for  the plane-wave metric (\ref{kk}),
  there are additional
 isometries associated with the scaling of the light-cone coordinates
\be \la{sca}
u\rightarrow \ell u\ , \ \ \ \ \ \ \ \ \
v\rightarrow \ell^{-1} v   \   .
\ee
The
special choice  of $\lambda$ allows us also
to determine the function $a(u)$
explicitly.
 The solutions of  (\ref{kkeq}) are qualitatively  different
in the three cases: (i)  $0<k<{1\over4}$,
(ii) $k>{1\over4}$
and (iii) $k={1\over4}$.

The
 Killing vectors of the metric (\ref{kk}) for $0<k<{1\over4}$ are
\bea
T&=&\partial_v
\nonumber
\\
X_i&=& u^\nu \partial_i-\nu u^{\nu-1} x_i
\partial_v
\nonumber
\\
\tilde X_i&=& u^{1-\nu} \partial_i-(1-\nu) u^{-\nu} x_i
\partial_v
\nonumber
\\
D&=&u \partial_u -v \partial_v
\nonumber
\\
R_{ij}&=&x_i \partial_j- x_j \partial_i~, \la{nuin}
\eea
where $D$ is the generator associated with the scaling symmetry and
\be \la{saq}
\nu\equiv {1+\sqrt{1-4k}\over 2}\ .  \ee
 We shall mostly focus on the
case where $0<k<{1\over4}$.
 It is this range that appears
 in the Penrose limits
of the
 FRW metrics, Dp-branes ($3<p<7$)  and
 of the fundamental  string background \ci{blau,sekk,Rya}.
In the latter case  $k={3\over 16}$  and thus  $\nu={3\over4}$.

It is straightforward  to compute the Lie bracket algebra of the Killing vectors
for $0<k<{1\over4}$ to
find
that the non-vanishing commutators are as follows:
\begin{eqnarray}
[D,T]&=&T
\nonumber
\\
\left[ X_i , \tilde X_j\right ]&=&(2\nu-1) \delta_{ij} T
\nonumber
\\
\left[ D, X_i \right]&=&\nu X_i
\nonumber
\\
\left[ D,\tilde X_i \right]&=&(1-\nu) \tilde X_i
\nonumber
\\
\left[ R_{ij}, X_k \right]&=& X_i\delta_{jk}-X_j\delta_{ik}
\nonumber
\\
\left[ R_{ij},\tilde X_k \right]&=& \tilde X_i\delta_{jk}-\tilde
X_j\delta_{ik}
\nonumber
\\
\left[ R_{ij}, R_{kl} \right]&=& \delta_{jk} R_{il}-\delta_{ik}
R_{jl}
+\delta_{il} R_{jk}-\delta_{jl} R_{ik}~.
\la{alg}
\end{eqnarray}
The algebra of isometries of the metric in (\ref{kk}) is  similar
to that
of the BFHP  plane-wave.
The metric \rf{typ}  with $\lambda={k \ov u^2} $
(i.e. \rf{kk})  and that of BFHP plane-wave
 have the same number of isometries. The two isometry
algebras also contain
a Heisenberg subalgebra with $n$ position and $n$ momentum
generators. They both  also have an external generator of automorphims
which however acts differently on the rest of generators in the two cases.
In the BFHP case, the external automorphism of the Heisenberg
algebra rotates the $X_i$ Killing vectors to $\tilde X_i$ ones
and vice versa. It is a compact generator. In the case of
\rf{lam}
 we are investigating here
the external automorphism $D$
rotates $X_i$ generators to themselves and acts similarly on
the $\tilde X_i$
generators. It also rotates the central generator of the Heisenberg algebra $T$.
It is a non-compact generator.

Next, let us   consider the case where $k>{1\over4}$. The Killing vectors
 $T,D$ and $R_{ij}$ are the same as those
in the case of the metric (\ref{kk}) with $k<{1\over4}$ above.
Solving (\ref{kkeq}) for real $a(u)$,
we find that the Killing vectors $X_i$ and $\tilde X_i$
 can
be expressed as
\bea
X_i&=& u^{{1\over2}} \cos(\gamma\ln u)\partial_i - x_i u^{-{1\over2}} \big({1\over2}
\cos(\gamma \ln u)-\gamma \sin(\gamma \ln u) \big)
\partial_v
\nonumber
\\
\tilde X_i&=& u^{{1\over2}}  \sin(\gamma\ln u)\partial_i-
 x_i u^{-{1\over2}} \big({1\over2}
\sin(\gamma \ln u)+\gamma \cos(\gamma \ln u) \big)
\partial_v\ , \la{tamm}
\eea
where $\gamma={\sqrt{k-\four}}$.
The Lie bracket algebra of the Killing vectors in this case is
\begin{eqnarray}
[D,T]&=&T
\nonumber
\\
\left[ X_i , \tilde X_j\right ]&=&\gamma \delta_{ij} T
\nonumber
\\
\left[ D, X_i \right]&=&{1\over2} X_i-\gamma \tilde X_i
\nonumber
\\
\left[ D,\tilde X_i \right]&=&{1\over2} \tilde X_i+\gamma X_i\ .
\end{eqnarray}
The  commutators involving the generators $R_{ij}$ are  the same as in \rf{alg}.
 $D$ is again an outer automorphism of the Heisenberg algebra.

Finally, in the  $k={1\over4}$ case,
the Killing vectors
 $T,D$ and $R_{ij}$ are the same as for
 $k<{1\over4}$ and $k>{1\over4}$ above.
 To find the analogue
of the Killing vectors $X_i$ and $\tilde X_i$, we observe that the
two independent
solutions of
(\ref{kkeq}) for $k>{1\over4}$ are
$a=u^{{1\over2}}$ and $a=u^{{1\over2}} \ln u$.
 Then
\bea
X_i&=& u^{{1\over2}} \partial_i - {1\over2} x_i u^{-{1\over2}}
\partial_v
\nonumber
\\
\tilde X_i&=& u^{{1\over2}} \ln u  \partial_i-
 x_i u^{-{1\over2}} \big({1\over2}
 \ln u+1 \big)
\partial_v\ .
\eea
The Lie bracket algebra of the Killing vectors in this case is
\begin{eqnarray}
[D,T]&=&T
\nonumber
\\
\left[ X_i , \tilde X_j\right ]&=&- \delta_{ij} T
\nonumber
\\
\left[ D, X_i \right]&=&{1\over2} X_i
\nonumber
\\
\left[ D,\tilde X_i \right]&=&{1\over2} \tilde X_i+ X_i\ .
\end{eqnarray}
The rest of the algebra is again as in \rf{alg}.

In all the above cases the number of Killing
vectors   is   $2+ {1\over2} d (d+3)$.
Therefore,   a 10-dimensional
 metric
(\ref{kk}),  $d=8$, admits a group of isometries
of dimension fourty six.

\subsection{Homogeneous space structure
}

Let us now  proceed to show that the space-time of the plane-wave (\ref{kk})
is a homogeneous Lorentzian space.
It is
 worth  starting with summarizing some of the
elementary properties of homogeneous spaces. Let $G/H$ be a homogeneous
space. The Lie algebra ${\bf g}$ of the group $G$ decomposes as
${\bf g}={\bf h}+{\bf m}$, where ${\bf h}$ is the Lie algebra of the
subgroup $H$ and ${\bf m}$ is the rest which is identified with the
tangent space of $G/H$ at the origin, with
$[{\bf h}, {\bf h}]\subset {\bf h}$
and $[{\bf m}, {\bf m}]\subset {\bf h}+{\bf m}$.
If $[{\bf h}, {\bf m}]\subset {\bf m}$,
then $G/H$ is called {\it reductive};
$G/H$ space is called {\it symmetric} if it is reductive and
$[{\bf m}, {\bf m}]\subset {\bf h}$.

Let $G/H$ be a reductive homogeneous space.
Consider a local section  $s$ of $G\rightarrow G/H$. Then we write
$s^{-1} ds= e+\omega=e^m t_m+\omega^a t_a$, where $\{t_m\}$ is a basis in ${\bf m}$
and $\{t_a\}$ is a basis in ${\bf h}$.
As  is well known,   $e$ is the left-invariant frame
on $G/H$ and $\omega$ is the canonical
 connection. The structure equations are
\bea
{\cal R}:&=&d\omega+\omega\wedge \omega=-e\wedge e|_{\bf h}\nonumber
\\
{\cal  T}:&=&De:=de+\omega\wedge e-e\wedge \omega=-e\wedge e|_{\bf m}
\eea
where ${\cal R}$ is the curvature of the canonical connection and $\cal T$ is the torsion.
The torsion vanishes for symmetric spaces. A metric is invariant under the left
action of $G$ on $G/H$ provided
 it is associated with an $H$-invariant
quadratic form on ${\bf m}$, i.e.  a quadratic form $B$ such that
$B([t_a, t_m], t_n)+ B(t_m, [t_a, t_n])=0$. The metric on $G/H$ is constructed
by using the quadratic form $B$ and the invariant frame $e$.

To demonstrate that the space-time with the
plane-wave metric (\ref{kk}) is homogeneous, we have to identify
the Lie algebra of the subgroup $H$ and compute the frame $e$ and
canonical connection $\omega$. We shall begin with the case $0<k<{1\over4}$.
 The Lie algebra ${\bf g}$ of the group
$G$ is generated by the elements $X_i, \tilde X_i, D, T$ which are
 taken to satisfy
the Lie commutators
\bea
\left[X_i, \tilde X_j\right]&=&- (2\nu-1) \delta_{ij}T , ~~~
~~~~~~~\left[D, T\right]=-T
\nonumber
\\
\left[D, X_i\right]&=&-\nu X_i ,~~~~~~~~~~~~~~~
\left[D, \tilde X_i\right]=-(1-\nu) \tilde X_i\ .
\label{lierel}
\eea
We have used the same letters to denote the generators of ${\bf g}$ and
the Killing vectors of (\ref{kk}). Note the sign difference in the
commutator of the Killing vectors  and that of the generators of ${\bf g}$
(cf. \rf{alg}).

Next,  we define  ${\bf h}=<\alpha X_i+\beta  \tilde X_i>$,
where $<...>$ denotes a linear span,
i.e.  $h_i=\alpha X_i+\beta \tilde X_i$,    for some
non-vanishing constants $\alpha,\beta$ which we shall specify later.
  Then we define ${\bf m}=<m_i, D, T>$,
where we take
$m_i=-[h_i, D]$. The non-vanishing commutators of ${\bf g}$ in terms
of this new basis are
\bea
\left[h_i, m_j\right]&=& \delta_{ij}T,~~~~~
~~~\left[D, T\right]=-T \nonumber
\\
\left[D,h_i\right]&=&m_i,
~~~~~~~~\left[D, m_i\right]=-k h_i- m_i\ ,
\label{streq}
\eea
where we have chosen the normalization $\alpha\beta=-{1\over 1-4k}$.
It is clear that this decomposition of ${\bf g}$ is reductive.
Because the second term in the left-hand-side of the last
commutator is not zero, this decomposition of ${\bf g}$ is associated
with a homogeneous space rather than a symmetric one.
Choose the local  section $s$  as
\begin{equation}
s=e^{{x^2\over 2}T} e^{x^i h_i} e^{x^i m_i} e^{w D} e^{v T}
\end{equation}
and write $e=e^T T+e^D D+e^i m_i$
and $\omega=\omega^i h_i$, where $v, x^i, w$ are local coordinates.
Then we find that
\bea
e^T&=& e^{w} dv-{k\over 2} x^2 dw, ~~~ e^i=dx^i, ~~~ e^D=dw
\nonumber\\
\omega^i&=&dx^i-k x^i dw\ .
\label{fram}
\eea
It is easy to verify that the structure equations for the homogeneous
space   are satisfied using (\ref{streq}).
The curvature of the canonical connection
is ${\cal R}^i=-k dx^i\wedge dw$.
An $H$-invariant quadratic form on ${\bf m}$ can be constructed by setting
\be
B(D, T)=1, ~~~~~~~~~~~~~~~~B(m_i, m_j)=\delta_{ij}~.
\label{qform}
\ee
The associated invariant metric on   $G/H$ is
\be
ds^2=2 e^D e^T+ \delta_{ij} e^i e^j
\ee
or explicitly
\be\label{newcor}
ds^2=dw (2 e^{w} dv-k x^2 dw)+ ds^2(\bE^d)\ .
\ee
Changing the coordinates by setting $u=e^{w}$, we recover the
metric (\ref{kk}). We conclude, therefore,
that the space-time  corresponding to  the plane-wave metric  (\ref{kk})
is a {\it homogeneous} space.

Next,  consider the case of  $k>{1\over4}$. The Lie algebra ${\bf g}$ of
the group $G$ is again generated by $<T,D,X_i, \tilde X_i>$ satisfying
the Lie bracket relations
\bea
\left[X_i, \tilde X_j\right]&=&\gamma T\delta_{ij},  ~~~~~~~~~~\left[D, T\right]=T
\nonumber
\\
\left[D, X_i\right]&=&-{1\over2} X_i +\gamma \tilde X_i, ~~~~~~~~~~~
\left[D, \tilde X_i\right]=-{1\over2} \tilde X_i-\gamma X_i\ .
\eea
i.e. those for  the Killing vectors with a sign changed
(where again  $\gamma={\sqrt{k-\four}}$).
Then we identify the generators $h_i$ of the Lie algebra ${\bf h}$ of $H$ as
$h_i=\alpha X_i+\beta \tilde X_i$. Writing ${\bf g}={\bf h}+{\bf m}$, the tangent
space at the origin of the coset
is then
 ${\bf m}=<T,D, m_i>$,  where $m_i=-[h_i, D]$=
$-({\alpha\over2}+\gamma
\beta)X_i-({\beta\over2}-\alpha\gamma)\tilde X_i$. The Lie
bracket relations  of ${\bf g}$ corresponding  to this
decomposition can be rewritten as in (\ref{streq}) provided we choose
$\gamma^2 (\alpha^2+\beta^2)=1$ and allow $k>1/4$.
Because of the last commutator in (\ref{streq}), the space is again
 homogeneous rather than symmetric.
Using the frame (\ref{fram}) and the quadratic form  (\ref{qform}) and
 changing the coordinates
as in the case $k<{1\over4}$ above, we conclude that
 the space-time with metric
(\ref{kk}) for $k>{1\over4}$ is homogeneous.  The canonical connection is
$\omega^i=dx^i-kx^i dw$ and the canonical
curvature is ${\cal R}^i=-k dx^i\wedge dw$.

It remains to investigate the case of  $k={1\over4}$.
 The Lie algebra ${\bf g}$ of
the group $G$ is again generated by $<T,D,X_i, \tilde X_i>$ satisfying
\bea
\left[X_i, \tilde X_j\right]&=& \delta_{ij}T,  ~~~~~~~~~~\left[D, T\right]=T
\nonumber
\\
\left[D, X_i\right]&=&-{1\over2} X_i,  ~~~~~~~~
\left[D, \tilde X_i\right]=-{1\over2} \tilde X_i- X_i\ ,
\eea
i.e. the commutation relations
 of the Killing vectors with a sign changed.
Then we identify the generators $h_i$ of the Lie algebra ${\bf h}$ of $H$ as
$h_i=\alpha X_i+\beta \tilde X_i$. Writing ${\bf g}={\bf h}+{\bf m}$, the tangent
space at the origin of the coset is ${\bf m}=<T,D, m_i>$,  where
$m_i=-[h_i, D]=-({\alpha\over2}+ \beta)X_i-{\beta\over2}\tilde X_i$.
The brackets of ${\bf g}$ according to this decomposition can be written as
in in (\ref{streq}) provided we choose $\beta^2=1$ and set $k=1/4$.
Again, because of the last commutator in (\ref{streq}), the space is
 homogeneous rather than symmetric.
Using the frame (\ref{fram}) and the quadratic form  (\ref{qform}) and
after changing coordinates
as in the other two cases  above, we conclude
that the space-time with metric
(\ref{kk}) for $k={1\over4}$ is homogeneous.
 The canonical connection is
$\omega^i=dx^i-{1\over4}x^i dw$ and the canonical
curvature is ${\cal R}^i=-{1\over4} dx^i\wedge dw$.

\newsection{Global structure}

\subsection{Group--theoretic method}

Homogeneous Lorentzian spaces are not always geodesically
complete. The simplest example of that is two-dimensional
Minkowski space-time with one light line removed. Since the
structure of the homogeneous plane wave (\ref{kk}) is similar to
the Minkowski space-time,
we shall start with reviewing  the Minkowski space
example in some  detail. Write the two-dimensional Minkowski
metric in light-cone coordinates as
$ds^2=2 du dv$. Suppose we
remove from the space the light line located at $u=0$. The
remaining space is topologically $\bR^*\times \bR$,
where   $\bR^*=\bR-\{0\}$.  In addition
$\bR^*\times \bR$ admits the following group action of
isometries: $(u, v)\rightarrow (\ell u, \ell^{-1} v+a)$ where
$\ell\in \bR^*$ and $a\in \bR$. The group is the
semidirect product of $\bR^*$ with the group of translations
$\bR$, i.e.  $\bR\rtimes  \bR^*$. It is easy to verify that this group
action is transitive and the little group at every point is the
identity. So $\bR^*\rtimes  \bR$ is, in fact,  a group but clearly
incomplete with respect to the Minkowski metric. Of course we can
add back  the light line  we have removed  to recover the whole
two-dimensional Minkowski space which is complete. This
light line can be thought of as a special orbit $\bR=
 (\bR^*\rtimes \bR)/\bR^*$ of the group
 $\bR\rtimes \bR^*$ acting now on the
two-dimensional Minkowski space.

To describe  the homogeneous structure of the plane wave (\ref{kk})
with $0<k<{1\over4}$
 globally, we consider the group multiplication on
 $G^+=\bR^+\times \bR^d \times \bR^d\times \bR$ as follows:
 \bea
 (\ell_1, x_1, y_1, v_1)(\ell_2, x_2, y_2, v_2)&=&
 \nonumber\\
 \Bigl(\ell_1 \ell_2, \ell_1^{-\nu} x_2+x_1, \ell_1^{-(1-\nu)} y_2+y_1,
  \ell_1^{-1} v_2
&+&v_1-
 {2\nu-1\over 2} ( \ell_1^{-(1-\nu)} x_1 y_2- \ell_1^{-\nu}x_2 y_1)\Bigr)
 \nonumber
\eea
where $\bR^+= \{r\in\bR: r>0\}$ and $\nu={1\over2}(1 +
\sqrt{1-4k})$. It is straightforward to see that the above multiplication
is associative with  identity  $(1,0,0,0)$ and
 inverse  $(\ell, x,y, v)^{-1}=(\ell^{-1}, -\ell^\nu x, -\ell^{1-\nu} y,
-\ell v)$. The Lie algebra of $G^+$ is $\bR\oplus \bR^d\oplus \bR^d\oplus \bR$.
 The left-invariant frame is
\bea
e^D&=& \ell^{-1} d\ell
\nonumber\\
e^{X_i}&=& \ell^\nu dx^i
\nonumber\\
e^{\tilde X_i}&=& \ell^{1-\nu} dy^i
\nonumber\\
e^T&=&\ell dv-{2\nu-1\over2} \ell (y_i dx^i-x_i dy^i)
\eea
We choose the subgroup $H=\bR^d=\{(1, q_1 (1-\nu) x, q_2 \nu x, 0)\in G^+\}$,
where $q_1 q_2=1$.
This normalization for $q_1, q_2$ has been chosen so that the left-invariant
frame satisfies the Maurier-Cartan equations associated with the Lie algebra
relations (\ref{lierel}).
The left invariant one-forms along the subgroup are
$e^{h_i}=p (1-\nu) e^{X_i}+ q \nu e^{\tilde X_i}$.
A global section of the coset space $G^+/H$ is $s=(\ell, q_1 x, q_2 x, v)$.
The homogeneous metric associated with the quadratic form $B$  (\ref{qform}) is
\begin{equation}
ds^2=2 d\ell dv+ (q_1 \ell^\nu+q_2 \ell^{1-\nu})^2 dx^idx^i\ .
\label{metros}
\end{equation}
As we shall demonstrate in the next section, this is
the form of the metric ({\ref{kk}) in Rosen coordinates.

 To describe the whole plane-wave space-time, we
define the following group  multiplication
on  $G^*=\bR^*\times \bR^d \times \bR^d\times \bR$:
$$
 (\ell_1, x_1, y_1, v_1)(\ell_2, x_2, y_2, v_2)
 =
 \Bigl(\ell_1 \ell_2, |\ell_1|^{-\nu} x_2+x_1, |\ell_1|^{-(1-\nu)} y_2+y_1,
  $$
\be
|\ell_1|^{-1} v_1+v_2-
 {2\nu-1\over2} (|\ell|^{-(1-\nu)} x_1 y_2-|\ell|^{-\nu} x_2 y_1)\Bigr)~.
 \label{gract}
 \ee
 Clearly,  the group $G^*$ is disconnected.
 The left-invariant one-forms are
 \bea
e^D&=& \ell^{-1} d\ell
\nonumber\\
e^{X_i}&=& |\ell|^\nu dx^i
\nonumber\\
e^{\tilde X_i}&=& |\ell|^{1-\nu} dy^i
\nonumber\\
e^T&=&|\ell| dv-{2\nu-1\over2} |\ell| (y_i dx^i-x_i dy^i)
\eea
We also identify the $H=\bR^d$ subgroup of $G^*$ as in the case $G^+$ above, and
the plane wave space-time can be identified as $G^*/H$. This is a disconnected
space. Topologically,  it is to be identified with $\bR^d\times \bR^2$
after removing a null hyperplane located at $\ell=0$. The homogeneous metric is
\begin{equation}
ds^2=2 \ell^{-1} |\ell| d\ell dv+ (q_1 |\ell|^\nu+q_2 |\ell|^{1-\nu})^2 dx_i dx_i \ .
\label{whomet}
\end{equation}
It remains  to investigate the possibility of gluing the null hyperplane
back into the space-time. This can be modeled by taking $G^*$ to act
on $\bR\times \bR^d\times \bR^d \times \bR/H$ using (\ref{gract}),
where $H$ acts diagonally on the
$\bR^d$ subspaces with weights $q_1, q_2$.
 The hyperplane is the  orbit $(0, x,y,v)H$ under $G^*$.
This will extend the plane-wave space-time into the whole $\bR^d \times \bR^2$ space.
The homogeneous metric (\ref{whomet}) is,  however,  singular along the
null hyperplane.

\subsection{Rosen coordinates and
conformal compactification}

To conformally compactify the homogeneous plane-wave space
(\ref{kk}), it is convenient to use  Rosen coordinates as in
(\ref{metros}).  It  will be  useful later to describe the
transformation of a
 more general class plane-wave metrics  \rf{typ},
i.e.
\begin{equation}
ds^2=2 du dv-\lambda(u)  x^2 du^2+ dx^i dx^i \ . \label{typl}
\end{equation}
from Brinkmann to Rosen coordinates.
The   required change of coordinates  is $(u,v,x) \to
(u,\tv,\x)$, where
\be \la{giv}
v= \tv + \ha  h(u) \x^i \x^i\ , \
\ \ \ \ \ x^i = a(u) \x^i \ , \ \ \ \ \ \  h= - a a'\ , \ \ \ \ \
\ a''(u)= - \l (u) a(u)  ,
\ee
which lead to
\beq \la{opse} ds^2 = 2d\u
d\tv  +  a^2 (\u)  d\x^i d\x^i \ .
\ee
Note that this metric and
thus \rf{typl} are {\it conformally flat}.

In the special  case of \rf{kk} with $\l={k\ov u^2}$ and assuming that
$u>0$, we have
$a''= - {k\ov u^2}a$
\be \la{yyy}
a= q_1
u^{\n} +  q_2  u^{1-\n}  \  , \ \ \ \ \ \n  = \ha(1 + \sqrt{  1
-4 k} )\  ,  \ \ \ \ \ \  0 < k < {1 \ov 4}
\ee
and
\be \la{yyyy}
a=  u^{1/2}(q_1   + q_2 \ln u ) \  , \ \ \ \ \ \ \ \ \ \ \ k= { 1
\ov 4} \  .
\ee
The singularity at $u=0, x=0$ in the Brinkmann coordinates is mapped
to the hyperplane $u=0$ in the Rosen coordinates
(this can also be
seen by looking at  the location where
  timelike geodesics end, cf. Appendix C). 
For $k > { 1 \ov 4}$ and $u>0$,
one gets an
oscillating solution (cf. \rf{tamm})
 \be
  a =   u^{1/2}   [ q_1 \cos ( \g \ln u)
 + q_2 \sin ( \g \ln u) ] \   ,  \ \ \ \ \ \ \   \ \g= \sqrt{ k -
{\textstyle{1\ov 4}} }  \ .
\ee
In what follows  we  shall focus
in the case where $0<k<{1\over4}$, i.e. $\ha  < \nu < 1$.

 For the region
$u<0$ and assuming that $\l(u)=\l(-u)$, we set $\ww=-u$ and write (\ref{typl}) 
as
\begin{equation}
ds^2=-2 d\ww dv-\lambda(\ww)  x^2 d\ww^2+ dx^i dx^i \ . \label{typla}
\end{equation}
Then  performing  the analog of the transformation \rf{giv} as 
\be \la{giva}
v= \tv + \ha
h(\ww) \x^i \x^i\ , \ \ \ \ \ \ x^i = a(\ww) \x^i \ , \ \ \ \ \ \  h=
a a'\ , \ \ \ \ \ \ a''(\ww)= - \l (\ww) a(\ww)  ,
\ee
we find that 
\beq
\la{opsen} ds^2 = -2d\ww d\tv  +  a^2 (\ww)  d\x^i d\x^i \ .
\ee
In terms of $u$ we thus  have 
\beq \la{opsea} ds^2 = 2du d\tv  +  a^2
(-u)  d\x^i d\x^i \ .
\ee
Observe that for $k<{1\over4}$ the
metric for $-\infty<u<+\infty$ takes the form of (\ref{whomet})
after substituting $\tv\rightarrow -\tv$ for the region $u<0$.
The transformation from Brinkmann to Rosen coordinates described
above (\ref{giv}) and
(\ref{giva}) can be inverted away from the region that $a(u)$ vanishes.

There is  a freedom in the choice of a transformation
 from the Brinkmann
to Rosen coordinates parameterized by  two integration
constants $q_1$ and $q_2$  because the equation 
for $a(u)$ in \rf{giv} is second order. 
For the homogeneous plane wave (\ref{kk}),
the spacetime in Brinkmann coordinates is diffeomorphic to that
in Rosen coordinates away from $u=0$.


We shall use this freedom to patch 
 the regions $u>0$ and $u<0$ in the
context of a conformal compactification  below.
The conformal compactification
to Einstein static universe is most easily
described for the case $q_1=0$ and $q_2\not=0$, 
so in what follows we shall concentrate on  this case. The rest
of the cases will be presented in Appendix B.
After a rescaling of the $\x$ coordinates
we can set $q_2=1$ in which case we find 
\be \la{simp}
a(u)=u^{1-\nu} \ .
\ee
It is
another remarkable property of the model \rf{kk} that it takes
such a simple form in the Rosen coordinates.

We shall further express (\ref{opse}) and
(\ref{opsea}) in coordinates where the metric 
is explicitly  conformal to Minkowski
metric (we shall refer to them as ``conformal'' coordinates).
In particular, we find
\be \la{zes}
ds^2=\  \B({\z}) \ (2d{\z}  d\tv+d\x^i d\x^i)\ , \ 
\ee
\be \  \B(\z) = a^2(u) \
,\ \ \ \  \ \ \ \ \ \ \ \    \ d\z= {du \ov a^2(u)}\ . \ee
In the  case of the plane wave with $a(u)$ in \rf{simp}
  (i.e. for $q_1=0,  q_2=1$)  we  find for $u >0$
\be
\z=  { 1 \ov 2 \n -1}  {  u^{2 \n -1} } \ , \
\ \     0 < \z < \infty
 \ , \ \ \         \ \
\B(\z) = [(2\n-1) \z]^{ { 2- 2 \nu \ov 2 \n -1}}    \ .       \la{szza}
\ee
Note that the plane wave metric \rf{gene}  is conformally flat for 
all isotropic plane waves, i.e. having $A_{ij} = \lambda(u) \delta_{ij}$. 
In the present case of \rf{kk} with $\lambda (u)=k/u^2$,
both $a(u)$ in \rf{opse}  and $\Sigma (\z)$ in \rf{zes} 
are simply powers of their arguments. This
 leads to an  extra scaling symmetry -- rescaling  $u$ or $\z $ combined
with an 
appropriate 
rescaling of  other coordinates 
is an isometry of the metric.

The analogous 
 transformations for remaining values of $(q_1,q_2)$ 
 can be thought of as  different conformal  embeddings
of the homogeneous plane wave into (d+2)-dimensional Minkowski
spacetime, see Appendix B. 
In \rf{szza},
 the singularity
at $u=0$ in Rosen coordinates is mapped to $\z=0$ in the Minkowski
coordinates and the region $u=+\infty$ in Rosen coordinates is mapped
to $\z=+\infty$ in conformal (Minkowski) coordinates, i.e. 
 the image of the  $u>0$ spacetime in
Rosen coordinates is the $\z > 0$  part of
the  Minkowski space 
 with the singularity
located at $\z=0$.

For  $u < 0$,  we find
\be \z= - { 1 \ov 2 \nu -1}  {  (-u)^{2 \n -1} }
\ , \ \ \     -\infty < \z < 0
 \ , \ \ \         \ \
\B(\z) = [- (2 \n -1) \z ]^{ { 2- 2 \nu \ov 2 \n -1}}  
\ .  \label{szzai}
\ee
Again,   the singularity at
$u=0$ in Rosen coordinates is mapped to $\z=0$ 
and the  $u=-\infty$ region --  to $\z=-\infty$ region 
in conformal  coordinates. 

The ``null cosmology'' interpretation of the homogeneous plane wave 
spacetime with $ -\infty < u < \infty$  (see Section 4.2)
is thus of 
 a universe that undergoes collapse to  the singularity and then expands
in light-cone time.

To summarize, the image of the homogeneous
plane wave space  with $-\infty <u<+\infty$ under the transformations
(\ref{szza}) and (\ref{szzai}) is {\it conformal to the 
 Minkowski space with the
hyperplane $\z=0$ removed}. 
As a result,  the Penrose diagram
can be most easily constructed  in these coordinates.

To do the conformal
compactification of the homogeneous plane wave
 to the Einstein static universe
we are to  further compactify the Minkowski metric. For this we use
 (\ref{szza}) and  the following change of coordinates. First,  set
$\z= t+y$ and $\tv=\ha (-t+y)$ and then write the Euclidean space
metric $ds^2(\bE^{d+1})=dy^2+ d\x^i d\x^i$ in angular coordinates
$ds^2(\bE^{d+1})=dr^2+r^2 ds^2(S^d)$, where $ds^2(S^d)=
d\theta^2+\sin^2\theta\  ds^2(S^{d-1})$. Next,
 write $v'=t+r$ and $w'=t-r$ and in addition $\tan \rho=v'$
and $\tan \sigma=w'$. After all these transformations, and
setting  $ \varphi= \r+  \s, \ \psi= \r - \s$,  we find that
\begin{equation}
ds^2= C(\varphi,\psi,\theta)
 \big[-d\varphi^2 + d\psi^2 +  \sin^2 \psi\
(  d\theta^2+\sin^2\theta\  ds^2(S^{d-1}) )  \big] \ , \la{poi}
 \end{equation}
where $ 0 \leq \psi \leq \pi$ and $0 \leq \theta \leq \pi$,
\be
 C(\varphi,\psi,\theta)
= (4\nu-2)^{2-2\nu\over 2\nu-1} {  (\sin \varphi +\sin\psi \cos\theta\bigr)^ {2-2\nu\over
2\nu-1} \over \bigl(\cos\varphi +\cos\psi \bigr)^{2\nu\over
2\nu-1}} \ , 
\ee
and we have used that
$$
\z=2 {\sin\varphi+\sin\psi\cos\theta\over \cos\varphi+\cos\psi} \ . 
$$
 The  conformal boundary is at the points where
the conformal factor $C$ is infinite, i.e.  the equation for the
conformal boundary is
 \begin{equation}
\cos\varphi +\cos\psi=0 \ . \la{opl}
\end{equation}
The singularity at $\z=0$ is where  $C=0$, which implies
\begin{equation}
\sin \varphi +\sin\psi  \cos\theta=0 \ . \la{oopl}
\end{equation}
The conformal boundary of the homogeneous plane wave is thus that
of the Minkowski spacetime, i.e.
it  is generically $d+1$ dimensional but it
has special points. If  $\sin\psi=0$, the conformal boundary
collapses to points located at $\cos\varphi\pm 1=0$.

In general,  the singularity is described by a (d+1)-dimensional
subspace in the Einstein static universe and it is mostly
spacelike. However,  for $\cos\theta=\pm1$, it becomes a null line.
Since crossing the singularity changes the sign of $\z$, we associate
the region (II) above the singularity with $\z>0$ and the region (I) below the
singularity with $\z<0$.
 The Penrose diagram of the
homogeneous plane wave can be drawn in 3-dimensional space 
 with
coordinates $\varphi, \psi, \theta$. A generic point in such a
diagram is a $(d-1)$-sphere. It is more instructive though  to
draw the standard Penrose diagrams in two dimensions with coordinates
$(\psi, \varphi)$ parameterized with the angle $\theta$. Again,  a
generic point in these diagrams is a $(d-1)$-sphere. There are
infinitely  many such diagrams,  but it turns out that most of them
have similar properties regarding the relative locations of the
singularity and the conformal boundary. Plots of 
various Penrose diagrams are given in appendix D.


\newsection{
Homogeneous plane-wave metrics\\ 
\ \ as string-theory backgrounds}

\subsection{The metric-dilaton   model}
To embed the metric \rf{kk} or,  more generally, \rf{typ}
into string theory
we need to compensate its non-zero Ricci tensor $R_{uu}=  \l(u) d$
by a  contribution of other background fields.
The simplest option is to include a  $u$-dependent dilaton field.
 The resulting background
 may be viewed as a plane-wave analog of
a metric-dilaton cosmological
 background
(see below).
The most symmetric  ansatz is thus given by
\beq \la{best}
ds^2 = 2d\u d\v - \l (\u) x^2 d\u^2 + dx^i  dx^i\ ,\ \ \ \ \ \ \ \ \
\phi  = \phi(\u) \ ,
\eeq
where $
i=1,...,d
$  and $d \leq 8$. In what follows,  we do not
 mention the  additional free ``spectator'' coordinates
which should complement $x^i$ to  8 transverse bosonic coordinates because
they do not affect our arguments.
The (exact) conformal invariance condition
 $R_{\m\n} = - 2 D_\m D_\n \p,$
i.e. $R_{\u\u} = - \ha \del^2_x K = - 2 \del^2_\u \p$
with $K= - \l(\u) x^2$
 then implies  (prime is derivative over $u$)
\be \la{kop}
 \phi''(u)  = - {d\over 2}\  \l(\u) \ .
\ee
In general, one  simple solution   is
\be \l=\l_0=\const > 0 \ ,  \   \ \ \ \
\phi = \p_0 -  m ^2  u^2 \ , \ \ \ \ \  m^2 = \ha d \l_0  \ .   \ee
A remarkable feature of the corresponding string model  is
that the effective
string coupling $e^\p = g_0 e^{ - m^2  u^2}$
 is small everywhere
if it small at $u=0$, i.e. if  $g_0= e^{\p_0} \ll 1 $.
As in the R-R 5-form model of \ci{mets} based on the
BFHP solution \cite{blabla}
here the bosonic  string modes with momentum $p^u \not =0$
  have mass  $\sqrt \l_0  p^u$
  (the fermionic modes  remain massless)\foot{The role of the
  dilaton $ \sim u^2$ (i.e. $\sim \tau^2$ in the \lc gauge)
   will be to contribute a constant
  term to the stress tensor  that will cancel
  the conformal  anomaly coming from  the  bosonic mass terms
  (in the R-R model of \ci{mets}  this anomaly was cancelled by the
  fermionic  mass term contribution).
Let us note that in general any  sigma model satisfying
$R_{\m\n} = D_\m W_\n + D_\n W_\m$
(for example, the one
 defined by the plane-wave metric \rf{gene})
is  already scale-invariant (on-shell, or modulo field redefinition).
The specific dilaton contribution ensures Weyl
invariance of the resulting 2-d  theory as required
for a string background (see also section 6 below).}
are thus  confined to the  small $u$ region. As a result,
 the theory is always in the weakly-coupled regime:
even the  massless modes with $p^u=0$ are  weakly interacting.

Another   special case -- the one
that we are  primarily interested in here --
is  \rf{kk}, i.e.    $\l(\u)= {k\ov \u^2}$.
We shall assume that $k > 0$ in order to have a  positive
 mass term in the \lc  gauge action
 as well as the vanishing string coupling at $u=0$.
In this  case  there is an additional
scaling symmetry in $\u,\v$  already mentioned above.
 It follows then from \rf{kop}
that
\be \la{pou}
\phi =\phi_0  - c \u  + {1\over 2} d k    \ln \u \ . \ee
Thus the total string background is not invariant under
the scaling symmetry.
The linear dilaton  term (with an arbitrary constant  $c>0$) ensures that
the string coupling $e^\phi $ is regular not only at $u=0$
 but also at $\u= \infty $.

Note the solution \rf{pou}
 for $\p$ ``spontaneously'' breaks
the $u \to - u$ symmetry of the equation \rf{kop}.
In writing \rf{pou}  we assumed $u>0$.
To define the solution both at $u >0$ and $u <0$ we may replace
$\ln u$ by $\ha \ln u^2$.  However,
the linear  term represents a problem:  if we keep   it   as
$ - c u$ at $u <0$ then the string coupling blows up at $u = - \infty$;
if we replace $ - c u$  by $ - c |u|$,  that
would mean introducing an additional (``domain-wall'' or
better shock-wave type)
 $\delta(u)$  term in $\l(u)$    supporting
the  solution of \rf{kop} for $\p$  at $u=0$.
An alternative ``regularization'' of this model that
allows one to  maintain the symmetry $\p(-u)= \p(u)$
of the dilaton function and thus to smoothly
continue the solution from $u > 0$   to $ u < 0$ region
 will be discussed in section 4.2 below.
An interesting feature of this model is that the
issue  of singularity of the metric at $u$ is thus
effectively connected  to  the behaviour of the dilaton
at large $|u|$.

The string model we shall study below is
 defined by the following
two-parameter $(k,n)$ family of plane wave backgrounds
\beq \la{op}
ds^2 = 2d\u d\v -  {k \over \u^2} x^2 d\u^2 + dx^i dx^i \ ,
\eeq
\beq\la{opp}
e^{2\p} =   \u^{ k d} \   e^{-  2\u} \ , \ \ \ \ \ \ \ \ \         u > 0 \ . \eeq
Here we used that the  constant $c$ in \rf{pou} can be set to one 
 by rescaling
$\u,\v$ and shifting the constant value of the dilaton.
The value of the
 constant $k$ cannot  be  changed
by rescaling of $\u,\v$ and is thus
an important characteristic  of the model.
As it has been  already noted in \ci{blau} and demonstrated
 in detail in section 2,
the string-frame metric \rf{op}
describes a Lorentzian homogeneous space.

The special case of \rf{op},\rf{opp} with
$d=8 ,  \ k = {3\ov 16} $
corresponds\ to the Penrose limit
of the fundamental string background \ci{blau}.\foot{Depending on how one takes
the Penrose limit one may or may not \ci{blau}
get a  linear term in the dilaton \rf{opp}.
This linear term is,  however, crucial to have string coupling small at
large $u$,  in agreement with the  regularity of string coupling
$e^\p$ for the original asymptotically flat fundamental string
 solution
one starts with.}
The metric \rf{op}
is also a Penrose limit  of the (spatially-flat) cosmological
FRW  metric \ci{blau} (for a  4-dimensional FRW
metric  $d=2$).\foot{If the  FRW scale factor
is $a(t)= t^\beta,$ \ $\beta= {2 \ov 3} (1+ \g)^{-1}$
where $0\leq \g\leq 1 $ is the constant in the matter equation of
state  $p= \g \rho$, then $k= {\beta \ov (1+ \beta)^2} = { 6(1+ \g) \ov
(5+ 3 \g)^2}$.
The exotic matter case
with $\g=-{1\ov 3}$, i.e. $\b=1$,  corresponds\ to $k= {1\ov 4}$.}

\bigskip

Let us now mention  some
  other   related plane wave backgrounds
with $\u$-dependent dilaton. For example,  we may generalize
the BFHP solution \ci{blabla}
supported  by 5-form flux by promoting the metric coefficient
and the 5-form coefficient to functions of $\u$ and adding a
$\u$-dependent dilaton
$$
ds^2 =2 dudv - \l(u) x^2 du^2 + dx^i dx^i , \ \ \  $$ \be\la{kiu}
F_5 = 2 \f(u)\ (1+ *) du \wedge  dx_1 \wedge ...\wedge dx_4 \ ,
\ \ \  \ \ \ \ \ \   \p=\p(u)  \ . \ee
Then  the conformal invariance condition
$R_{\m\n} = -2 D_\m D_\n \p + { 1 \ov 24}   e^{2\p} (F_5^2)_{\m\n}$
 gives
\be \la{iop}
\l = - { 1 \ov 4}  \p''   +   e^{2\p} \f^2 \ . \la{uuo}
\ee
This is a generalization of \rf{kop} (here $d=8$)
to the case of a non-zero $F_5$-form, and a generalization of
BFHP plane wave  (where $\l,\f =$const)
 to the case of $\p\not=$const.

One obvious solution  has the same metric \rf{op}
with $\l= {k\ov u^2}$, trivial dilaton  $\p=0$  and
$\f= {\sqrt k\ov u}$, i.e. 
 the same  background as in  \ci{blabla}
but with non-constant $\l$ and $\f$.

Another simple  special solution is
$
\l=0, \     \f=\f_0 = \const ,   \ e^{2 \p} = { 1 \ov 2 \f_0  u}.$
Here the string-frame metric is flat, but the dilaton is non-constant
(similar example with flat metric and NS-NS 3-form and dilaton
was discussed in the past in \ci{klimt}).  
The corresponding  superstring  theory is exactly solvable
(in the \lc gauge on the cylinder or in the covariant approach \ci{berk})
but the string coupling  blows up near $u=0$.

To avoid the strong-coupling  singularity  we may include also
the
  $\l= {k \ov u^2} $ term in the metric.
Then the solution near  $u=0$ will be as  in the $\f=0$ case
\rf{pou}, i.e.
$ \p_{u\to 0}   \to  4k \ln u $, $ e^{2\p} \to  u^{8k} \to 0$.
This  model is thus a  simple  dilatonic deformation of the
BFHP plane wave, or, alternatively,  an  $F_5$-deformation
of the metric-dilaton model considered above.
  Here the supersymmetry preserved by  the background
is reduced from  maximal  (32 supercharges) to   1/2  
(16 supercharges),
as usual for a generic  plane wave.

\subsection{``Null cosmology'' interpretation }

As was discussed in the preceding  section,
the metric \rf{op} takes a very simple form \rf{opse},\rf{yyy}
in  the Rosen coordinates.
We shall assume that $ 0 < k < \four$, i.e. $ 1 >   \nu  > \ha$.
With the choice of  $q_1=0, q_2=1$ in \rf{simp} we have
\beq \la{ops}
ds^2 =2 d\u d\tv  +  \u^{2\mu} d\x^i d\x^i \ , \ \ \ \ \ \ \ \ 
\m\equiv  1- \nu  \ . 
\eeq
In the  flat space  case $k=0, \nu=1,  \mu = 0$.
Another   special  case is $k={1\ov 4}$ when   $\nu =\mu=\ha $.
The corresponding metric
 $ds^2 = 2dud\tv + u\ d\x^i d\x^i$
 and dilaton $\p= - cu +  {1\ov 4} \ln u$   (here we set $d=2$)
 may be viewed as a  flat spatial section  analog
 of the 4-d string background  with $S^2$-sections
  defined  by the
  metric $ds^2 = 2dud\tv + u\ ds^2(S^2)$
 and the  dilaton $\p= v +  {1\ov 4} \ln u$  which was
 considered in \ci{ttd}.

The  metric \rf{op},\rf{ops}
  may  be considered as a ``plane-wave analog''  of the simplest
   spatially flat cosmological metric\foot{In our case $ 0 < \mu < \ha$  so this is a
subluminal expansion  with $ a'' < 0$.}
\be \la{cosm}
 ds^2 = - dt^2   +  a^2(t)   d \x^i d \x^i \ ,
\ \ \ \ \ \ \ \ \ a(t) = t^{\mu} \ .
 \ee
While in the standard cosmological metric \rf{cosm}
the evolution of fields
is described by second-order differential equations
in time $t$, in the  case of \rf{ops}
the evolution effectively takes place in the ``null time'' $u$
and is described, after the Fourier transform in $\tv$,
by  first order differential equations in $u$.
It is thus natural to refer to  the background
defined  by \rf{ops} as  ``null cosmology''.

The  key  question is whether  string theories  defined
on  the backgrounds
 \rf{ops} and \rf{cosm} are somehow related or  have  some
common properties.  The obvious difference is  that
 while  the ``null cosmology''
metric \rf{ops} supported  by the dilaton \rf{opp}
 is an exact string solution, the metric \rf{cosm}
 with $\m= {1 \ov \sqrt d}, \ d=D_{crit}-1$  supplemented
  by the dilaton \ci{muel}
 $\p = \p_0 + \k \ln t$, $ \k =4 \mu(1-\mu) d $
 should receive corrections of all orders
  in $\a'$.
  Furthermore, while the string 2-d  equations can be solved
  for the  sigma model  defined by \rf{ops} and thus
  string theory can be quantized explicitly  in the \lc gauge
  (as will be explained below),
 it is not clear how to do this  directly for the
 cosmological model \rf{cosm}.

One   relation  between the $2+d$ dimensional plane-wave
  metric \rf{ops}  and $1+d$ dimensional cosmological  metric \rf{cosm}
  is via a Penrose limit. The  plane-wave metric \rf{ops}
  is the cosmological metric \rf{cosm} ``boosted to the speed
  of light''  in  one extra  compact
 direction $y$. This is similar to
  the relation between the Schwarzschild metric and
  the Aichelburg-Sexl or shock wave  metric.\foot{This is a different
  Penrose limit of the  cosmological metric than the one
  considered in \ci{blau}  where the $y$-direction was not added and
  the boost was in the radial spatial  direction.}
  Indeed, adding $dy^2 $ term to \rf{cosm},
  changing the coordinates
  \be \la{pen}
  t=u,\ \ \  \ y= u +  \epsilon^2 \tv, \ \ \
  \ x^i = \epsilon   \x^i , \ \ \ \ \ \ \epsilon \to 0 \ , \ee
and taking the limit $\epsilon \to 0$
  we get the string model defined by  \rf{ops}
  with rescaled string tension ($\a' \to \epsilon^2
  \a'$).\foot{This
rescaling   explains in particular
 why $\a'$ corrections  present for the cosmological
 metric should disappear for  its plane-wave  descendant
 \ci{sfet}. Let us note also that the Penrose limit of the Mueller solution
  \ci{muel} corresponds\ to the plane-wave metric
with the special value of  $k= {1\ov \sqrt d} ( 1 - {1\ov \sqrt d}) $.
  The reason for this restriction is that before the Penrose limit the
  Einstein equations impose more constraints on the  metric than after
  the limit.
  In particular, the 5-d plane wave associated with 4-d
  isotropic ``critical'' Mueller solution has
  $d=4, \  k={1\ov 4}$.}

 The relationship demonstrated above  is formal. In standard  cosmology
there is no natural  reason for introduction of one 
 extra dimension  and study of the
 sector of states  with large  momentum  along it, i.e. 
 the  states that probe the plane-wave geometry
 \rf{ops}. Still,  the simplicity and
  exact solvability of the  model based on \rf{ops}
  makes
 it a natural starting point  for an  investigation of strings in
 {\it curved non-static }  backgrounds.

In general, starting with  a cosmological  background
\be \la{jjj}
 ds^2 = - dt^2 +   g_{ij} (t)  d\x^i d\x^j \ ,  \ \ \ \  \ \ \ \ \
 \p= \p(t) \ ,
\ee
 adding a spectator  dimension $y$ and  then
taking  the Penrose limit
\rf{pen},  we get the plane-wave metric
\be \la{mee}
ds^2 = 2dud\tv + g_{ij}(u) d\x^i d\x^j  \ . \ee
To ensure an embedding into
string theory  this metric  should be supported by the dilaton
  $\p(u)$ subject to
one equation \foot{For completeness,
let us mention that the flat  models
(null orbifold and null brane)
considered in \ci{liu}
were described by \rf{mee}  with $d=1, \ g_{11} = u^2$ and compact
$x_1$, and  with  $d=2, \ g_{11} = 1,\ g_{12} = R, \  g_{22} = R^2 + u^2$
with compact ($2\pi$ periodic)
 $x_1,x_2$. }
\be \la{onno}
- \ha g^{ij} g''_{ij} + { 1\ov 4} g^{ij} g^{mn}
g'_{im} g'_{jn}  + 2 \p''=0 \ .  \ee
For example, starting with  the inflationary (de Sitter) metric
$g_{ij}(t) = e^{2 m t} \d_{ij}$  we find that
the corresponding plane wave is supported by the dilaton
$ \p = \p_0 - cu + { 1 \ov 4} d m^2  u^2 $.
Here  the string coupling grows  at large $u$
  but this
 may  be possible to change  by adding extra
background fields.\foot{The Brinkmann
form of the corresponding plane-wave metric
is $ ds^2 = 2du dv  + m^2  x^2 du^2  + dx^i dx^i$.
Here  the coefficient
of the  second term here is $u$-independent
but has a ``wrong''  sign, i.e.
the mass term in the \lc gauge is
``tachyonic''.
Thus here the fluctuations are  not confined  near $x=0$
as in the model of \ci{blabla}
but rather are  repelled to infinity (see  \ci{maro,sing}
for a discussion of similar examples).
This is reminiscent of a  distinction
 between the AdS and dS spaces.}

More specifically, we  may  look for ``null cosmology'' analogs of
pre-big bang cosmology  backgrounds \ci{GV}.
The suggestion in  \ci{GV} is
to start with the metric-dilaton system only, and
assume that  at $t=\pm \infty$  the   cosmology is simple
and nearly flat  (and weakly-coupled, at least at  $t=-\infty$)
while in between (i.e. near $t=0$)
 string $\a'$ corrections should smooth out the
singularity usually
present  in all cosmological solutions of this type.
The cosmological
metric-dilaton  system solves the leading-order string
equations only for a specific scale factor
$a(t)$  and dilaton   $\p(t)$ subject to two separate equations,
with generic singularity at $t=0$.
At the same time,
in  the  corresponding
``null cosmology'' set-up,
  there is only one (exact in $\a'$)
 equation relating the
two functions
$a(u)$  and $\p(u)$ -- eq.\rf{onno} with $g_{ij}= a^2(u)  \d_{ij}$,
i.e.
$-   a'' d  + 2 a \p''=0$
(which is equivalent to \rf{kop} in view of \rf{giv}).
Thus it is possible in principle to
choose a solution so that to avoid the singularity at $u=0$
keeping dilaton and thus the string coupling
regular and  small everywhere.
That  may    produce  a more regular  solution
 than  the one
in \rf{op},\rf{opp} (where we need to restrict $u$ to be
positive), but the down side will be the  lack of
explicit solvability
at string-theory level.

Indeed, it is easy to find   examples of such regular backgrounds.
One is the direct 1-parameter  generalization of the
$\l(u) = { k \ov u^2}$ model \rf{op}
\be \la{repa}
\l(u) = { k^2 \ov u^2 + s^2} \ ,
\ee
where $s$ is an arbitrary constant.
Here  the components of the curvature
are regular at $u=0$:   $R_{iuju} =  { k^2 \ov u^2 + s^2} \d_{ij}$.
The corresponding  dilaton
 that solves
 $ \p'' = -  {  d k   \ov 2(u^2 + s^2)}  $
  is  (here we set the integration constant
in the linear dilaton term to zero)
 \be \la{ddi}
e^{2\p} = e^{2\p_0} ( 1 + { u^2 \ov s^2} )^{ \ha d k  }
\exp ( - { d k \ov 2 s }\  u\   {\rm arctan} { u \ov s} )
\ .
\ee
For $u\to \pm \infty$ we have  $e^{2\p} \to 0$, so the
string  coupling is small  everywhere.
 In  the limit $s\to 0$ we indeed recover the expression
\rf{pou} (with specific coefficient of the linear term
$c= {d k \pi \ov 4 s}$).
Note that here $\p(u)=\p(-u)$; this suggests that a  natural
continuation of the model \rf{op},\rf{opp} to $u< 0$ region is indeed to
 replace $u$ by $-u$ there.
We may  thus view the  background
 \rf{repa}, \rf{ddi}
as defining a regularized version of a string model
for  the solution \rf{op},\rf{opp}.

Another similar regular  background  corresponds to
\be \la{anop}
\l(u) = { k \ov (u^2 + s^2)^2} \ .
\ee
Note that for  $k = { 2 s^3\ov \pi}$ and  $s \to 0$
 this $\l(u)$ is a
regularised   $\delta$-function.
The  dilaton here looks  very simple
\be \la{iui}
e^{2\p} = e^{2\p_0} \exp (
 - { dk  \ov  2 s^3 }\  u \ {\rm arctan} { u \ov s} )  \ ,
\ee
i.e. the  string coupling can again be made
   small everywhere -- from  $u=-\infty$ to  $u=+\infty$.

In what  follows we shall concentrate on the model \rf{op},\rf{opp}
due to its explicit  solvability. Note that its extra scaling symmetry
is not shared by the above regular models.\foot{The regularized ``$\delta(u)$''-model
\rf{anop} may  be possible to solve explicitly
in the limit $s \to 0$. Similar  shock wave
model  with  $K = A_{ij}(u)  x^i x^j=  - \l(u) ( x_1^2 - x^2_2) $
and  thus no dilaton  was discussed in \ci{hs,DS,DSS},
and its different ``regularization'' -- in \ci{JN}. }



\newsection{Scalar  field theory in plane-wave background:\\
point-particle quantization}

Before  proceeding  to quantize string theory in the metric--dilaton
background  \rf{op},\rf{opp}, it is instructive to
consider first the point-particle  limit,
i.e.  the quantum theory of a scalar relativistic  particle
propagating in
this background.
This may be viewed as an infinite tension limit of the
corresponding first-quantized string theory,
with the particle representing the ``lightest''  point-like
 state  of the string (a
massless supergravity mode in the case of a superstring).

The standard  covariant quantization of a relativistic particle
  leads\ to the Klein-Gordon equation in the corresponding curved background.
Using the isometry  of our plane-wave
 background generated by $T=\del_v$,
it is possible to relate the covariant quantization
of this  system to the  quantization in the light-cone gauge.
The \lc    quantization of
a relativistic particle  on a plane wave metric
\rf{typ},\rf{best}
reduces to a harmonic oscillator problem  with a
time-dependent frequency. Solution of similar  models
 have  been previously discussed  in \ci{hs,gim}.

\subsection{Covariant Klein-Gordon equation}

The leading-order equation for the space-time
field  $\Phi$ representing  a massive scalar string mode
in a metric-dilaton  background
is described by the standard  action
 \be \la{ttt}
 S = \int d^{D} x \ e^{-2\p} \sqrt { G}
 \big( G^{\m\n} \del_\m \Phi \del_\n \Phi  + m^2 \Phi^2  +  c_3 \Phi^3 + ... \big)\ ,
 \ee
 where $G_{\m\n}$ is the string-frame metric
and we included a cubic interaction term.
Defining the  new field $\td \Phi$ as
 \be \la{red}
 \td \Phi = e^{-\p} \Phi \ , \ee
 we can rewrite  \rf{ttt}, using integration by parts,  as
 \be \la{tttt}
 S = \int d^{D} x  \sqrt { G}
 \big[ G^{\m\n} \del_\m \td \Phi \del_\n \td \Phi  + \td  m^2(x)
 \td \Phi^2  + c_3   g_{s}(x) \td \P^3 + ... \big]\ ,
 \ee
with
 \be \la{dew}
  \td  m^2(x) = m^2  -  \DD^2 \p + G^{\m\n} \del_\m  \p \del_\nu \p  ,
\ \ \ \ \ \ \ \   g_{s}(x)  = e^{\p }  \ , \la{dima}
  \ee
  where $\DD$ is the covariant derivative with respect to
 the  metric $G$.
  In the case of the plane-wave metric  \rf{best}  and
  the dilaton depending  only on $u$
  we have
 $ D^2 \p =0, (\del \p)^2 =0 $, and thus the redefined mass is
the same as the original constant one:
 $ \td  m(x) = m$.
This implies   that the {\it free}  field  Klein-Gordon equation
\be
[- { 1 \ov e^{-2\phi}\sqrt{G}} \del_\mu \big(e^{-2\phi} \sqrt{G} G^{\mu\nu }  \del_\nu )
+ m^2 ]\P=0
\
\ee
expressed in terms of the redefined \rf{red} field $\td \P$
\be \la{rew}
[- { 1 \ov \sqrt{G}} \del_\mu \big( \sqrt{G} G^{\mu\nu }  \del_\nu )
+ m^2 ]\td \P=0
\
 \ee
will depend only on the
string-frame metric  but {\it not} on the
dilaton.\foot{Related remark is
 that while the original string field $\Phi$  would have
the dilaton factor $e^{-2\phi}$ in the corresponding measure,
the normalization condition for the redefined field $\td \P$ does
 {\it not} involve the dilaton.}
 However, the dilaton does influence
 the tree-level
interaction terms through the effective string coupling factor $g_{s}$
which   depends on $u$.
 Thus for the background in \rf{opp} we are interested in
the interactions of redefined  $\td \P$-fields will be
{\it suppressed }  near $u=0$.

In general, there is  a question about  potential strong back reaction
on the geometry near the $u=0$ singularity
(as was the case in the null orbifold case \ci{liu}).
As a step  towards clarifying this issue
 let us solve the KG  equation \rf{rew} explicitly
using different choices of  coordinates.

As usual, the general real solution of the KG equation may be 
written as 
 $\td \P = \sum_k [ \a_k \vp_k  (x) +  \a^*_k \vp_k  (x) ]$
where $\{\vp_k,\vp^*_k\}$
is a complete set of special solutions
normalized according to $ \int d^{D-1} x \ \sqrt{-G} G^{0\mu}
( \vp^*_k \del_\mu \vp_{k'} -  \vp_{k'} \del_\mu \vp^*_k )
= - i \delta_{kk'}$ which guarantees $[a_k, a^+_{k'}] =  \delta_{kk'}$
(and thus particle interpretation) 
after the quantization. In the present case the role of time-like
Killing vector is played by $ \del_v$ and so 
$ G^{0\mu}\del_\mu \to G^{u\mu}\del_\mu =  \del_v $.  The explicit form of the basis 
$\{\vp_k,\vp^*_k\}$ will depend on a choice of coordinates
(and boundary conditions).

\subsubsection{Rosen coordinates}

The Klein-Gordon \rf{rew}
can be readily solved for the general plane-wave metric
\rf{gene} written in { Rosen} coordinates \rf{opse},\rf{mee}, i.e.
 $ds^2 = 2 du d \tv  + g_{ij}(u) d \x^i d \x^j$,
where it takes the form
\be \bigg[ \del_\tv \del_u +
 {1 \ov \sqrt{ g(u)} } \del_u( \sqrt{ g(u)}  \del_\tv)
  + g^{ij} (u) \del_i \del_j  - m^2\bigg]
 \td \P =0 \ . \ee
It is straightforward to show that 
\ci{klim,GAV}
\be \la{sla}
\td \P(u,\tv, \x^i) = \int dp_\tv d^d p_i \
e^{i p_\tv \tv } e^{ip_i \x^i}\   \chi  (u; p_\tv, p_i) \ , \ee
\be \la{sool}
\chi (u; p_\tv, p_i)
=   {1 \ov (\sqrt{ g(u)}\ )^{1/2}}\
 {\rm exp} \big( -  { i  \ov 2 p_\tv}
   [ m^2  u + \z^{ij}(u)  p_i p_j ] \big) \  F(p_\tv, p_i) \ , \ee
where
\be   \z^{ij}(u) \equiv   \int^u du' g^{ij}(u') \ ,\ \ \
\ \ \ \ \ g(u) = \det\ g_{ij}  \ .  \la{pya} \ee
 $F$ is an arbitrary function
of the $d+1$ conserved  ``momenta'' $p_\tv$ and $p_i$  corresponding to
the ``linear'' isometries
of the plane-wave background.
In general, this solution (more precisely,  the original field
$\Phi = e^\p \td \Phi$ in \rf{red})  may be used to determine  the corresponding
string vertex operator (cf. \ci{JN}).

For the  scalar equation in plane-wave background it is
natural to define
the KG scalar product
 at null surface $u=\const $ as \ci{gibbons,sing}
\be
( \td \P, \td \P') = i \int d\tv \ d^d \x \ \sqrt{ g(u) } \
( \td \P^*\ \del_\tv \td \P' - \del_\tv \td \P^*\ \td \P' ) \ .
\la{scala}
\ee
Then
$$
( \td \P, \td \P') = \int  dp_\tv d^d p_i \ \sqrt{ g(u) } \
2 p_\tv \
 \chi^* (u; p_\tv, p_i) \chi' (u; p_\tv, p_i)  $$
\be
= \int  dp_\tv d^d p_i \ 2 p_\tv \   F^*(p_\tv, p_i)  F'(p_\tv, p_i)
 \ , \la{sala}
\ee
i.e. the scalar product does not indeed depend on $u$
and ``Fourier modes'' in \rf{sla},\rf{sool} are
$\delta$-function normalized (with extra measure 
factor $p_\tv$).

In the present case  of the metric
\rf{opse} we have  $g_{ij}= a^2(u) \d_{ij}, \ \
g= a^{2d}, $  where $a(u)$
is given by \rf{yyy} or (for the simplest choice $q_1=0, q_2=1$) 
by  \rf{simp},\rf{ops}   with $u>0$.
We get  $ \z^{ij}(u) = \delta^{ij} \z(u)$ where $\z(u)$
was already found   in  \rf{szza}.
Thus \foot{The expression in the case of $q_1=1, q_2=0$ in \rf{yyy}
is obtained by replacing $\nu \to 1- \nu$.}
\be \la{kolt}
\chi (u; p_\tv, p_i)
= {  1 \ov  u^{(1-\nu) d/2}}\
 {\rm exp} \big( -  { i  \ov 2 p_\tv}  [ m^2  u  +
 {  u^{2\nu -1}   \ov 2 \n -1 } p^2_i  ] \big) \  F(p_\tv, p_i)  \ .
  \ee
As $u \to 0$ this function exhibits  singular behaviour
because  of 
the overall factor $ [ g(u)]^{-1/4}$ \ ($0 < 1-\nu  < \ha $). 
The choice of  $\nu=1$ corresponds to the flat space  case.
Note that one 
 can readily rewrite the solution \rf{kolt} in Brinkmann  coordinates using
the transformation \rf{giv}, i.e. (for $a= u^{1-\nu}$) \ 
$\tv= v + {1-\nu \ov 2 u} x_i^2 \ , 
\ \  \x^i = u^{\nu-1} x^i$. 

One   feature of the   Rosen coordinates  is that  the metric 
\rf{opse} does not
  reduce to the flat one at $u \to \infty$,
so that   $p_i, p_v$ do not have the interpretation
of momenta of asymptotical  plane-wave states at infinity.\foot{This problem would not appear  if $a(u)$ in the 
Rosen metric \rf{opse} were approaching 1 at large $u$.}
This raises  the question about the choice of the Fourier mode functions
$F(p_\tv, p_i)$ that should correspond to  natural asymptotic states.

In order to try to see how the singular behavior 
of \rf{kolt} shows up in
 gauge invariant physical quantities,  let us  compute the value of the $n$-point 
local vertex function in the action \rf{ttt},\rf{tttt}
evaluated on a classical solution.
 To probe  possible singular behavior at $u=0$, it is enough
to consider a wave packet moving in a direction orthogonal
to the $u$ direction,
carrying some distribution of momentum $p_\tv$ and having $p_i=0$.
Using (\ref{kolt}}) with
$F(p_\tv,p_i)=f(p_\tv )\delta^{(d)}(p_i)$,
and integrating over $u,  \tv$ and $\x^i$   we obtain
for a $n$-point term in  \rf{tttt}
\bea
\langle
\tilde \Phi ^n\rangle
&\equiv &\int d^Dx\ \sqrt{G}\  e^{(n-2)\phi }\ \tilde\Phi^n
\nonumber\\
&=&
 \Gamma(\a )  \int \prod_{r=1}^n [ dp^r_\tv \ f(p^r_\tv )] \
 \delta(\sum^n_{s=1} p_\tv^s) \
[n-2 +\ha  i  m^2 \sum_{s'=1}^n {(p^{s'}_\tv)^{-1}  } 
 ]^{-\a }\ ,
\la{this}
\eea
where
\be 
\a =1 - \ha d (n-2)\nu^2\ .\ee
We have used the expression for the dilaton in \rf{opp}.
{}For sufficiently large $n$ and  generic value of $\nu $ in the interval
${1\over 2}<\nu<1$ the parameter   $\a $ is negative. As a result, 
the integral over $u$ in the first line of \rf{this}
is formally divergent at $u\to 0$ where
the integrand behaves as $u^\a $.
In obtaining \rf{this}
 we have adopted the  analytic continuation prescription  implied by
 the definition  of the  $\Gamma $-function. As a result, we got 
a regular expression. It remains to see whether this
prescription is  consistent at the level of the 
full quantum theory.
 We shall return to the  issue of back reaction
 later in  this section.

Let us note that the above $\Gamma $-function prescription  could not 
be used in the null orbifold case \cite{liu}. The reason is that there
the  basic functions
contain factors of $1\ov \sqrt{u}$, leading to the  expression for 
 $n$-point vertex 
with a $\Gamma $-function  term  whose argument is a negative integer.
In addition, in our present example there 
is  the $\exp(- u)$ term in the
 effective string coupling in \rf{opp}
that regularizes the integral at infinity.


\subsubsection{Conformally-flat coordinates}
A simple,  yet remarkable property of {\it massless}
 KG 
equation in a plane-wave background  with a conformally-flat 
metric like the one under the discussion here 
is that it 
reduces to the KG equation in flat Minkowski space. 
The reason is simply the vanishing of the curvature scalar for 
the conformally-flat plane-wave metric.

To see this explicitly in the present context, we shall use (\ref{ttt}) and
the expression of the homogeneous plane wave
metric in conformal  coordinates given in (\ref{zes})
i.e. 
$ds^2=\Sigma(\z) (2d\z d\tv+ d\x^2)~,
$
where  $\Sigma$  is given in (\ref{szza}).
After the redefinition\foot{We shall
use subscript 0 to indicate a massless field.}
 $\tilde \Phi_0= e^{-\phi} \Sigma^{{d\over4}}\Phi_0$,
the quadratic part of the action (\ref{ttt}) becomes that of a
free massless  scalar field in flat space.

Since the homogeneous plane wave is conformal to the Minkowski space
with either $\z>0$ or the hyperplane $\z=0$ removed, it is convenient
to quantize the theory  in light-cone coordinates instead
of the standard Minkowski ones. For this we  choose $\z$ as the 
light-cone time, and  have to find
an  invariant measure on the constant $\z$ slices.
We shall consider  the case $\z>0$. The presence
of the $\z=0$ hypersurface breaks the Poincar\'e group of the Minkowski
space to a subgroup generated by the infinitesimal transformations
(cf. \rf{nuin}) 
\bea
T&=&\partial_\tv~,~~~~~~X_i=\partial_i~,
~~~~~~D=\z\partial_\z-\tv\partial_\tv~,~~~
\nonumber\\
\tilde X_i&=& \x_i \partial_\tv-\z \partial_i~,~~~~~~
R_{ij}=\x_i \partial_j-\x_j \partial_i~.
\eea
The Poincar\' e group generators $\x_i \partial_\z-\tv\partial_i$  and
 $\partial_\z$ 
do not appear because their orbits include the  $\z<0$ values.
The measure on the light-cone time slices should be invariant
under the above transformations; it can be chosen as 
\be
d\mu(p)={d^{d+2}p \over (2\pi)^{d+2}} \delta(p^2) \theta(-p_\z)~.
\ee
The presence of the last factor  is necessary for the light-cone energy
to be positive. 
The most general solution of the Klein-Gordon
equation for the field $\tilde \Phi$ is the standard flat-space one
\be \la{ffll}
\tilde \Phi_0 =\int d\mu(p) \bigg[
a(p) e^{i (p_\z \z+p_\tv \tv+ \pp_i \x^i)}
+ a^\dagger(p)  e^{-i (p_\z \z+p_\tv \tv+ \pp_i \x^i)}\bigg]~.
\ee
The theory can be quantized in the standard way leading
to the commutation relations $[a(p), a^\dagger(p')]= 2 p_\tv  
(2\pi)^{d+1} \delta(p_\tv-p'_\tv) \delta^d(\pp_i-\pp_i')$.
The  Fock space is constructed in the standard way
by acting with creation operator $a^\dagger$ on vacuum $\vert 0 \rangle$ 
which is annihilated by the  operators $a(p)$, i.e. 
$a(p) \vert 0 \rangle=0$. 
After normal ordering the energy of the vacuum vanishes, i.e. 
$
\langle 0 \vert  \int d\tv d^d\x\ T_{\tv\tv} \vert 0 \rangle=0~.
$

\subsubsection{Brinkmann coordinates}

In contrast to Rosen coordinates, the Brinkmann coordinates are
global and manifestly ``asymptotically-flat''.
Below we shall  solve  the KG equation directly in the Brinkmann 
 coordinate system,  establishing also  a connection with the \lc gauge  framework
which will be the only  formalism  available to us  in the
case of the  string theory. 
 For that reason,  the discussion of the particle
case in the \lc frame 
  is a natural preparation for the study of the string case.

The  basis of states that will be used below 
 in the \lc  treatment will be different
from the one of the 
Fourier modes \rf{kolt} 
in Rosen coordinates.  The basis in the Hilbert space will
not be labeled  by continuous parameters $p_i$ as in \rf{sla}
but rather it will be
 constructed using oscillator-type creation operators, much like
in for BFHP plane-wave case  in \ci{mets}.
This basis will  not be smoothly related to plane-wave  basis in flat space,
and  that seems to be a key   feature of the present model:
the massless particle (zero-mode) sector of the theory
does not resemble the one in flat space, despite the fact  that the 
metric \rf{kk} approaches flat space metric at large $u$.
Different choices of bases  (adapted to Rosen or conformal or  Brinkmann
coordinates)
correspond to different definitions of observables in this
time-dependent  geometry.

Using { Brinkmann} coordinates,
the explicit form of the  massless
  Klein-Gordon equation for the background
  \rf{op},\rf{opp}  is
\bea \la{kgg}
(2\partial_u \partial_v + {k\ov u^2} x^2\partial_v^2
+\delta^{ij}
\partial_i\partial_j)\td \P_0=0~.
\eea
Performing  the  Fourier transformation with respect to  $v$
\bea \la{kerr}
\td \P_0(v, u, x)=\int d p_v \  e^{i p_v v} \psi (u, x; p_v )
\eea
  we find
\bea \la{abo}
\big[2i p_v \partial_u - {k\ov u^2} x^2 (p_v)^2 +\delta^{ij}
\partial_i\partial_j\big]\ \psi (u,x; p_v ) =0~.
\eea
The dependence on the coordinate $v$ drops out of  the Klein-Gordon
equation because of the isometry generated by the Killing vector $T=\del_v$.
Renaming  the coordinate $u$ as  $\t$  
\be \la{ioo}
u={p_v \tau }\ ,  \ee
we can rewrite the  equation \rf{abo}
 as the  standard  Schr\" odinger equation
\bea
i \partial_\t\psi(\t,x; p_v ) =   {1\over2}\big(-\delta^{ij}
\partial_i\partial_j+  {k\over \t^2} x^2 \big)\ \psi (\t,x; p_v) \
\label{sch}
\eea
for a non-relativistic harmonic oscillator with a  {\it time-dependent}
frequency.
Note that eq.  \rf{sch} does not explicitly depend on the value of the momentum $p_v$:
this is  due to the scaling symmetry \rf{sca} of our  metric.

The  solution of  the Klein-Gordon equation \rf{kgg} thus reduces to the
solution of  the time-dependent  Schr\" odinger equation \rf{sch}.
It is easy to see that the latter  is precisely
the equation which can be derived by quantizing a massless  relativistic particle
in the light-cone gauge in  the plane-wave metric   \rf{op}.
Indeed, using the capital letters $(U,V,X^i)$
 to denote the particle coordinates as functions of
 world-line time $\t$,
 the Lagrangian
of the particle theory in the light-cone gauge $U=p^u \t$ \ (cf. \rf{ioo},
 $p^u=p_v$)
can be written as
\begin{equation}
L={1\over2} (\partial_\t X^i \partial_\t X^j \delta_{ij}- {k\ov \t^2}
 X^2)~.
\label{lcl}
\end{equation}
 The  corresponding Hamiltonian is  then $H={1\over2} (P^2+{k \ov \t^2}
 X^2)$. After the  quantization, the Schr\"odinger equation
associated with this  Hamiltonian  is given by (\ref{sch}).

\subsection{Light cone  quantization}
Let us now study the solution
of the KG equation in Brinkmann coordinates
in the ``light-cone'' representation  \rf{abo} or \rf{sch}.
Before proceeding to solve the specific
Schr\"odinger equation
(\ref{sch}) let us  make some general remarks
on the quantum mechanical  problem of a
harmonic oscillator with time-dependent frequency,
\be \la{basa}
i\partial_\t  \vert \Psi  \rangle=\hat H \vert \Psi  \rangle~ \ , \ \ \ \ \ \ \ \ \ \
\hat H={1\over2} [\hat P^2+\omega^2(\t) \hat X^2]. \ee

\subsubsection{Harmonic oscillator with time-dependent frequency}

Eq. \rf{basa}  is solvable
by the method  developed in
\cite{QM} which was  discussed   in a similar context in \ci{gim}.
The main point of the method is to construct a {\it basis} in the
Hilbert space of the system such that the
operator $i\partial_\t-\hat H$ is {\it diagonal}.
In fact, we shall show that it is proportional to a unit operator.
Working in the Schr\"odinger picture, where the position
and momentum operators are time-independent, we define the operators
(here for simplicity we consider the case of a single oscillator coordinate)
\be \la{aaaa}
\hat A(\t)=i(\vv^* \hat P-\partial_\t \vv^* \hat X), ~~~~~~~~~\hat A^\dagger(\t)
=-i(\vv \hat P-\partial_\t \vv \hat X) \ ,
\ee
where $\vv(\t)$ is a complex solution of the classical equations of motion
satisfying the Wronskian condition \be \la{wro}
\vv\partial_\t\vv^* -\vv^*\partial_\t \vv=i~.
\ee
For constant frequency $\vv= { 1 \ov \sqrt{ 2 \omega}} e^{-i \omega  \t}$.
Using  this condition and the canonical commutation
relations, one can show that  $\hat A, \hat A^\dagger$ satisfy
\be [\hat A, \hat A^\dagger]=1 \ , \ \ \ \ \ \ \ \
i\partial_\t\hat A=\left[\hat H,\hat A\right]\ , \ \ \ ~~~~~~~
i\partial_\t\hat A^\dagger=\left[\hat H,\hat A^\dagger\right]~.
\label{stnew}
\ee
The required (normalised)  basis in the Hilbert space is defined
as the standard Fock space basis at given $\t$
with  $\hat A$ and $\hat A^\dagger$  interpreted
as the annihilation and creation operators, i.e.
\begin{equation} \la{bnb}
\vert \ell, \t \rangle = \frac{1}{\sqrt{\ell!}}
\Bigl(\hat{A}^{\dagger} (\t) \Bigr)^\ell \vert 0, \t \rangle\ ,\ \ \ \ \ \ \
\hat{A} (\t) \vert 0, \t \rangle = 0 .
\end{equation}
It is easy to see that the operator $i\partial_\t-\hat H$ is diagonal in the above basis.
Indeed,  for the $\vert 0, \t \rangle$ state we have
\be
i\partial_\t (\hat A(\t)\vert 0, \t \rangle)=(i\partial_\t \hat A(\t)) \vert 0, \t \rangle
+\hat A(\t)(i\partial_\t\vert 0, \t \rangle)=0~.
\ee
Using the equations (\ref{stnew}) and the definition of the state $\vert 0, \t \rangle$,
we find that
\be
\hat A ( (i\partial_\t-\hat H)\vert 0, \t \rangle)=0
\ee
and therefore, provided that $\vert 0, \t \rangle$ is unique,
 the state $(i\partial_\t-\hat H)\vert 0, \t \rangle$ should be
 proportional
to the state $\vert 0, \t \rangle$. We set
\be\la{lll}
(i\partial_\t-\hat H)\vert 0, \t \rangle=\Lambda(\t)\vert 0, \t \rangle~,
\ee
for some function $\Lambda$ of $\t$.
We remark that the operator
$(i\partial_\t-\hat H)$ is not necessarily self-adjoint and so $\Lambda$
may be complex.
We shall proceed to show that $(i\partial_\t-\hat H)$ is diagonal on  the rest
of the basis $\{\vert \ell, \t \rangle\}$. We shall first show this
for $\ell=1$ and then  use the induction argument. For $\ell=1$, we
have
\bea
(i\partial_\t-\hat H)\vert 1, \t \rangle&=& (i\partial_\t-\hat H)
 \hat A^\dagger\vert 0, \t \rangle
\nonumber \\
&=&
(i\partial_\t \hat A^\dagger)\vert 0, \t \rangle +
 \hat A^\dagger i\partial_\t\vert 0, \t \rangle-
\hat H\hat A^\dagger\vert 0, \t \rangle\nonumber
\\
&=& \hat A^\dagger (i\partial_\t-\hat H)\vert 0, \t \rangle
= \Lambda(\t) \vert 1, \t \rangle~.
\eea
Suppose now that the same relation  holds\ for the state  $\vert k-1, \t \rangle$,
and let us show that it holds\ then for   $\vert k, \t \rangle
=   {1\over \sqrt{k}} \hat A^\dagger\vert k-1, \t \rangle$. Indeed,
\bea
(i\partial_\t-\hat H)\vert k, \t \rangle&=& {1\over \sqrt{k}}
(i\partial_\t-\hat H) \hat A^\dagger\vert k-1, \t \rangle
\nonumber \\
&=&{1\over \sqrt{k}}
(i\partial_\t \hat A^\dagger)\vert k-1, \t \rangle + {1\over \sqrt{k}}
\hat A^\dagger i\partial_\t\vert k-1, \t \rangle- {1\over \sqrt{k}}
\hat H\hat A^\dagger\vert k-1, \t \rangle\nonumber
\\
&=& {1\over \sqrt{k}}\hat A^\dagger(i\partial_\t-\hat H)\vert k-1, \t \rangle
= \Lambda(\t) \vert k, \t \rangle~.
\eea
This demonstrates  that the Schr\"odinger operator  is  indeed
 proportional to a  unit operator when acting on  the basis \rf{bnb}.

Using the completeness  relation,  the most general solution of the
 Schr\"odinger equation is  then ($c_\ell$=const)
 \be \la{whah}
  \vert \Psi  \rangle =\sum_\ell c_\ell   \vert \Psi_\ell
  \rangle  =\sum_\ell c_\ell  e^{ i \g_\ell(\t)}
\vert \ell, \t \rangle \ , \ee
\be
 \la{wha}
\g_\ell(\t) =
\int^\t ds\ \langle \ell, s \vert i \frac{\partial}{\partial s}
- \hat{H} (s) \vert \ell, s \rangle
 = \int^\t ds\ \Lambda(s) \ ,
\ee
where   the phase $\g_\ell(\t)$ does not actually depend on $\ell$.

Note that in coordinate space representation the basis \rf{bnb} has the form
(see, e.g., \ci{kimm})
\be\la{coor}
\langle x \vert \ell, \t \rangle
= (\sqrt{2 \pi} 2^\ell \ell!  |\X(\t)|)^{-1/2} \
({\X(\t) \ov \X^*(\t)})^\ell \ {\rm H}_\ell ( { x \ov \sqrt 2 |\X(\t)|})
\ \exp [ {i \ov 2 } {\del_\t \X^*(\t) \ov \X^*(\t)} x^2  ] \ ,
\ee
where ${\rm H}_\ell$ is the Hermite polynomial.

The advantage of this basis is that  various
expectation values (which are the main observables in the time-dependent
cases like the present one) can be readily computed.
In particular, to compute the expectation values of the Hamiltonian
operator in the basis $\{\vert \ell, \t \rangle\}$, it is convenient to express
it in terms of the operators  $\hat A$ and $\hat A^\dagger$. We find
\be \la{heh}
\hat H={1\over2} \left[(\partial_\t\vv)^2+\omega^2 \vv^2\right] \hat A^2
+\ha \left[(\partial_\t\vv^*)^2+\omega^2 \vv^*{}^2\right]
 (\hat A^\dagger)^2+
(|\partial_\t \vv|^2+\omega^2 |\vv|^2) (\hat A^\dagger
\hat A+{1\over2})~.
\ee
Expectation values of the Hamiltonian operator in the basis $\{\vert \Psi_\ell\rangle \}$
of solutions of the Sch\"odinger equation can be expressed
in terms of  expectation values
of $\hat H$ in the basis $\{\vert \ell, \t \rangle\}$
as follows
\be
\langle \Psi_\ell \vert \hat H \vert \Psi_m \rangle
=  \exp(-2\int^\t ds\ {\rm Im}\ \Lambda(s)) \ \langle \ell, s \vert \hat H \vert m, \t \rangle~.
\ee
Since $\hat H$ is quadratic in the creation and annihilation operators
$\hat A$ and $\hat A^\dagger$, we conclude that
\bea
\langle \Psi_\ell \vert \hat H \vert \Psi_\ell \rangle &=&
\exp(-2\int^\t ds\ {\rm Im} \Lambda(s))~(|\partial_\t \vv|^2+\omega^2 |\vv|^2)
(\ell+{1\over2}) \ ,
\nonumber
\\
\langle \Psi_{\ell} \vert \hat H \vert \Psi_{\ell+2} \rangle &=&{1\over2}
\sqrt{(\ell+2) (\ell+1)}
\exp(-2\int^\t ds\ {\rm Im}\ \Lambda(s))~[(\partial_\t\vv)^2+\omega^2 \vv^2]\ ,
\la{tam}
\eea
with  $ \langle \Psi_{\ell+2} \vert \hat H \vert \Psi_{\ell} \rangle
=  \langle \Psi_{\ell} \vert \hat H \vert \Psi_{\ell+2} \rangle^*$
and  the remaining  matrix elements being zero.


\subsubsection{Solution of the  Schr\"odinger equation}


Let us now return to the problem of solution of
the  Schr\"odinger equation \rf{sch}  corresponding to the special case of
\rf{basa} with
$$ \omega^2(\t) = { k \ov \t^2} \ . $$
We begin by constructing the classical solutions for  the equations of motion
$\partial_\t^2X^i+{ k \ov \tau^2} X^i=0$ corresponding to the
Lagrangian
(\ref{lcl}). This is essentially the same problem
that  we have solved already above (in sections 2,3)
in order to determine the
Killing vectors and the Rosen form of the space-time metric (\ref{kk}).
For $0<k<{1\over4}$,
the solutions can be expressed as (cf. \rf{yyy})
\be \la{soll}
X^i(\t) =k^i \t^\nu    + \td k^i \t^{1-\nu}
\ , \ \ \ \ \ \ \ \
\nu={1\over2} (1+\sqrt{1-4k}) \ ,  \ \ \ \ \ \t > 0 \ ,
\ee
where $k^i, \td\k^i$
are real constants.
 The momenta are $P^i(\t)=\partial_\t X^i$.
We  actually need to  find  a  complex basis of the  classical solutions
 which
satisfy the Wronskian condition \rf{wro}.
To do this we observe that at   $\t=\t_0$, the system is a collection of
harmonic oscillators with frequency
\be \la{www}
\omega_0=\omega(\t_0) = {\sqrt{k}\over \t_0} ~. \ee
Then, we follow the analogy  with the creation/annihilation operators
of a harmonic oscillator and
 define the complex constants
\be
\alpha^i={\omega_0 X^i(\t_0)+i P^i(\t_0)\over \sqrt{2\omega_0}}\ ,  \ \   \   ~~~~~
\alpha^*{}^i={\omega_0 X^i(\t_0)-i P^i(\t_0)\over \sqrt{2\omega_0}}\ .
\ee
Expressing  $\beta$ and $\td \beta$ in terms of $\alpha$ and $\alpha^*$
we  can rewrite  \rf{soll} as
\be
X^i(\t)= \vv(\t) \alpha^i+ \vv^*(\t) \alpha^i{}^*\ ,
\ee
where
\be
\vv(\t)={i\over  \sqrt{2\omega_0} (2\nu-1)} \left[-\big(\sqrt{ k}-i(1-\nu)\big)
\big({\t\over \t_0}\big)^\nu+ (\sqrt{k}-i\nu) \big({\t\over \t_0}\big)^{1-\nu}\right]\ .
\ee
It is easy to check  that $\vv$ does satisfy
the Wronskian condition \rf{wro}.
  Using
$\vv$ and $\vv^*$, we then define the operators $\hat A^i(\t), \hat A^i{}^\dagger$
as in \rf{aaaa}, i.e.
\be
\hat A^i(\t)=i(\vv^* \hat P^i-\partial_\t \vv^* \hat X^i) \  , \  \
~~~~~~\hat A^i{}^\dagger(\t)
=-i(\vv \hat P^i-\partial_\t \vv \hat X^i)
\ee
The Hamiltonian  $\hat H$ expressed
 in terms of $\hat A^i$ and $\hat  A^i{}^\dagger$
is given by \rf{heh}, i.e.
\be \la{hek}
\hat H= c(\t) \hat A^2 +c^*(\t)  \hat A^\dagger{}^2+ b(\t)
 (\hat A^\dagger \hat A+{d\over2})\  ,
\ee
where we suppressed the summation over  the
index $i=1,...,d$   of  oscillators with the same frequency
   and
$$
c(\t)=-{1\over 2 (1-4k) k} \bigg[ \nu (\sqrt{k}-i (1-\nu))^2 ({\t\over \t_0})^{2\nu-2}
+ (1-\nu) (\sqrt{k}-i\nu)^2 ({\t\over \t_0})^{-2\nu}
$$  \be  +\ i\ 4k \sqrt{k}
({\t\over \t_0})^{-1}\bigg] \ ,
\label{kkka}
\ee
\bea
b(\t)={\omega_0\over 2 (1-4k)} \left [({\t\over \t_0})^{2\nu-2}
+({\t\over \t_0})^{-2\nu}-8k ({\t\over \t_0})^{-1}\right] \ .
\label{kkkb}
\eea
Note that $b(\t_0)=\omega_0$ and $c(\t_0)=0$ in agreement with our initial condition.

To find the explicit form
of the  general solution \rf{wha} of the Schr\"odinger equation, we
 shall first determine the state $\vert 0,\t \rangle$
defined by $\hat A^i \vert 0,\t \rangle=0$.
In the position space representation we have (cf. \rf{coor})
\be
(\vv^*\partial_i-i\partial_\t \vv^* x^i) \psi_0 (x,\t) =0 \ , \ \ \
\ \ \ \   \psi_0(x,\t) \equiv   \langle x\vert 0,\t \rangle \ ,
\ee
which leads to
\be
\psi_0(x,\t)
=\bigl[{\sqrt{k} (2\nu-1)^2\over \pi f(\t)}\bigr]^{d\over 4}
\exp\bigg[{-(2\nu-1)^2 \sqrt{k}+i
 k [({\t\over \t_0})^{2\nu-1}+({\t\over \t_0})^{1-2\nu}-2]\over 2 f(\t)} \ x^2\bigg]
 \ee
 where
 \be
 f(\t)\equiv  \nu ({\t\over \t_0})^{2-2\nu}+ (1-\nu) ({\t\over \t_0})^{2\nu}\ .
 \ee
The factor in front of the exponent arises from the normalization
condition  $\langle 0,\t  \vert 0, \t \rangle=1$.

Other  basis functions are given by the expressions in \rf{coor},
so that the solution of the Schr\"odinger equation in coordinate
representation then follows from
\rf{wha}. To find the phase factor there
it  remains to compute the function
$\Lambda(\t)=\langle 0,\t \vert i\partial_\t-\hat H  \vert 0, \t \rangle$
in \rf{lll}.
Using   that  $A \vert 0, \t \rangle=0$ we get
\be
\langle 0,\t \vert\hat H  \vert 0, \t \rangle={d\omega_0\over 4(1-4k)}
\left [({\t\over \t_0})^{2\nu-2}
+({\t\over \t_0})^{-2\nu}-8k ({\t\over \t_0})^{-1}\right]\ .
\ee
In addition, we have
$$\langle 0,\t \vert i\partial_\t \vert 0, \t \rangle
$$
\be =-{d \omega_0\over 4 (1-4k) f(\t) }
\Bigl[ 1-8k-(1-\nu) ({\t\over \t_0})^{4\nu-2}-\nu ({\t\over \t_0})^{2-4\nu}
+4k ({\t\over \t_0})^{1-2\nu}+4k ({\t\over \t_0})^{2\nu-1}\Bigr] ~.
\ee
Thus we find
\bea
\Lambda(\t)={d \omega_0\over 2 (1-4k) f(\t) } \left( 4k-1+2k (2\nu-1) [
  ({\t\over \t_0})^{1-2\nu}-   ({\t\over \t_0})^{2\nu-1}]\right)
~.
\eea
Observe that ${\rm Im} \Lambda=0$.
Its   integral is
$$
\int^\t ds \Lambda(s)
={d \sqrt k \over 2 (1-4k)  }
\bigg[ {  4 k -1 \ov ( 2 \n -1) \sqrt{ \nu ( 1-\n)} }\
{\rm arctan} ( \sqrt{ \nu^{-1}-1}\ ({\t\ov \t_0})^{2\nu-1} )
$$
\be \la{ine}
+ { k \ov \n ( \n-1) }
( [2 \n ( 2 \n -1)  +2 ]\ln {\t\ov \t_0} + \ln [(1-\n)
 ({\t\ov \t_0})^{4\n-2} + \n ] )  \bigg] + \const \ ,
\ee
where $ 0 < k < { 1 \ov 4}$ ($\ha < \nu < 1$) and $\omega_0 = \sqrt k/\t_0$.
For example,
for $k=1/4$ we get $\nu=\ha $ and then
$\int^\t ds \Lambda(s)
=- {d \ov 4 } \ln {\t\ov \t_0}$.

As in \rf{tam} we  can also  compute
the expectation values of the Hamiltonian operator $\hat H$.
Now the quantum number $\ell$ is a ``vector''
$\ell=(\ell_1, \dots, \ell_i, \dots,  \ell_d)$ so
that the basis of solutions in \rf{whah} is 
$\vert \Psi_\ell\rangle =\vert
\Psi_{(\ell_1, \dots,  \ell_d)}\rangle $.
 In particular,  we find that
\be \la{kj}
\langle \Psi_\ell \vert \hat H \vert \Psi_\ell \rangle =
(\sum_{i=1}^d \ell_i +{d\over2}) \ b(\t)
\ , \ee
\be
\langle \Psi_{(\ell_1, \dots, \ell_i+2, \dots \ell_d)} 
\vert \hat H \vert \Psi_\ell \rangle=
\langle {\Psi_\ell} \vert \hat H \vert
{\Psi_{(\ell_1, \dots, \ell_i+2, \dots \ell_d)} \rangle^* =
\sqrt{(\ell_i+2) (\ell_i+1)}}\
 c(\t) \ .
\ee
where $c$ and $b$ are given in (\ref{kkka}) and (\ref{kkkb}),
respectively. We observe that the expectation values of the light-cone 
Hamiltonian operator
diverge at $\tau\rightarrow 0$ as well as $\tau\rightarrow \infty$.

\bigskip

Let us now comment on an  interpretation of these results.
The set of functions  $\psi_\ell (\tau,x) \equiv  \langle x| \Psi_\ell(\tau) \rangle $
represents a basis \rf{whah}
in the space of solutions of the Schr\"odinger equation \rf{sch},
and thus of the original KG equation \rf{kgg},\rf{kerr}.
This basis (labelled by the natural quantum numbers $\ell_i$)
corresponds to a different set of physical states (different choice of
 boundary conditions)  as compared to the
Fourier mode basis \rf{sool} used in the solution in Rosen coordinates.
The expectation value of the \lc  Hamiltonian \rf{kj}
or $E=\int d^d x\  \psi^* (\tau, x) H(\tau)  \psi(\tau,x)$
 may be related to the value of the free part of the scalar Lagrangian
\rf{ttt} or the target-space energy evaluated on the corresponding
solution of the KG equation.

Qualitatively, this combination, multiplied by second 
power of the effective string  coupling, i.e. by 
$e^{2\p(u)}$, 
will appear as a source in the Einstein equations 
for the string-frame  metric (note that $E$ is built out ``redefined'' fields 
in \rf{red}).\foot{Let us note that sources terms produced by a bilinear of a given state 
in equations for other 
string modes (described by ``redefined'' fields in \rf{red}) 
will be multiplied by a single power of $e^\p$ 
(see \rf{tttt}) and thus will be more singular at small $\tau$.}
At large $\tau$ (large $u$) the dilaton \rf{opp} decays  exponentially
and thus suppresses further the expectation value  $E$.
At small times  we get
$e^{2\phi}\sim \tau^{kd}$  and $ E \sim  \tau^{-2\nu}$ (see \rf{kj},\rf{kkkb})
so that  the condition  of small back reaction  appears to be
$ kd \geq  2 \nu$. Assuming $d=8$ and  $0<k\leq {1\over4}$
this is satisfied if  
${1\ov 4}  > k \geq { 3 \ov 16}$, i.e. $\ha < \nu \leq {3 \ov 4}$.
 \foot{Let us note for 
comparison that if we would  construct
$E$ (i.e. energy or  stress tensor)
 as a bilinear of the solution \rf{kolt}  in Rosen coordinates, then 
$E \sim u^{- 8 (1 - \nu)}$ and thus the condition for 
finiteness of $e^{2\p} E$  would be 
$ k \geq 1- \nu$,  which  is never satisfied.} 


Classically, 
a growth of  the energy density near $u=0$   may be  attributed to 
  the focusing of null and time-like geodesics
near $u=0$ in   the  plane wave background.
 This  may  potentially
produce  a large gravitational
back reaction.  As we have just seen, 
  a  large
back reaction near the singularity $u=0$ may be {\it suppressed }
 due to  vanishing of  the {\it string coupling} there.
Nevertheless, this conclusion depends on
the definition of observables and/or choice
of physical states and 
 deserves  further clarification.

\newsection{Solution of
string-theory model }

Our aim here is to
solve the first-quantized
superstring model corresponding to the background
\rf{op},\rf{opp}  using \lc gauge.
The \lc gauge  action is quadratic in ``transverse''
bosonic and fermionic coordinates for any plane-wave
metric \rf{gene}, or, in particular,  for any function $\l(u)$
in \rf{best}, and thus is formally solvable.
By ``solution'' of first-quantized string model
 here  we mean solving explicitly the classical
equations, performing the canonical quantization and
writing down the expression for the \lc Hamiltonian
in terms of creation and annihilation operators, allowing one
to study  time evolution of expectation values.\foot{A  preliminary  
study of the same string model  appeared in \ci{sekk}, 
where the classical equations were solved and the 
general expression for the \lc Hamiltonian was written down.
Here  we  diagonalize the \lc 
Hamiltonian, compute the vacuum energy (accounting for the dilaton 
contribution),   and discuss  calculation of 
some  physical quantities, aimining at understanding 
 their behavior near the 
singularity.}

What distinguishes the model \rf{op}  with $\l= {k \ov u^2}$
 (in addition to its remarkable  scale invariance \rf{sca})
is that the corresponding expressions
are  more explicit  and analytically controllable
than for other potentially  interesting  choices of $\l(u)$
(like the ones in \rf{repa},\rf{anop}).

We  shall assume that the parameter $k$ is restricted to
 $ 0 < k \leq \four$.
The parameter $d$ equal to  the number of ``massive''  scalars
in the \lc action explicitly 
appears in   the dilaton \rf{opp},
but the dilaton coupling does not
enter   the classical string equations and
 its only  role is to cancel
the quantum conformal anomaly (see below).

We will not explicitly
discuss the contributions of
  the fermion modes: in the \lc gauge
the fermionic fields
 are the standard  massless   GS fermions,
 and  their  inclusion is straightforward.
Indeed, as shown in \cite{RUST}, for any  pp-wave background the
fermion part of GS action in the
light-cone gauge is always quadratic in the
fermions.
The only possible
non-trivial coupling of fermions to the background in the \lc gauge
 is  through the generalized  covariant derivative.
In the present case  there are no p-form background
 fields and the gravitational connection term is trivial in the bosonic
\lc gauge,  i.e. the  covariant derivative reduces to  the flat one. Thus there are
 8 massless GS fermion modes (left and right components).

Since the bosonic fields have $u$-dependent mass terms,
 it is clear that here there is  no global world
sheet supersymmetries.\foot{These typically arise when there are extra
``supernumerary" supersymmetries \ci{cvet} which are
``orthogonal'' to the light-cone gauge condition, i.e. the
Killing spinors obeying $\Gamma^u\epsilon\neq 0$.}
The absence of world-sheet supersymmetries in GS action
indicates, in particular,
that the number
of unbroken space-time
supersymmetries in the present background must be  16, as in a generic
plane wave. This can  be easily
 seen directly from the dilatino and
gravitino transformation laws. The former leads  to the condition
$ \Gamma^m \del_m \phi \ \epsilon =0 $. Since in the present case
 $\phi=\phi (u)$, we find
$\Gamma^u\epsilon =0$, leaving 16 unbroken supersymmetries.
The gravitino transformation law then implies that  $ \epsilon$
should be constant.

Before proceeding to the solution of this string theory
on a cylinder in the \lc gauge let us make a  brief comment
on its 1-loop (torus) partition function  $Z_1$.
To define the partition function one should start with the
standard covariant path integral
representation for it. Then  it is easy to argue
that for any pp-wave model the value of $Z_1$ is the same as in the flat
space case.  Indeed, integrating over the $v$-coordinate in the path integral
gives the delta-function constraint $ \del^a \del_a u=0$.
If we formally define  (by an analytic continuation)
 the pp-wave sigma model
on a euclidean 2-torus,  this constraint  will imply
(assuming $u$ is a non-compact coordinate) that  $u=\const$. Then
the rest of the path integral becomes trivial.
  In the case of the Minkowski signature
 in the target space  it  is
 more natural  \ci{liu,bala,craps} to  define the sigma model on
a  Lorentzian 2-torus
 $ds^2 = (d\s_1 + \tau d\s_2) ( d\s_1 + \bar \tau  d\s_2) $,
where $\s_a$ are periodic,  $\s_a \equiv \s_a + 1 $, and
the moduli parameters $\tau$ and $\bar \tau$  are real and  independent.
Here again the equation  $ \del^a \del_a u=0$ has  only
$u=\const$ as its solution.
The same argument should   apply also
 at higher loops, i.e. at higher genera.
In the present case of supersymmetric plane wave, 
 the vanishing of the partition function 
(and of the 1-point functions on a torus) follows also from the residual 
16 supersymmetries preserved by the background.

\medskip

\subsection{Classical equations and canonical quantization
}

In this section we shall use the capital letters $U,V, X_i$
to denote the 2-d scalar fields\ representing the  bosonic
string coordinates.
The bosonic part of the string  action   in the light-cone gauge
 \be U=2\a' p^u \tau  \ee
is \foot{Here we assume that $\tau$ and $\s$ are dimensionless while
 the string coordinates
(and $\sqrt{\a'}$) have dimension of length.
The 2-d metric is $\eta_{ab} = (-1,1)$.}
\begin{equation}
I=- { 1 \ov 4 \pi \a'}\int d\t \int ^{\pi}_0
d\s (\partial^a X^i \partial_a X^j \delta_{ij}  +
{k \ov \tau ^{2}} X^2_i)\ . \la{modr}
\end{equation}
Note again the cancellation of the $p^u= p_v$ dependence as
in the particle case \rf{lcl}.
The equations of motion are
\begin{equation} \la{ska}
(\partial_\tau ^2-\partial_\sigma^2 ) X^i+   {k \ov \tau ^{2}} X^i=0\ .
\end{equation}
Expanding in Fourier modes in $\s$, we get an infinite collection 
of oscillators with time-dependent frequencies.\foot{As was already mentioned 
earlier, similar equation appears in the study of scalar perturbations in cosmology,
$\vp''_p  + [ p^2 - {a''\ov a} ] \vp_p =0$, 
where $p$ is spatial momentum and $a(t)$ is scale factor.
For $a(t)= t^\mu$ we get the effective frequency (mass term) as 
$ p^2 +  { k\ov t^2}$, $k= \m (1-\m) $. We are grateful to G. Veneziano for a discussion
of that point.}
The general solution of \rf{ska} 
is given by\footnote{
One can also solve the equations in Rosen coordinates. In this case, one
has to solve $ (\del^2_\tau + { \beta \ov \tau} \del_\tau
  -\del^2_ \s) X^i_{\rm rosen} =0$,
with  $\beta=2(1-\nu )$.
The general solution is the same as in \rf{ska} up to a factor $\tau^{-\beta/2}$.
{} Note the resemblance to the equation of a damped harmonic oscillator
 (with friction coefficient proportional to $\beta/\tau_0$
in a small interval of time around some $\tau_0$).
}
\bea
X^i(\s ,\tau ) &=& x_0^i(\tau ) + {i\over 2} \sqrt{2 \a '}
\sum_{n=1}^\infty {1\over n} \bigg[ Z(2n\tau ) \big( \a_n^i e^{2 in \s }
+ \tilde \a_n^i e^{-2 in \s } \big)
\nonumber\\
&-& Z^*(2n\tau ) \big( \a_{-n}^i e^{-2 in \s }
+ \tilde \a_{-n}^i e^{2 in \s } \big) \bigg]\ ,
\label{gesol}
\eea
where
\beq
Z(2n\tau )\equiv e^{-i {\pi\over 2} \nu } \sqrt{\pi n \tau }\ \big[
J_{\nu -{1\over 2} }(2n\tau ) - i Y_{\nu  -{1\over 2} }(2n\tau ) \big]\ ,
\ \ \ \ \ \ \
\nu \equiv  {1\over 2} (1 + \sqrt{1-4 k}) \ , \la{zez}
\eeq
\beq \la{xxpp}
x_0^i(\tau )=  {1\over \sqrt{2\nu-1} }\big( \td x^i \ \tau^{1-\nu }
 + 2\a' \td p^i \ \tau^{\nu } \big) \ ,\ \ \ \ k\neq {1\over 4}\ ,
\eeq
\beq
x_0^i(\tau )= \sqrt{\tau }( \td x^i  + 2\a' \td p^i \ \log \tau )\ ,\ \ \ \ k={1\over 4}\ ,
\eeq
\be \la{nuui}
\td x^i={\sqrt {\a'  }\over \sqrt {2 }}(a^i_0+ a^{i\dagger}_0)\ ,\ \ \ \
\td p^i={1\over i \sqrt {2\a' }  }(a^i_0- a^{i\dagger}_0)\ .
\ee
Here
$J_{\nu  -{1\over 2} }(z)$  and $Y_{\nu  -{1\over 2} }(z)$ are the
 usual Bessel functions. Asymptotically,
\beq \la{taks} 
Z(2n\tau )\cong e^{-2in\tau } \big[1+O(\tau^{-1}) \big]\ ,
\eeq
so that for large $\tau$
 the oscillator part of (\ref{gesol}) reduces to that of
the flat-space theory.

This asymptotic ``flatness'' behaviour is not shared by
 the zero mode part of the string coordinate \rf{gesol}: 
 it never reduces
to $x_{0\ \rm flat}^i(\tau )= \td x^i  + 2\a' \td p^i \  \tau  $,
 in any coordinate system.\foot{In Rosen coordinates \rf{opse}
 there is a translational mode
$x_{0\ \rm rosen}^i(\tau )= \td x^i  + 2\a' \td p^i \
 \tau ^  {2\nu -1} $ which grows for large $\tau$.}
This 
 is a direct consequence of the {\it scale-invariance} of the 
equation \rf{ska} restricted to the zero-mode
($\s$-independent) part, i.e. the  invariance under $\tau \to a \tau$, 
\ $a=$const.\foot{
The same  ``non-flatness'' property of the zero-mode solution 
is found also in non-scale-invariant  cases of $\l(u)= {k \ov u^n}$
with $ n < 2$.} 
 As a result, the Fock-space vacuum for the zero-mode part
is   different from the flat-space zero-mode vacuum at
 all scales.

As was mentioned in section 2, the parameter $k$ must be non-negative in 
order to have a regular string coupling at $u=0$ (as well as
a positive mass-squared term for  $X_i$  in the light cone gauge).
There are two special cases:

\smallskip

\noindent a) $k=0$: \ \  this is the flat space, and $Z, Z^*$ reduce to
 plane waves $e^{\mp 2in\tau } $.


 \smallskip

\noindent b)  $k={1\ov 4}$: \ \ this is a limiting value, where the solution
depends\ on $J_{0}, \ Y_{0}$ Bessel functions.

 \smallskip

\noindent
{}For higher values of $k$, the parameter $\nu $ becomes imaginary, and the Bessel
functions have a singular (infinitely oscillatory)
behavior at $\tau =0$.\footnote{
The distinction between models with $k<1/4$ and $k>1/4$ has the
same origin as in the case of the $k/r^2$ potential in quantum mechanics.}
We will restrict our  discussion to the case
of  $0<k < {1\over 4}$, corresponding to ${1\over 2}<\nu <1$.

The requirement that $X^i$ are real functions implies
\beq
(\a_n^i)^\dagger = \a_{-n}^i\ ,\ \ \ \ (\td \a_n^i)^\dagger = \td \a_{-n}^i\
\eeq
The canonical momenta $\Pi^i$  and the total momentum carried by the string are 
given by
\beq
\Pi^i(\s ,\tau )= {1\over 2\pi\a '} \partial_\tau X^i 
\ , \ \ \ \ \ \ \ \ \ \ 
p^i_0(\tau )=\int_0^\pi d\s\ \Pi^i={1 \ov 2\a'}\ \dot x_0^i(\tau ) \ . 
\eeq
Using the recursion relation for the  Bessel functions, we get:
\beq
\partial_\tau Z(2n\tau )= {\nu \over \tau }\ Z(2n\tau ) - 2n W(2n\tau )
\label{deriv}
\eeq
with
\beq
W(2n\tau ) \equiv  e^{-i{\pi\over 2} \nu }
\sqrt{\pi n \tau }\ \big[ J_{\nu  +{1\over 2} }(2n\tau ) - i
Y_{\nu  +{1\over 2} }(2n\tau ) \big]\ .
\label{zert}
\eeq
Thus
\beq
\Pi^i = {1\over 2\pi \a' }  {\nu \over \tau }
(X^i-x_0^i(\tau )) +\hat \Pi ^i\ ,
\eeq
where
\beq
\hat \Pi ^i ={\dot x_0^i(\tau )
\over 2\pi \a' } -{i \over \pi \sqrt{2\a '}}
\sum_{n=1}^\infty \bigg[ W(2n\tau )
\big( \a_n^i e^{2 in \s }
+ \tilde \a_n^i e^{-2 in \s } \big) - W^*(2n\tau )
\big( \a_{-n}^i e^{-2 in \s }
+ \tilde \a_{-n}^i e^{2 in \s } \big) \bigg]\ .
\eeq
Next, we  need to  impose the  canonical commutation relations
\beq
[\Pi ^i(\s ,\tau ) ,X^j(\s ',\tau )]=-i \delta^{ij} \delta (\s-\s' )\ ,
\ \ \ \ \ \ [X ^i(\s ,\tau ) ,X^j(\s ',\tau )]=0\ .
\label{canon}
\eeq
These are ensured by assuming the standard  commutators for the modes
\beq
[\a_n^i, \a_m^j]=n \delta^{ij} \delta_{n+m}\ ,\ \ \ \ [\td \a_n^i, \td \a_m^j]
=n \delta^{ij} \delta_{n+m}\ ,\ \ \ \ \ \ \  [\a_n^i, \td \a_m^j]=0\ .
\label{commu}
\eeq
{}For the zero-mode part, we find
\beq
[a_0^i,a_0^{j\dagger }]=\delta^{ij}\ ,
\eeq
which implies
\beq
[\td x^i, \td p^j]=i \delta^{ij}\ \ \ \ \ {\rm and}\ \ \ \
[ x^i_0(\tau ),  p^j_0(\tau )]=i \delta^{ij}\ .
\label{zcommu}
\eeq
To check the above commutation relations of the mode operators, we note that
\bea
[\Pi^i, X^j] = [\hat \Pi^i, X^j] &=& -{i\over\pi} \delta^{ij}+  {1\over 2\pi }
\delta^{ij}\sum_{n=1}^\infty \big[ Z(2n\tau )W^*(2n\tau )
- Z^*(2n\tau ) W(2n\tau ) \big]
\nonumber\\
&& \times \ (e^{2in(\s -\s') } + e^{-2in(\s  -\s ') })\ .
\eea
At first sight, this  looks non-trivial and
 time-dependent. Note, however, that
\beq
 ZW^*
- Z^*W =2i \pi n\tau [J_{ \nu -{1\over 2} }(2n\tau )
Y_{\nu +{1\over 2} }(2n\tau )-J_{\nu +{1\over 2}  }(2n\tau )
Y_{\nu -{1\over 2} } (2n\tau )]
= -2 i\ ,
\label{relacion}
\eeq
where we have used a standard relation for the  Bessel functions.
As a result, we verify   the canonical commutation relations
in (\ref{canon}).

\subsection{Light-cone Hamiltonian}

The \lc Hamiltonian of the model is given by ($p^v=p_u, \ p_v= p^u$)
\beq \la{hhh}
H= - p_u ={1\over  8\pi {\a'^2 } p_v} \int_0^{\pi }d\s \bigg(\del_\t
 X_i\del_\t X_i+\del_\s X_i\del_\s X_i+  {k\over \tau^2}X_i^2 \bigg) \ .
\eeq
Inserting the expansion of $X^i$ in terms of mode
operators into $H$,  we obtain
\be \la{mst}
H={p_s^2\over 2p_v} +{1\over  \a ' p_v} {\cal H} \ ,
\ee
where for generality we have included the contribution of $8-d$ spectator dimensions
with zero-mode momenta $p_s$ and
\beq
{\cal H}= {\cal H}_0(\tau)+{1\over 2}\sum_{n=1}^\infty\big[ \Om_n (\tau)
(\a^i_{-n} \a^i_{n} +\td \a^i_{-n}\td \a^i_{n})- B_n(\tau )
\a^i_n \td \a^i_n - B_n^*(\tau ) \a^i_{-n} \td \a^i_{-n} ] \ .
\label{hhqq}
\eeq
Here
\beq
\Om_n (\tau )=(1+{\nu \over 4\tau ^2 n^2})|Z|^2+
|W|^2  - {\nu \over 2 n\tau }(ZW^*+Z^*W)\ ,
\label{aaqu}
\eeq
\beq \la{aqua}
B_n(\tau )=  (1+{\nu\over 4\tau^2n^2}) Z^2  + W^2 
 - {\nu \over \tau n} ZW\ .
\eeq
$Z(2n \tau )$ and $W(2n \tau )$ are defined in \rf{zez} and \rf{zert}.
It will be useful in what follows to know how   the functions
$\Om_n (\tau ), B_n (\tau )$ in \rf{aaqu},\rf{aqua}
 behave at large and small $\tau $.
Using the asymptotic expansions of the Bessel functions, we obtain the following
expressions for $\Om_n (\tau)$ and $B_n(\tau )$ at large $\tau $ or
$n\tau\gg 1$:
\beq
\Om_n (\tau )=2+{k\over 4n^2\tau^2 }-{k^2\over 64 n^4\tau^4}+{k^2(2+k)\over 512 n^6\tau^6}+... \ ,
\label{aaq}
\eeq
\beq
B_n(\tau )=k\ e^{-4in\tau }\ \left( - {i\over 8n^3\tau ^3}+{1\over 32n^4\tau^4 }(3+2k)+...\right)\ .
\label{bbq}
\eeq
{}For $n\tau \ll 1$, one has
\beq
Z(2n\tau ) \cong -i e^{-i {\pi\over 2}\nu }\ {\sqrt{\pi }\
(n\tau )^{1-\nu }\over
\cos(\pi \nu )\Gamma ({3\over 2} -\nu ) }\ ,\ \  \ \ \
B_n(\tau )\cong - e^{-i\pi\nu }\ \Om_n (\tau )\ ,
\label{falta}
\eeq
\beq
\Om_n (\tau )\cong {\pi \over  (n\tau)^{2\nu }\cos^2(\pi\nu ) }\
\left( {1\over [\Gamma({1\over 2}-\nu )]^2 }
+{\nu\over 4  [\Gamma({3\over 2}-\nu )]^2 }+
{\nu\over  \Gamma({3\over 2}-\nu )\Gamma({1\over 2}-\nu ) }\right)\ .
\label{eeww}
\eeq

The zero-mode part ${\cal H}_0(\tau )$ in \rf{hhqq} is
the same as the point-particle Hamiltonian \rf{hek} in 
Section 5.2.
In terms of $x_0^i(\tau )$  and  $p_{0i}(\tau )$ in \rf{xxpp} it is
given by
\be
{\cal H}_0(\tau )={\a' \over 2}\ \big[ (p_{0i})^2+ {k\over 4{\a '}^2\tau^{2}} (x_0^i)^2\big]\ .
\ee
In parallel with the discussion of the particle case in section 5.2, where
it was shown  how the covariant Klein-Gordon  equation reduces
(in the sector with fixed $p_v$)
to the time-dependent Schr\" odinger equation, here we will have the
 string wave functional
(of \lc string field theory, see \ci{GSW})  satisfying the
time-dependent
functional Schr\" odinger equation (cf. \rf{sch})
\be
i \del_\tau |\Psi(\tau; p_v) \rangle =  H  |\Psi(\tau; p_v) \rangle
 \ .  \la{evol}
\ee

Expressed
in terms of the modes  $\a_n ,\ \td \a_n$ and $ a_0,\ a_0^\dagger$ in \rf{nuui}
the Hamiltonian \rf{hhqq}  is non-diagonal.
The treatment of the zero-mode part
is again  the same as  in section 5.2.
In the non-zero mode part, there are non-diagonal
terms proportional to $B_n(\tau ),\ B^*_n(\tau )$.
The evolution of generic states made out of  $\a_n ^i, \ \td\a_n ^i $
acting on the vacuum
is thus non-trivial.
In principle, it can be studied using
time-dependent perturbation theory at large $\tau $, where the
light-cone Hamiltonian has the form $H=H_0+\tilde H(\tau )$, with
$\tilde H=O(\tau^{-2})$, but even 
 this  simpler problem is  complicated by the non-diagonal form of the
 Hamiltonian.

Below we will
find a new set of  modes in terms of which the Hamiltonian becomes 
{\it diagonal}.
Before doing that, let us
note that the Hamiltonian \rf{mst} is diagonal
 in the special subspace
of the full  Fock space which is
obtained by acting on the vacuum
by  products of  $\a_n^i ,\ \td\a_n^i$ which
are antisymmetric under the exchange of the ``right'' and ``left'' moving  modes,
$\a_n^i \leftrightarrow \td\a_n^i$.
These are the states of the oriented closed string whose
quantum  wave functional $\Psi\big( X^i(\s )\big)$
is antisymmetric under $\s \to -\s $.
  Indeed, consider the expectation value of the
non-diagonal term
$ \sum_n [B_n(\tau )\a_n^i \td \a_n^i  - B_n^*(\tau ) \a_{-n}^i \td \a_{-n}^i ] $
in \rf{hhqq} between two of such states.
We may  commute  $\delta_{ij} \a_n^i \td \a_n^j 
$ to the right until it gets to
$|0,p_v\rangle $. What remains as a result of the commutation are
terms with antisymmetric indices contracted with $\delta_{ij}$, which 
 vanish.
Similarly,  we may commute the term  $\delta_{ij}\a_{-n}^i \td \a_{-n}^j $
to the left until it annihilates $\langle 0,p_v|$.
The time evolution of such states is,  however,  non-trivial, because
these states are not eigenstates of the Hamiltonian.
For example, the state
\be
|\Psi \rangle ={\cal N}\ b_{ij}  \a_{-n}^i \td\a_{-n}^j |0,p_v\rangle
\ , \ \ \ \ \ \
b_{ij}=-b_{ji} \ , \ \
\ \ \ \ \ {\cal N}={1\over n}  (b_{ij}b^{ij})^{-1/2}\ ,
\ \ \ \ \         \langle \Psi |\Psi \rangle =1\           \la{staa}
\ee
 has  energy 
\beq
{\cal E}_n(\tau )\equiv
\langle \Psi | H(\tau ) |\Psi \rangle = {1\over \a'p_v} \ n\ \Om_n (\tau )\ .
\eeq

Returning to the problem of diagonalising the Hamiltonian, 
let us 
introduce a  new set of  {\it time-dependent} string modes $\aa^i_n, \td \aa^i_n$
 defined by
\bea
{i\over n}( Z(2n\tau )  \a_n^i -  Z^*(2n\tau ) \td \a_{-n}^i ) &=&  {i\over\sqrt{w_n} }
( e^{-2i w_n\tau } \aa_n^i(\tau ) -  e^{2i w_n\tau } \td \aa_{-n}^i(\tau ) )\ ,
\nonumber\\
{i\over n}
( \dot Z(2n\tau )  \a_n^i -  \dot Z^*(2n\tau ) \td \a_{-n}^i )
&=&  2\sqrt{w_n}
( e^{-2i w_n\tau } \aa_n^i(\tau ) +  e^{2i w_n\tau } \td \aa_{-n}^i(\tau ) )\ ,
\label{defimo}
\eea
 where
\beq
w_n\equiv w_n(\tau )=\sqrt{n^2+{k\over 4\tau^2} }\ ,
\label{frecu}
\eeq
with similar relations defining $\aa_n^\dagger =\aa_{-n}$ and
 $\td \aa_n^\dagger = \td \aa_{-n}$.
It follows then that
\bea
\aa_n^i(\tau ) &=& \a_n^i \ f_n(\tau ) +\td\a_{-n}^i\ g_n^*(\tau )
\ ,\ \ \ \ \
\aa_n^{i\dagger}(\tau ) = \a_{-n}^i \ f_n^*(\tau ) +
\td\a_{n}^i\ g_n(\tau ) \ ,
\\
\td \aa_n^i(\tau ) &=& \a_{-n}^i \ g_n^*(\tau ) +\td \a_{n}^i\ f_n(\tau )
\ ,\ \ \ \ \
\td \aa_n^{i\dagger}(\tau ) = \a_n^i \ g_n(\tau ) +\td\a_{-n}^i\ f_n^*(\tau )\ ,
\label{zemod}
\eea
where
\beq
f_n(\tau )=\haf e^{2iw_n\tau }\big[ Z(2n\tau )+{i\over 2w_n}\dot Z(2n\tau ) \big]\ ,\ \ \
g_n(\tau )=\haf e^{-2iw_n\tau }\big[ -Z(2n\tau )+{i\over 2w_n}\dot Z(2n\tau ) \big]\ .
\label{funci}
\eeq
Using the commutation relations (\ref{commu}) for  $\a_n, \td \a_n $ \rf{commu} 
and the properties (\ref{deriv}), (\ref{relacion}) of the Bessel functions,
we find
\beq
\big[ \aa ^i_n(\tau ), \aa^{j\dagger }_m (\tau )\big] =\delta_{nm} \delta^{ij}\ ,
\ \ \ \
\big[ \td \aa ^i_n(\tau ), \td \aa^{j\dagger }_m (\tau )\big] =\delta_{nm} \delta^{ij}\ ,
\eeq
$$
\big[ \aa ^i_n(\tau ), \td \aa^{j\dagger }_m (\tau )\big]=0\ .
$$
In terms of these  variables, the mode expansion of $X^i(\tau )$ in  \rf{gesol} 
becomes
\bea
X^i(\s ,\tau ) &=& x_0^i(\tau ) + {i\over 2} \sqrt{2 \a '}
\sum_{n=1}^\infty {1\over \sqrt{nw_n(\tau) } } \bigg( e^{-2iw_n\tau }
\big[ \aa_n^i (\tau)  e^{2 in \s }
+ \tilde \aa_n^i(\tau)  e^{-2 in \s } \big]
\nonumber\\
&-& e^{2iw_n\tau } \big[ \aa_{-n}^i(\tau)  e^{-2 in \s }
+ \tilde \aa_{-n}^i (\tau) e^{2 in \s } \big] \bigg)\ ,
\label{gesolt}
\eea
Inserting this expansion
into the Hamiltonian \rf{hhh}, we find
\beq
{\cal H}(\tau )=
{\cal H}_0(\tau )
+\sum_{n=1}^\infty
w_n(\tau )\big[
\aa^i{}^\dagger_n  (\tau) \aa^i_n(\tau)  + \td \aa^i{}^\dagger_n(\tau)
  \td \aa^i_n (\tau) \big]+ \he(\tau)\ ,
\la{bbb}
\eeq
where  $w_n$ is defined  in \rf{frecu}
and $\he(\tau ) $ is a ``normal ordering'' c-function
discussed below in section 6.3.

Thus the Hamiltonian \rf{mst},\rf{bbb}  became diagonal.
In terms of the new modes, its  bosonic part
looks like   the Hamiltonian of a free massive 2-d
field theory with a constant mass replaced by
 an effective  time-dependent
mass  $\omega(\tau )= {\sqrt k \ov \tau}$.
To obtain the full superstring
 Hamiltonian, one just needs to add to \rf{bbb} the standard
contribution of {\it massless}
free GS fermionic  modes.

As in the point-particle case in Section 5.2, constructing string states
 using $\aa_n$ and $\aa^\dagger_n $  as annihilation and creation operators, 
it is,  in principle,  straightforward to find  the solutions of 
the corresponding infinite set of 
Schr\"odinger equations \rf{evol}. The wave functions will thus
contain extra time-dependent phase factors, which, however, will 
cancel in  simplest expectation values (cf. \rf{kj}).


\subsection{UV finiteness and the  role of  dilaton }

The \lc gauge model \rf{modr}, \rf{hhh}
has an interesting  feature of being
scale-invariant but not 2-d Lorentz invariant. Indeed, 
despite  the presence of the
mass term for $X_i$,  the Lagrangian is invariant
under $ \tau \to a  \tau, \ \sigma \to a \sigma, \ \ a=$const.
 This classical symmetry is, however, expected to be
broken by  quantum  corrections
 producing a logarithmic
divergence proportional to the mass term,
i.e. $ { k\ov \tau^2} \ln \ep$.
Indeed, the vacuum expectation value of the \lc Hamiltonian \rf{bbb}
given by the naive expression for the normal-ordering  c-function
$\he_0$ in \rf{bbb} is logarithmically divergent:
\beq
\he_0(\tau )= d\sum_{n=1}^\infty [ w^{(B)}_n(\tau)  - w^{(F)}_n]
=
 d\sum_{n=1}^\infty
 \big(\sqrt{n^2+{k\over 4\tau^2}}-n\big)=
{ d k \ov 8 \tau^2} \sum_{n=1}^\infty { 1 \ov n}  + ... \ .
\label{nort}
\eeq
Here  $w^{(B)}_n= w_n$ in \rf{frecu}
and  we included the free GS fermion contribution
with $w^{(F)}_n =n$. As a result,   there is a standard
 cancellation of the  power
divergence in the vacuum energy, but there is  remaining logarithmic
divergence.

In the case of the BFHP plane wave model 
this logarithmic divergence was cancelled against
a similar one coming from the fermionic mass terms  \ci{mets}.
The latter originated from the non-zero 5-form background
that ensured that the plane-wave background solved
the string equations of motion.
The same would happen again if we would consider
the non-constant mass analog of the model of \ci{mets}
corresponding to  the solution of \rf{uuo}
with $\l=\f^2 = {k\ov u^2}$ and zero dilaton.
The  \lc Hamiltonian corresponding to this model
is an obvious analog of \rf{bbb}  with
both bosonic and fermionic oscillator terms multiplied by the same
$w_n(\tau)$ functions and $\he=0$.

 The UV finiteness
of the \lc  2-d theory should be
 of course correlated with
 the 2-d conformal (Weyl) invariance
of the original covariant string sigma model.
In the present case of \rf{op},\rf{opp}
 the role of an additional background that
cancels the conformal anomaly due to  the non-vanishing
Ricci tensor of the plane-wave metric \rf{best} is played by the
dilaton field.
Since the  dilaton is known to contribute to the expression for
 the 2-d stress tensor of the string model,
this implies  that the definition of the \lc  Hamiltonian
should be adjusted so that to cancel an  apparent
divergence in \rf{nort}.
More precisely, this divergent term is cancelled
by a singular field redefinition while the
role of the dilaton contribution is to
cancel the   finite Weyl anomaly term
which acompanies this divergence (see below).
Then the  ``dilaton-corrected''
 finite expression  for $\he(\t)$ in \rf{bbb}  should  be
\beq
\he(\tau )=  d\sum_{n=1}^\infty
 \big(\sqrt{n^2+{k\over 4\tau^2}}-n -{  k \ov 8 n  \tau^2} \big)=
{d}
\sum_{r=2}^\infty {\sqrt{\pi }\ \zeta(2r-1) \over 2\ r!\ \Gamma({3\over 2}-r)}
({k\over 4\tau^2})^r \ ,
\label{grande}
\eeq
where we have used that
$\sqrt{n^2+{k\over 4\tau^2}}=\sum_{r=0}^\infty {\sqrt{\pi }
\over 2r!\Gamma({3\over 2}-r)} ({k\over 4\tau^2})^r { 1 \ov n^{2r-1}}$.
It is possible to show
that $\he(\tau)$ is  negative-definite, and
   $\he(\tau ) \to -\infty $ at $\tau \to 0$.

\bigskip
Let us elaborate a bit more  on the  cancellation of
the logarithmic divergence  and the role of the  dilaton contribution.
Any covariant 2-d sigma model with metric whose  Ricci tensor
of the form $R_{\m\n}= D_\m W_\n + D_\n W_\m$ is formally
(1-loop) scale-invariant
on a flat 2-d background:
the  logarithmic divergence $\sim  \ln \ep\  R_{\m\n} \del^a X^\m \del_a X^\nu$
 vanishes on-shell
or can be cancelled by a (singular) field redefinition, i.e. by a  2-d
field renormalization\footnote{Such
redefinitions should not of course
change target-space diffeomorphism-invariant
observables which
should thus be finite
(see \cite{hps} and refs. there
 for a discussion
of field renormalizations in 2-d sigma models).}
\ $X^\mu \to X^\mu  +  \a' \ln \ep\ W^\mu (X)$.
In the present case of the plane-wave metric \rf{best}
we have only one non-zero component
$R_{uu}= \  d \ \l(u)$, i.e. $W_u(U) = \ha d  \int^U du \l (u) $. The
 required redefinition is  then: \
 $V \to V  + \ha d \a'  \ln \ep\ \int^U du \l(u)$, or,
 for $\l = { k \ov u^2}$,
$ V \to   V - \ha d \a' \ln \ep \ {k \ov U}  $.
 This redefinition of the $V$-coordinate should lead
 to the presence of a ``counterterm'' in the \lc
Hamiltonian \rf{bbb},\rf{nort}
 that ensures its finiteness (note that the redefinition applied to
$ \del_\tau V \del_\tau U $  produces, in the \lc gauge  $U=\a' p^u \tau$,
a term  $\sim \ln \ep\  { k \ov \tau^2}$ ).

In the string-theory context, the condition on the sigma-model
is not simply  scale invariance but (in general, for non-compact
or Lorentzian target spaces)
 stronger requirement
 of Weyl (conformal) invariance of a  theory
defined on a curved 2-d background. That implies a relation
between the Ricci tensor and the dilaton, i.e.
$W_\m= \del_\m \p$
(to 1-loop order in general, but to all orders in the present case;
for details see, e.g., \ci{call,TTT,hsp}).
In general, the dilaton couples to the 2-d curvature as
\be  I = -{ 1 \ov 4 \pi \a' } \int d^2 \xi
\sqrt{- g}   [ G_{\m\n}(X)  \del^a X^\m \del_a  X^\n +
\a' R^{(2)} \phi (X) ] \ , \la{kik} \ee
and thus modifies the expression  for  the
 2-d stress tensor
by a finite  term $\del_a \del_b \p - \eta_{ab} \del^2 \p$
whose role is to cancel the corresponding conformal anomaly
coming from the sigma model part.
In the present case where  $\p'' = - {d\ov 2} \l(u)$ (see \rf{kop}),
we get,  in the \lc gauge,
$(T_{\tau \tau })_\phi \sim {k \over {\t}^2}$, etc.
The Weyl invariance implies also the finiteness of the
expectation value of stress tensor components.
In  the  conformal gauge
$g_{ab} = e^{2 \r} \eta_{ab}$, and using
 a  {\it covariant} 2-d  world-sheet regularization
($ e^{2 \r} |\Delta \xi|^2 > \epsilon^2$) the dependence on
$\r$  and on the 2-d UV cutoff $\ep\to 0$ should
be   correlated (they should effectively appear in a
combination $\ep\ e^{-\r}$). Thus, the finite dilaton
 contribution  cancelling  the Weyl
anomaly should  be  accompanied also by a
divergent counter-term, cancelling the associated
logarithmic  2-d UV divergence
(the one which may be cancelled by  a field
renormalization as discussed above).
The final result is  the finite expression \rf{bbb},\rf{grande}
for the \lc Hamiltonian operator.

 Apart from
the role of the dilaton
in the Weyl anomaly cancellation, it also
determines  the effective interactions
of the properly  normalized string modes
(see \rf{ttt}--\rf{rew}).

\subsection{Evolution of a classical rotating string }

To get a better understanding of dynamics of strings in the
our plane wave background it is useful
to  consider first some  simple examples of classical solutions.

In flat space, a rigid string rotating in a plane represents a state
on the leading Regge trajectory with maximal  angular momentum
for a given energy.
In the light-cone gauge, it  is described by the solution
\be
U=2\a'  p_v \tau,\ \ \ \ \ \ \ V=2\a' p_u \tau \ ,  \ \ \ \ \ \ \ \
L^2 = - 2 \a'^2 p_u p_v \ , \ee
\be
 X \equiv X_1 + i X_2 = L  \ e^{-2i \tau}\ \cos (2\sigma )   \ ,
\label{flta}
\ee
where $X_1$ and $X_2$ stand for Cartesian coordinates of  a transverse
2-plane and we used the constraint to fix the value of $L$.
The center of mass can be at rest ($p_u=p_v$) or moving  on a line,
and the string is rotating in this frame.

The analogue of this  rotating string solution
in the present plane-wave background
is:
\beq
 U=2\a'  p_v \tau  \ , \ \ \ \ \ \  \ \ \
X = \ L \  Z(2 \tau)\  \cos (2\s ) \ , \ \
\label{classi}
\eeq
 with  $V$ determined by solving the constraint\footnote{
One gets
$\del_\tau  V(\s,\tau )={L^2\over\a' p_v}\big[ ZZ^*({\nu\over 4\tau^2}\cos^22\s-\sin^2 2\s)-WW^*\cos^22\s)+{\nu\over 2\tau }(ZW^*+WZ^*)\cos^2 2\s \big]$,
so that the string trajectories are non-trivial.}
and $Z$ defined in \rf{zes}. 
{}At $\tau \to \infty $, the solution (\ref{classi}) reduces
to the flat-case one (\ref{flta}). In general, it represents a rotating
string  whose effective length $
L_{\rm eff}=L \  \big| Z(2\tau )\big|\
$
shrinks with time as the incoming front wave is approaching.
For very small $\tau $, the effective length rapidly goes to zero as
$\tau^{1-\nu }$ before a period can be completed (see eq.~(\ref{falta})).\foot{It is interesting to
 note that in Rosen coordinates (obtained by multiplying (\ref{flta})
by $\tau^{\nu-1} $) the effective length is regular at $\tau =0$. However, the
solution in
Rosen coordinates has the  disadvantage of not reducing to
 flat-space solution at $\tau=\infty $.}

The light-cone energy of the rotating string (\ref{classi}) can be computed by
inserting the solution (\ref{classi}) into (\ref{hhh}), or simply by using
the values of $\a_1,\ \td\a_1$,  $ \a_{-1}$ and $ \td\a_{-1 }$ corresponding
to the above solution in the  eq.~(\ref{hhqq}). We get
\beq
H=-p_u (\tau) ={L^2\over  4{\a' }^2 p_v}\ \Om_1(\tau)\ ,
\label{ergios}
\eeq
where we used (\ref{deriv}).
Note that the non-diagonal term of (\ref{hhqq}) proportional to $B_1(\tau )$ and
$ B_1^*(\tau )$ has cancelled
out from (\ref{ergios}).
The light-cone energy $H$ of this state is thus time-dependent
and  is  determined  by eqs. (\ref{aaq}), (\ref{eeww})
with $n=1$:
 it is approximately  constant at $\tau\to \infty $, and it
increases as $\tau ^{-2\nu }$ as $\tau \to 0$.

As in flat space, the solution has two  integrals of motion --
$p_v$ and the angular momentum $J$ corresponding the symmetries of our metric
under shift of $v$ and rotations in transverse space,
but $p_u$ is no longer conserved.\foot{Note that   the
  integral of motion corresponding
to the invariance of the metric under the scaling symmetry,
$u' = \ell u, \ v' = \ell^{-1} v$ is trivial (equal to zero)
because of the  constraint.}
The value of the  angular momentum  is
\beq
J={1\over 2\pi \a' } \int _0^\pi d\s (X_1\dot X_2- \dot X_1 X_2)={ L^2\over 2\a' } \ ,
\eeq
where we used (\ref{deriv}) and the relation (\ref{relacion}).

It is interesting to note that we find the analogue  of the standard  leading
Regge trajectory relation
\be -2\a' p_u p_v = \a' (E^2 - p^2_y) =  2 J \ , \ee
$(E=p_0)$ but with an ``effective tension''
 function  $T= {1 \over 2\pi\a' }\cdot \ha  {\Om_1 (\tau)}$, i.e. 
\beq
-2\a' \ p_u(\tau)\ p_v \ = \ \Om_1(\tau ) J \ .
\eeq
At  $\tau \to \infty $, we have  $\Om_1\to 2$,  and  so  we recover
 the standard
flat-space Regge relation.
As we go back in time to the region of small $\tau $, the energy
of this physical  state gradually grows until it diverges
as $\tau \to 0$ (where the  ``effective Regge slope'' goes to zero).
%

One may wish to  compare this solution
with a similar one in the case of BFHP plane-wave background
where $\l(u)=\l_0$. Here
 the rotating string solution is given by
\beq  U=2\a'  p_v \tau  \ , \ \ \ \ \
X = {L}\  e^{2 i \omega \tau}\  \cos (2\sigma ) \  ,  \ \ \ \ \ \
\ \ \ \omega=\sqrt{1+{\rm m}^2}\ ,\ \ \
{\rm m}=\a' \sqrt{\l_0}  p_v\ ,
\label{flzz}
\eeq
with $V$ again determined by solving the constraint.
This gives $V=2\a' p^v \tau\ ,\  \ 2{\a'}^2p_vp^v=-L^2$.
{}From the canonical momenta, for the above solution one has the relations,
$p^u=p_v\ ,\ p^v=p_u+{L^2{\rm m}^2\over 2\a' p_v}$.
Here $p_v,p_u$ and $ J$  are all conserved  and we find
$-p_u={L^2\over  2{\a' }^2 p_v} \ \omega^2$ and $
J={ L^2\over 2\a' }\ \omega ,
$
so that we get the direct analog of the flat-space  Regge relation
\be
-2\a' p_u p_v = \omega\cdot   2J\ ,
\ee
with  the ``effective tension'' is $T={\omega \over 2\pi\a' }$
being  increased compared to
the flat case.

\subsection{Quantum string mode creation}
Let us  now turn to  a study of properties of quantum strings.

It is well known that in
gravitational  pp-wave backgrounds\ there is no  particle creation \cite{gibbons}.
Indeed, the existence of a covariantly constant null Killing vector guarantees
a definition of a frequency which is conserved.
Similarly, there cannot be string creation, so one can consistently describe string
propagation in this background by using the usual
 first quantized formalism.

Nevertheless, as was shown in \cite{hs}, and further studied in
\cite{BROOKS,DS,JN}, there can be string mode creation.
In general, given a pp-wave background with asymptotically flat
regions at
$u=-\infty $ and $u=+\infty $, the time evolution
of a string which starts  in a given state at $u=-\infty $,
may be such that
the  string ends\ up at $u=+\infty $  in a different state. In particular,
the string could have extra internal excitations,
 produced by the interaction with the pp-wave background.

Passing  through the  singularity at $u=0$ obviously requires
some prescription. This will be discussed in the next subsection.
Here we will consider a  creation of string modes
as seen by  an observer in the ``in" vacuum $|0, p_v \rangle_0$ at $u=\infty $.
By the  ``in" vacuum we mean the
Fock space state which is annihilated by the operators
$\a_n^i ,\ \td \a_n^i $ in \rf{commu}.
We shall start with  the string in the
$|0, p_v  \rangle_0$ state  at  $u=\infty $
 and study how
 this state  should evolve back to  $u=0$.
Equivalently, one may reverse the orientation of time,
 given  the symmetry of the metric
under $u\to -u, \ v\to -v$,
and interpret this as an evolution from the
 ``in" vacuum at $u=-\infty $
to some excited  state at later time.

\smallskip

Let us  consider the expectation value of the ``oscillator
number'' operator that appears in the Hamiltonian \rf{bbb}:
\beq
\bar N_n (\tau )\equiv {}_0\langle 0, p_v|
\big( \aa_n^{i\dagger }\aa^i_n + \tilde \aa_n^{i\dagger }\td \aa_n^i \big) |0,p_v\rangle_0
=2d \ n\ g_n^*(\tau ) g_n(\tau ) \ ,
\label{parpa}
\eeq
where
$d$ is again the range of index $i$, i.e.
the number of ``massive'' 2-d coordinates $X^i$.
 We have used eq. \rf{zemod}.
Inserting the definition (\ref{funci}) of $g_n(\tau )$
and using  (\ref{relacion}) we find
\beq
\bar N_n(\tau )=  {d\ n^2\over 2 w_n^2}\ \big[n\Om_n (\tau )- 2 w_n(\tau )\big]\ .
\label{risu}
\eeq
Here $\Om_n (\tau )$ was  defined  in (\ref{aaqu}) and
$w_n$ -- in (\ref{frecu}).
The total number of created oscillator  modes is then
\beq
\bar N_T (\tau )=\sum_{n=1}^\infty \bar N_n(\tau)
= {d}\sum_{n=1}^\infty
{n^2\over 2w_n^2}\ \big[n\Om_n (\tau )- 2 w_n(\tau )\big]\ .
\label{werz}
\eeq
Note that
for any, no matter how small, $\tau$
 there will be an infinite number of
modes for which $n\tau \gg 1$,
 which  will thus behave essentially
like  massless modes (cf. (\ref{frecu})).
For them, one can use the asymptotic form (\ref{aaq}) of $\Om_n (\tau )$
which applies for $n\tau\gg 1$.
When $\tau \gg 1$, one can use the asymptotic form of (\ref{aaq})
 of $\Om_n (\tau )$ for all modes, including  the  $n=1$ one.

Inserting eq.~(\ref{aaq}) into (\ref{werz}), one finds a surprising cancellation
between the first three terms in  the expansion.
Only the cancellation of the leading term of (\ref{aaq}) (i.e. $2$) is
obvious: we know
that $g_n(\tau )$ must vanish at large $\tau $ (by construction,
since at $\tau \to\infty $ the definition (\ref{defimo}) implies
that the modes $\a_n,\ \td\a_n $ and $\aa_n, \ \td\aa_n $ are the same, modulo
normalization).
The resulting  expression at {\it large} $\tau$  is then
\beq
\bar N_T (\tau )\cong {d}\sum_{n=1}^\infty  {k^2\over 512\  \tau^6} \ {1\over n^5}
={d \ k^2 \zeta(5)\over 512\ \tau^6}\ ,
\label{wezz}
\eeq
i.e. is finite.
This also proves that the series is converging
 for {\it any} $\tau $:
as was pointed out above, for any given $\tau $, we can use the expansion
(\ref{aaq}) for all the modes with
$n\gg \tau^{-1}$. This produces 
the  convergent sum
 $\sum_{n=n_0}^\infty {1 \ov n^{5}}$, with $n_0\gg\tau^{-1}$.

Let us now investigate the behavior of $\bar N_T$ as we approach
the {\it  small} $\tau$ region, i.e.
$\tau \ll 1$. In this case,  for all $n\ll \tau^{-1}$ the
function $\Om_n (\tau )$ may  be approximated by the expression in
(\ref{eeww}), so that  the leading contribution is
\beq
\bar N_n (\tau ) \sim {\rm const} \cdot
\  \tau ^{1-\sqrt{1-4k} } \ .
\label{fret}
\eeq
Thus each individual contribution  in the sum over  modes
vanishes for sufficiently small $\tau $.

Nevertheless, the total number of created oscillating  modes
$\bar N_T(\tau )$ diverges as $\tau \to 0$.
This can be seen as follows.
Let $\tau=\epsilon,\ 0<\epsilon\ll 1$.  The number of excitations
$\bar N_n (\tau )$
of a given frequency $n$ must have a maximum as a function of $n$, since
it vanishes in the limits $n\ll \epsilon^{-1} $ (see eq.~(\ref{fret}))
and   $n \gg \epsilon ^{-1} $ (as $1/\tau^6$, see (\ref{wezz})).\footnote{
One can check numerically that there is only one maximum.}
{}For small $\tau $, the most important contributions to the
 sum come in fact from
the modes with $n\tau =O(1)$. Since $\Om_n (\tau )$ in \rf{aaqu}
is a function of $n\tau =O(1)$,
one has $n\Om_n (\tau )=O(n)$ in (\ref{risu}),  so that
 $\bar N_n(\tau )=O(n)$ for $n=O(\epsilon^{-1})$.
Thus $\bar N_T(\tau )$ picks up the main contribution from
the terms  $\bar N_n(\tau )$ with $n=O(\epsilon^{-1})$.
To compute $\bar N_T(\tau )$, one notes that the
 number of terms $\bar N_n$ with $n\tau = O(1)$ is also of order $n=O(\epsilon^{-1})$.
Therefore, $\bar N_T \sim n^2 \sim \tau^{-2}$ for small $\tau $.

In conclusion, we find  that the total number of  excited modes on
 a string state  which started in  the 
 ``in" vacuum $|0, p_v\rangle_0$
tends to infinity as we approach the $\tau =0$ point. 
We stress that
the origin of the divergence is in 
the  creation of very high frequency $n$ modes, with
$n\to\infty $ as $\tau \to 0$,
since, as explained above,
the creation of modes with  $n<\tau^{-1}$ is suppressed  at small $\tau $.


 This singularity is, however,  observer-dependent.
The state  $|0, p_v \rangle_0$ represents a vacuum only
at $\tau =\infty $, since it is defined  as being annihilated
 by {\it massless} 2-d scalar bosonic  modes (the mass
${\sqrt k\ov \tau}$ vanishes at $\tau= \infty$).
The singularity of $\bar N_T(\tau ) \to \infty $ at $\tau \to 0$ is 
analogous  to the creation of a divergent number of modes
near a horizon of a black hole as seen by a
Schwarzschild observer.
{} At the same time,
there
is no string mode creation
in the vacuum  state
annihilated by the operators $\aa_n^i,\ \td\aa_n^i$
in terms of which the Hamiltonian \rf{bbb} is diagonal.


\subsection{String transition through the $u=0$ singularity }

Let us now consider 
the evolution from
a free string state in the asymptotically flat
region at
$u=-\infty $  to the asymptotically flat region at
$u=+\infty $.
We shall assume that the complete
space-time  can be  obtained by 
patching together the regions with $u<0$ and $u>0$.

Some of the paths (i.e. geodesics) of {\it  point-like} particles 
cannot  smoothly traverse the singular point
at $u=0$,  and  excising this point makes  the space-time
  geodesically incomplete \cite{hs}.
In \cite{hs}, string mode creation in the evolution 
from $u=-\infty$ to
$u=\infty$ was discussed for a generic $\lambda (u)$
which goes to zero at large $|u|$ and is non-trivial
for finite $u$ with singularity in the middle.
In the case when $-\lambda (u)$ (i.e. an effective ``potential'' 
in the string  equation) has a form of  a  potential well
it was
argued that the number of excitations typically increase with the
depth of
the well, but the case of $\lambda (u)=k/u^2$ was not explicitly
considered.

Here we would like to  argue that strings 
may pass through  the singular
point $u=0$. We will adopt a natural prescription implied  by
an analytic continuation
of the Bessel functions.

We will consider two regions I and II, corresponding to $u\sim \tau<0$ and
$u\sim\tau >0$ respectively.
In region II, the solution for $X^i(\s ,\tau )$ is given by eq.~(\ref{gesol}),
with $\a_n^i,$ $\td \a_n^i ,$  $\a_{-n}^i$ and  $\td \a_{-n}^i$
representing the ``out" modes.
In region  I, the general solution is given by
\bea
X^i(\s ,\tau ) &=& x_0^i(\tau ) + {i\over 2} \sqrt{2 \a '}
\sum_{n=1}^\infty {1\over n} \bigg[ Z^*(-2n\tau ) \big( \beta_n^i e^{2 in \s }
+ \tilde \beta_n^i e^{-2 in \s } \big)
\nonumber\\
&-& Z(-2n\tau ) \big( \beta_{-n}^i e^{-2 in \s }
+ \tilde \beta_{-n}^i e^{2 in \s } \big) \bigg]\ ,
\label{graso}
\eea
where the ``in" modes $\beta^i_{n},\ \td \beta^i_{n},\ \beta^i_{-n},\ \td \beta^i_{-n} $ obey  similar commutation relations as the $\a_n^i $
in \rf{commu}. Here $Z$-functions are the same as in \rf{zez} and the
 zero mode part $x_0^i(\tau )$ is the same as in \rf{gesol}
with the  replacement of  the
``out" modes $a^i_0,\ a_0^{i \dagger}$ by the  ``in" modes   $b_0^i,\ b_0^{i \dagger}$.

The series expansion (\ref{graso}) is written so that asymptotically, 
at $\tau\to -\infty $, where (see \rf{taks}) 
\beq
Z^*(-2n\tau )\cong e^{-2in\tau } \big[1+O(\tau^{-1}) \big]\ ,
\eeq
it  reduces to the standard free string theory mode expansion
for $X^i(\s,\tau)$ (modulo a peculiarity of the 
  zero-mode part asymptotics mentioned below \rf{taks}).

Since the evolution is governed by linear equations,
the ``in" $\beta_n , \td \beta_n $ and ``out" $\a_n ,\td \a_n $
modes will be related by
a linear transformation:
\beq
\a_n =C_n  \beta_n + D_n \td \beta_{-n} \ ,\ \ \ \ \ \ \ \ \ \ \
\td \a_{-n} = \td C_n \beta_n + \td D_n \td \beta_{-n} \ .
\eeq
The most natural way to go from the
region I to the region II is by an analytic continuation of the
 Bessel functions in \rf{zez}.
The Bessel functions have a branch point  at $\tau =0$.
To make the transition from $\tau<0$ to $\tau>0$
we choose a contour which goes below the $\tau =0$ point
(or,  equivalently, we replace $k\ov u^2$ 
by $k\ov {(u-i\epsilon)^2}$).\foot{A  similar continuation
prescription was proposed in  an interesting paper \ci{tut} 
for  a  quantum field theory  in a compactified Milne-type   metric 
($ds^2 = - dt^2 + t^2  dy^2 + dx^2$,\  $  -\infty < t < \infty, \ y \equiv y+  L$) 
motivated by the cosmological model of \ci{setu}.}
As we shall show below, an
 alternative choice of the  contour
which  passes  above the $\tau=0$  point
would lead to an unphysical result.

Using the  formula \cite{grad}
\beq
H_\mu^{(2)}(e^{-i\pi } z)= -e^{i\pi \mu }H_\mu^{(1)} (z)\ , \ \ \ \ \ \ \
H_\mu^{(1,2)}(z)\equiv J_\mu(z)\pm i Y_\mu (z) \ ,
\eeq
we get
\beq
Z(e^{-i\pi }2n\tau )=\ Z^*(2n\tau )\ ,\ \ \ \ \ \ \ \
Z^*(e^{-i\pi}2n\tau )=
Z(2n\tau )\ ,
\eeq
where we have taken into account the
 extra factor of $\sqrt{\tau }$ in the definition of $Z(2n\tau )$
in \rf{zez}.
It follows  that
\beq
\a_n= \beta_n\ ,\ \ \ \ \ \ \ \   \td\a_{-n}=\td \beta_{-n} \ ,
\eeq
and  similar relations for the modes with $n\to -n $.
{}It is then natural to assume
  the same conditions for  the zero-mode part:
$$
a^i_0=b^i_0\ ,\ \ \ \ \ \ \ a^{i\dagger }_0=b^{i\dagger }_0\ ,
$$
even though  this produces  a discontinuity in the  time derivative.\footnote{
One may try to find a relation between ``in'' and ``out'' zero modes
by using the regularized background (\ref{repa}).
Then the zero-mode part is given in terms of hypergeometric functions and
extends from $\tau=-\infty $ to $\tau =\infty $. The matching at
$\tau \to \pm \infty $ with
$x_0^i(\tau )$  (given in terms of $a^i_0,\  a^{i\dagger }_0$,
and
$b^i_0 ,\ b^{i\dagger }_0$, respectively)
leads to a non-trivial Bogoliubov transformation between
$a^i_0,\  a^{i\dagger }_0$,
and
$b^i_0 ,\ b^{i\dagger }_0$. It  depends on the 
``regularization parameter''
$s$, and  is singular as
$s\to 0$, so it cannot be used to determine the relation between
the two sets of modes.}

With the above prescription, the 2-d ``S-matrix" is trivial,
 and there is
no mode creation. In particular,
\be
\langle 0_{\rm in}|\hat N_n^{\rm out} |0_{\rm in}\rangle =
\langle 0_{\rm in}| (\a_{-n}^i \a_n^i + \td \a_{-n}^i\td \a_n^i)|0_{\rm in}\rangle =0 \ ,  \ee
where $|0_{\rm in} \rangle $ is the vacuum state for the ``in" operators $\beta_n^i,\ \td\beta _n^i $.

If the contour passes the point $\tau=0$ from above,
one gets the relation
\beq
H_\mu^{(2)}(e^{i\pi } z)= 2\cos(\mu\pi )H_\mu^{(2)}
+e^{i\pi \mu }H_\mu^{(1)} (z) \ .
\eeq
This prescription would lead to a non-unitary
2-d ``S-matrix" connecting $(\a_n, \td\a_{-n} )$ with
$(\beta_n, \td\beta_{-n} )$  which does not conserve probabilities.

\medskip

In conclusion, there is a natural way to identify oscillating modes
which leads to trivial transition amplitudes in this sector. In the zero mode
sector, there is a discontinuity in the time derivative.
This may suggest that
some source is needed at $u=0$, which affects the string oscillations
only in a mild way, but takes care of the zero-mode part.
Needless to say,  issues related to
the singularity at $u=0$  deserve further study.

\section{Concluding remarks}

In an attempt to  address the issue of singularities of cosmological
or time dependent  backgrounds  in string theory,
here we have studied a simple solvable model
which has the  basic ingredients that are needed to mimic big-bang
cosmology: a ``null''  time 
 dependent scale factor and a singularity at $u=0$,
where tidal forces become infinite.
This is a mild type of singularity because all curvature
scalars vanish and some geodesics can be extended through it.
Nevertheless,  some other time-like geodesics 
of classical test particles
are incomplete,
making this simple spacetime a rather direct analog of 
a singular cosmology.
Having a solvable string model, one  may try to elucidate, in 
particular,
the evolution of quantum string states as they approach and pass through 
the  $u=0$ region.

Understanding string theory in a time-dependent background involves
some well known conceptual issues, such as how to  define observables in 
a situation with a singularity.
In particular, one may ask 
whether a consistent quantum theory can be defined by specifying 
initial conditions at the singularity
 and considering only the region   $ 0 < u < \infty$, or one must
 include
also the ``pre-big bang'' $ - \infty < u < 0$ part.
Some related  studies of string theory in simplest time-dependent dilatonic backgrounds
seem to favour \ci{craps}  the  latter option.
In the present case, the  existence of special geodesics which
can go through $u=0$  smoothly from $u<0$ to $u>0$ 
(see Appendix B  and also \cite{hub2}) is another indication
 that the full  quantum  theory should
be formulated in the extended   region $-\infty<u<\infty $.
Thus it is important to understand how to define a consistent 
string  theory in this case.
As we have shown in section 6,   there is indeed a natural way
to extend  oscillating string 
modes
 from $u<0$ to  $u>0$ region, using  the analytic continuation
property of the Bessel functions.

 In the zero mode
sector, however, there is a discontinuity in the time derivative, which suggests
that some source or ``domain wall'' is  needed at $u=0$.
This could resolve the problem of classical geodesic 
incompleteness in the following hypothetic way.
A closed string  coming from $u <0$ and approaching  $u =0$
 may be 
absorbed by the source (or ``brane''), which then gets excited
and re-emits the closed-string mode  at $u >0 $.
One  possible approach is to try to add an open string sector 
with boundary condition $\partial_\tau X^\mu|_{\tau=0}=0$.
Introducing a D-brane at $u=0$ seems   natural, given
the origin of the  metric \rf{kk} with $\l= { k\ov u^2}$ 
 as a Penrose limit of Dp-branes, with 
$u=0$ corresponding  to the original $r=0$ singularity of the Dp-branes 
 \ci{blau,sekk,Rya} (see also Introduction).
One technical problem with this idea  is
that then one cannot use the light-cone gauge (where
$\partial_\tau U\neq 0$), while 
it is not clear how to solve the model directly 
 in a covariant gauge.

As was recently discussed in \ci{liu,lial}, 
formulating string theory in singular
time-dependent backgrounds
may 
lead to a number of potential problems,
such as new divergences in loop diagrams and various  sorts of instabilities.
It would be interesting to study 
string scattering amplitudes for our present
plane-wave  background but this is much more complicated than in the flat
(orbifold)  case of \ci{liu} (computation of scattering amplitudes 
is a non-trivial problem already for the  $\l(u)=$ const or BFHP  plane wave).
Simplest scalar massless vertex operator is  related to the
 solution of the Klein-Gordon equation
discussed in section 5.\foot{It  simplifies  for special 
 states with $p_i=0$ moving along v-direction, so it may be interesting 
to try to study the  tree-level amplitudes for such
states.}

Comparing the present case 
to the  null orbifold model \ci{HS,liu},
 where  time dependence is introduced by a global identification
 and there is a static covering picture, 
 the present plane wave ``null cosmological''  model is closer
to a standard cosmology.
In particular, the  nature of the singularity at $u=0$ 
here is different from  the null orbifold case
(despite 
 similarity in  the form of the  metric in Rosen coordinates
($ds^2=dudv + u^2 d x^2$):
it is the compactness of $x$  that causes a singularity at $u=0$
in \ci{liu} 
 and  is related to 
multiplicity  of images in Cartesian coordinates of the orbifold.
In our case the singularity is related to the blow-up 
of the {\it curvature}, causing infinite tidal forces near $u=0$.
At the same time, as already mentioned above, 
 this singularity is still of a  different type compared to 
the standard cosmological one, as all curvature invariants 
vanish in the plane-wave case. 

Another  new feature of our model 
is  the vanishing of the string  coupling 
at the singularity,  implying  a reduction of  back reaction
 effects.\foot{The back reaction problem in the null orbifold case 
is somewhat different, being 
 related to multiple images of the fixed point in global coordinates:
as  pointed out by  Horowitz and Polchinski \ci{lial}, 
there an  introduction of a single particle makes the spacetime
to collapse.}
As discussed in  section 5,
the fall-off in $e^\p$ competes with  the
 focussing effect of geodesics, which typically leads to
configurations with infinite energy density near $u=0$.

A basic problem, however,  is  which are the proper 
 invariant quantities to ``measure''  back reaction effects
(for example,  one can always go to a frame where the energy 
discussed at the end of Section 5 is small).
 The situation may be analogous to the one when 
a small amount of matter is approaching the horizon of a black hole. 
In Schwarzchild coordinates, the energy would diverge near
the horizon, but in an inertial frame the energy is small,
 so strong back reaction effects are not expected.
More generally, there is a  well-known ambiguity in 
choice of  vacuum and observables in time-dependent backgrounds.
All these questions await further study in the string-theory context. 

\bigskip\bigskip
\bigskip\bigskip
\noindent
{\bf Acknowledgments}

We would like to thank M. Cvetic, J. Garriga, G. Gibbons, 
H. Liu, A. Recknagel, D. Robinson and  G. Veneziano
 for useful discussions
and help.
The research of  G.P. is partially supported by the PPARC grants
PPA/G/S/1998/00613 and PPA/G/O/2000/00451 and by the European grant
HPRN-2000-00122.
The work of A.A.T.  was supported in part
 by the grants DOE
DE-FG02-91ER40690,   PPARC SPG 00613,
INTAS  99-1590,  and by  the Royal Society
Wolfson Research Merit Award.

\setcounter{section}{0}
\setcounter{subsection}{0}

\appendix{A group manifold structure}

The plane-wave space-time (\ref{kk}) admits  a group manifold
structure. We should note from the start that the  metric \rf{kk}
will be left-invariant  only: the  corresponding
bi-invariant metric will be  degenerate (and will  not be
directly related to \rf{kk}).

To determine the group structure, it is sufficient to find the  frames
that satisfy
the Maurer-Cartan equations for a Lie group.
For this it is convenient to work with the plane wave metric as given in
(\ref{newcor}). Define the frames
\bea
e^0=dw
\ , \ \ \ \ 
e^1=e^{w} dv+\mu x^2 dw+ \rho x_i dx^i
\ , \ \ \ \ 
e^i= dx^i+ \sigma x^i dw~,
\eea
where $\mu$, $\rho$ and $\sigma$ are real constants which are to
be determined. It is natural  to
impose  the following Maurer-Cartan equations
\bea
de^0=0,~~~~~~~~~~de^1=e^0\wedge e^1,~~~~~~~~de^i=\sigma e^i\wedge e^0~.
\eea
Then we find that $
\sigma+\rho=0, ~~2\mu+\rho=0~.$
For the  metric
\be
ds^2=2 e^0 e^1+ \delta_{ij} e^i e^j
\label{nobi}
\ee
to be equal to the one given by (\ref{newcor}), we also find that
$
2\mu+\sigma^2=-k.$
Thus we get $\rho^2-\rho+k=0$. This has real solutions if $k\leq {1\over4}$.
As a result,  the   metric \rf{kk}  with $0<k\leq {1\over4}$
can be interpreted as left-invariant (but not bi-invariant)
metric on a  group manifold.

\appendix{Brinkmann, Rosen and conformal\\
 coordinates}
 

As we have explained there is freedom 
in the choice of coordinate transformation from
 Brinkmann
to Rosen coordinates parameterized by the the two integration
constants $q_1$ and $q_2$ because the equation 
for $a(u)$ in \rf{giv} is second order. 
The case where $q_1=0$ has already been presented in section 3.2.
For the case $q_2=0$ after a  rescaling of
the $\x^i$ coordinates we find
\be \la{simpaa}
a(u) = u^\nu \ .
\ee
Therefore the metric (\ref{kk}) in these Rosen coordinates is
\be
ds^2=2d u d\tv+ u^{2\nu} d\x^2~.
\ee
This metric can be easily be expressed in conformal coordinates
(see \rf{zes},\rf{zeq})
as
\be \la{zesss}
ds^2=\  \B({\z}) \ (2d{\z}  d\tv+d\x^i d\x^i)\ , \ 
\ee
where for $u >0$
\be
\z= -  { c \ov u^{2 \n -1} } \ , \ \ \  - \infty  < \z < 0 \ ,
\ \ \ \ \B(\z) =  (-{\z\ov c})^{ - { 2 \nu \ov 2 \n -1}}  \ , \ \ \ \ \ \
  q_1=1 \ , \ \ q_2=0
\ , \la{sz}
\ee
and
$$
c\equiv { 1 \ov 2 \nu -1}  \ .
$$
Observe that  the singularity
at $u=0$ in Rosen coordinates is mapped to $-\infty$ in the Minkowski
coordinates and the region $u=+\infty$ -- 
to $\z=0$ hyperplane in Minkowski coordinates.  This is unlike the
case with $q_1=0$ which we have discussed in section 3.2.
For $u < 0$,  we find
\be \z=  { c \ov
(-u)^{2 \n -1} } \ , \ \ \   0 < \z <  \infty \ , \ \ \ \ \B(\z) =
({\z\ov c} )^{ - { 2 \nu \ov 2 \n -1}}  \ , \ \ \ \ \ \
  q_1=1 \ , \ \ q_2=0
\ . \la{sza} \ee

For $q_1q_2 \not=0$, we can
normalize the constants $q_1,q_2$  so that $q_1 q_2=1$, 
which can
 be achieved by a further rescaling of the coordinates $\x^i$. So
 we can set
$ q_1= q$ and   $ q_2 = q^{-1}$. 
For $u > 0$
\be \z= {c
q^2 u^{2\nu -1} \ov q^2 u^{2\nu -1} + 1 }
 \ , \ \ \ \ \ \ \ \ \ \ \
  0<\z<c \ , \ \  0<u<+\infty \ ,  \la{zeez}
\ee 
where we have chosen the integration constant so that  to
have $\z(0)=0$. Similarly, for  $u<0$
 \be \z=- { c
q^2 (-u)^{2\nu -1} \ov q^2 (-u)^{2\nu -1} + 1 }
 \ , \ \ \ \ \ \ \ \ \ \ \
  -c <\z<0\ , \ \  -\infty<u<0\ .  \la{zeezb}
\ee
 Then the metric in the Rosen coordinates \rf{opsen}  written
in conformally-flat form \rf{zes} for $0<\z<{1\over 2\nu-1}$ has
$\B(\z)$ given by 
\begin{equation}
\B(\z) = b{\z^{2-2\nu\over 2\nu-1}\over  (c-\z)^{2\nu\over 2\nu-1}} \
,\ \ \ \ \ \ \ \ \ \
 b\equiv    (2\nu-1)^{-2} q^{-{4\nu\over 2\nu-1} }\  , 
 \label{stripa}
 \ee
while  for $ -{1\over 2\nu-1}<\z<0$, 
  \begin{equation}
\B(\z)= b{(-\z)^{2-2\nu\over 2\nu-1}\over  (c+\z)^{2\nu\over 2\nu-1}}
\ .
 \label{stripb}
 \ee
As a result,  the homogeneous plane wave in the
above coordinates is conformal to  a strip in $d+2$-dimensional
Minkowski space. The two boundaries of the strip are the
hypersurfaces located at the values $\z=c$ and $\z=-c$ of the
light-cone coordinate $\z$. The singularity is again a
hypersurface located at $\z=0$. We remark that the above form of
the metric \rf{zes} suggests a generalization of the type
$ds^2={p(\z)\over r(\z)} (2d\z d\tv+d\x^i d\x^i)$,  where $p(\z),
r(\z)$ are polynomials without common factors. The singularities
of the spacetime are at the roots of $p$ while the conformal
boundaries are at the roots of $r$.

 Let us mention also that it  is easy to construct the embedding of the 
homogeneous plane-wave space-time into the  flat space $\bR^{2,
2+d}$ with the metric
\begin{equation}
ds^2(\bR^{2, 2+d})=-dX_1^2-dX_2^2+ \sum_{i=1}^{d+2} dY^2_i  \ .
\end{equation}
 For example,  in the case $q_1 q_2=1$, the embedding equations are
 \bea
 \sum_i Y_i^2- X^2_1-X_2^2&=&0 \ ,
\nonumber \\
\Bigl[{1\over 4\nu-2} (X_2+Y_1)-(X_1+Y_2)\Bigr]^{2\nu}&=& {1\over
q^{4\nu} (2\nu-1)^{4\nu-2}} (X_1+Y_2)^{2 -2\nu}\ . \la{nous} \eea
Replacing $Y_i$ and  $X_{1,2}$  by the  spherical  coordinates,
 $Y_1=r\cos\psi, Y_2=r
\sin\psi \cos\theta, ...$, and $X_1=r\sin\varphi,
X_2=r\cos\varphi$, we recover the plane-wave metric \rf{poi}
written in the static Einstein universe coordinates.

\appendix{Geodesics}

The geodesics of the spacetime (\ref{kk}) in Brinkmann coordinates
 can be found explicitly.
It is easy to see that the equation for $v$  implies  that the
coordinate $u$ should be linear in an affine parameter $s$. Thus we have
$u=u_1 s  + u_0, $
where $u_0, u_1$ are constants.
Now there are two cases to consider. If $u_1=0$, then the
geodesics are as in flat space (for any $k$ in \rf{kk})
\be \la{uuos}
u=u_0 \ , \ \ \ \
x^i=x^i_1 s +x_0^i\ , \ \ \ \ \
v=v_1 s+v_0\ , \ee
where $x_1^i, x_0^i$, $v_1$ and $v_0$ are constants.
These geodesics are spacelike or null depending on whether $x^i_1\not=0$
or $x^i_1=0$, respectively.

In the case where  $u_1\not=0$, we find that
$$
u(s)= u_1 s + u_0 \ ,
$$
$$
x^i(s)= x_0^i u^\nu(s) + x_1^i u^{1-\nu}(s)\ , ~~~~~~~~~~
 \ha < \nu < 1\ \  (0<k<{1\over4}) \ , 
$$
\be
v= {\epsilon\over u_1}s -\ha [ {\nu} x_0^2 u^{2\nu-1}(s) 
  + {(\nu-1)}  x_1^2 u^{1-2\nu}(s)]  +v_0 \ ,
\label{geo}
\ee
where $x_0^i, x_1^i$ and $v_0$ are constants. For $\epsilon<0$ the geodesics
are spacelike, for $\epsilon>0$ the geodesics are timelike and for $\epsilon=0$
the geodesics are null.
Since for the case when $u_1=0$
 the geodesics are either spacelike or null,
all {\it time-like} geodesics are of the type (\ref{geo}).

Observe that for $u\rightarrow 0$, $x^i\rightarrow 0$.
Thus  the geodesics, and,  in particular,  the time-like ones 
with both 
$x_0$ and $x_1$ non-vanishing,
 focus at the line $u=0$,  $x^i=0$. 
As stressed in \cite{hub2}, the geodesics
 for which $x_1= x_0=0$, i.e. 
\be \la{ffp}
u= u_1 s + u_0 \ ,~~~~~ v= {\epsilon\over u_1}s +v_0 \ ,\ \ \ \ \ 
x^i=0 \ , 
\ee 
 go smoothly through $u=0$
and  are defined for any value of the affine
parameter $-\infty <s<+\infty$. 

In the special  case where $k={1\over4}$ and $u_1\not=0$, we find 
$$
u=u_1 s+ u_0 \ ,
$$
$$
x^i= x_0^i u^{1\over2} \log u+ x_1^i u^{1\over2}\ , ~~~~~~~~~~
\nu= \ha \ \ (k={1\over4})\ ,
$$
\be
v= {\epsilon\over u_1^2}s-{1\over8}x_0 \log u (  x_0 \log u
+ 2  x_1 + x_0 )  +v_0  \ .
\label{geob}
\ee
These geodesics are spacelike if $\epsilon>0$, null if  $\epsilon=0$
 or timelike if $\epsilon<0$.
As in the previous case, the geodesics, and in particular 
the time-like ones with $x_1$ and $x_0$ non-vanishing, focus at the 
line $u=0,\ x^i=0$ as $u\rightarrow 0$. 
If $u_1=0$ we get again \rf{uuos}. 
The geodesics with  $x_1=x_0=0$, i.e. $x^i=0$, 
are the same as in \rf{ffp}, i.e. 
can be extended through $u=0$.


One  can also find the geodesics 
 in Rosen  \rf{opse} and conformal \rf{zes}
coordinates where we have translational symmetry in $\x^i$ directions.
The geodesics of the metric \rf{opse} 
$ds^2 = 2 du d \tv  + a^2(u) d\x^2_i$ 
are given by
\be \la{rosen}
u=u_1 s+ u_0 \ , \ \ \ \ 
\x^i=\x_1^i \int^s ds' a^{-2}(s')+ \x_0^i \ , \ \ \ \ 
\tv={\epsilon\over 2 u_1} s-{\x_1^2\over 2 u_1} \int^s ds' a^{-2}(s')+ \tv_0
\ee 
for $u_1\not=0$.
We shall consider  the case of 
 $a(u)=u^{1-\nu}$ and $0<k<{1\over4}$, i.e. $\ha < \nu < 1$.
Then 
\be \la{ross}   u=u_1 s+ u_0 \ , \ \ \ 
\x^i={1\over u_1 (2\nu-1)} \x_1^i u^{2\nu-1}+ \x_0^i
\ ,\ \ \ \ 
\tv={\epsilon\over 2 u_1} s-{\x_1^2\over (4\nu-2) u_1^2} u^{2\nu-1}+ \tv_0
\ . 
\ee
The geodesics for $\epsilon<0$ are
time-like, for $\epsilon=0$ are null and for $\epsilon>0$ are space-like.
As $u\rightarrow 0$ these geodesics can end
at any $\x_0^i$,  unlike what happened in  the Brinkmann
coordinate case discussed  above. 
The geodesics for which $\x_1\not=0$ can not
be extended through $u=0$. However,  if $\x_1=0$,
i.e. $\x^i=\const$,  then the geodesics are 
\be
u=u_1 s+ u_0\ , ~~~~~~\tv={\epsilon\over 2 u_1} s+ \tv_0\ ,~~~~~~~\x^i=\x_0^i \ , 
\label{extn}
\ee
and can be extended smoothly through $u=0$ as in the Brinkmann coordinate
case above.
For $u_1=0$, i.e. $u=\const$,  we get
\be 
u=u_0\ , ~~~~\x^i= \x_1^i s+ \x_0^i \ , ~~~~~~
\tv=-{1\over2}(1-\nu)u_0^{1-2\nu} \x_1^2 s^2+ \tv_1 s+ 
\tv_0\ .
\ee
These are the geodesics parallel to the plane wave.
 These geodesics are
null if $\x_1=0$, otherwise they are space-like.

The geodesics of the metric in conformal coordinates
\rf{zes} 
$ds^2 = \B (\z) ( 2 du d \z  +  d\x^2_i)$ 
 are directly related to the above by redefining the coordinate 
$u \to \z$. Explicitly, 
\be \la{zeq}
\int^{\z(s)} d\z' \B(\z')=\z_1 s+ \z_0 \ , \ \ \ \ 
\x^i={\x_1^i\over \z_1}\z(s) +\x_0^i \ , \ 
\  \ \ 
\tv={\epsilon\over 2 \z_1} s-{x_1^2\over 2 \z_1^2} \z(s) + \tv_0 \ , \ee
for $\z_1\not=0$. 
In particular,  for $\B=({\z\ov 2 \nu -1})^{2-2\nu\over 2\nu-1}$ (see \rf{szza}), we find
\be 
\z={1\over (2\nu-1)^{2\nu-2}} (\z_1 s+\z_0)^{2\nu-1} \ , 
\ \ \ \ \  
\x^i={\x_1^i\over (2\nu-1)^{1-2\nu}} (\z_1 s+\z_0)^{2\nu-1}+ 
\x_0^i \ , 
\ee
\be 
\tv={\epsilon\over 2 \z_1} s-
{\x_1^2\over 2 (2\nu-1)^{1-2\nu} \z_1^2} (\z_1 s+\z_0)^{2\nu-1}+ \tv_0~.  \ee
The geodesics for $\epsilon<0$ are
time-like, for $\epsilon=0$ are null and for $\epsilon>0$ are space-like.
For $z_1=0$, we find
\be
\z=\z_0\ ,~~~~~~~ \x^i=\x_1^i s+ \x_0^i\ ,
\ \ \ \ \  \tv=-{1-\nu\over \z_0 (4\nu-2)} s^2+ \tv_1 s+ \tv_0~.
\ee
These are the geodesics parallel to 
the plane wave and they are null if $\x_1=0$;
otherwise they are spacelike.

It appears that there are no geodesics that can go through the singularity
in  conformal coordinates.
 This is due to the fact that the coordinate
transformation from Rosen to conformal coordinates is not $C^1$-differentiable
at $u=0$ and so the tangent vector of the geodesics at $u=0$ is not defined.
However, we know that for the geodesics (\ref{extn}) one can extend
 the affine parameter so that to reach the  $u<0$
values. Transformed  from Rosen to conformal coordinates 
 they are given by 
\be 
\z={1\over 2\nu-1} (u_1 s+u_0)^{2\nu-1}\ ,~~~ \ \ \ \ \ \ 
\tv={\epsilon\over 2 u_1} s+ \tv_0\ ,~~~~~~\x^i= \x_0^i~.
\ee

\appendix{Penrose diagrams}

Here we shall describe the Penrose diagram of homogeneous
plane wave (\ref{kk}) in more detail. The equations for the
conformal boundary and the singularity are \rf{opl} and \rf{oopl}, i.e.
$
\cos\varphi+\cos\psi=0
$
and
$
\sin\varphi+\sin\psi\cos\theta=0,
$
where $0<\psi, \theta<\pi$. As we have mentioned
in section 3,
 the Penrose diagram
of homogeneous plane wave spacetime is three dimensional but it is
better described by two dimensional diagrams for the 
coordinates
$(\psi, \varphi)$  parameterized
by the angle $\theta$.  In this description
the only special cases
arise when $\cos\theta=\pm1$.
If $\cos\theta\not=\pm 1$ in the
$(\psi, \varphi)$ coordinate system, the singularity is a curve
that joins $(0,0)$ with $(\pi ,0)$. For example,
 for $\theta={\pi\over 2}$, it
is the line joining $(0,0)$ and $(\pi ,0)$. For $\cos\theta<0$,
the Penrose diagram is given in Fig. 1 and for $\cos\theta>0$ the
Penrose diagram is given in Fig. 2. In fact,  solutions of the
equation for the singularity equation join the points $(0, 2n
\pi)$ with $(\pi, 2 n \pi)$; here we describe the case $n=0$. These
curves are generically spacelike.

\begin{figure}[hbt]
\vskip -0.1cm \hskip -1cm
\centerline{\epsfig{figure=
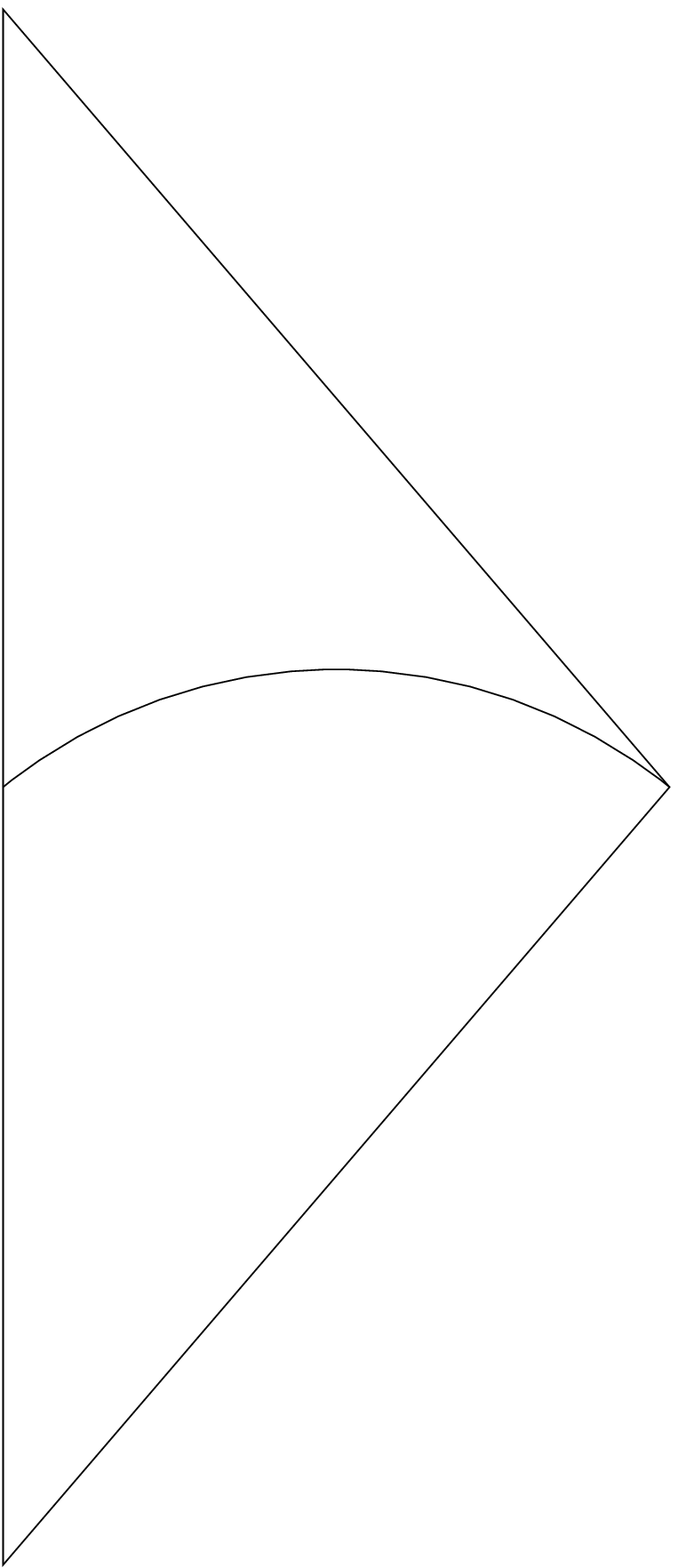,height=3truecm,width=2truecm}}
\vbox to 0pt{
\vspace{-4.0truecm}
\hspace{5.8truecm} $i^+$}
\vbox to 0pt{
\vspace{-3.7truecm}
\hspace{7.0truecm} $\ggg^+$}
\vbox to 0pt{
\vspace{-3.8truecm}
\hspace{6.0truecm} {\rm II}}
\vbox to 0pt{
\vspace{-3.8truecm}
\hspace{8.0truecm} $i^0$}
\vbox to 0pt{
\vspace{-3.5truecm}
\hspace{6.1truecm} {\rm I}}
\vbox to 0pt{
\vspace{-3.5truecm}
\hspace{7.0truecm} $\ggg^-$}
\vbox to 0pt{
\vspace{-3.4truecm}
\hspace{5.8truecm} $i^-$}
\vspace{-3.4truecm}
\caption{\footnotesize 
{}}
\end{figure}

\begin{figure}[bt]
\vskip  1cm \hskip -1cm
\centerline{\epsfig{figure=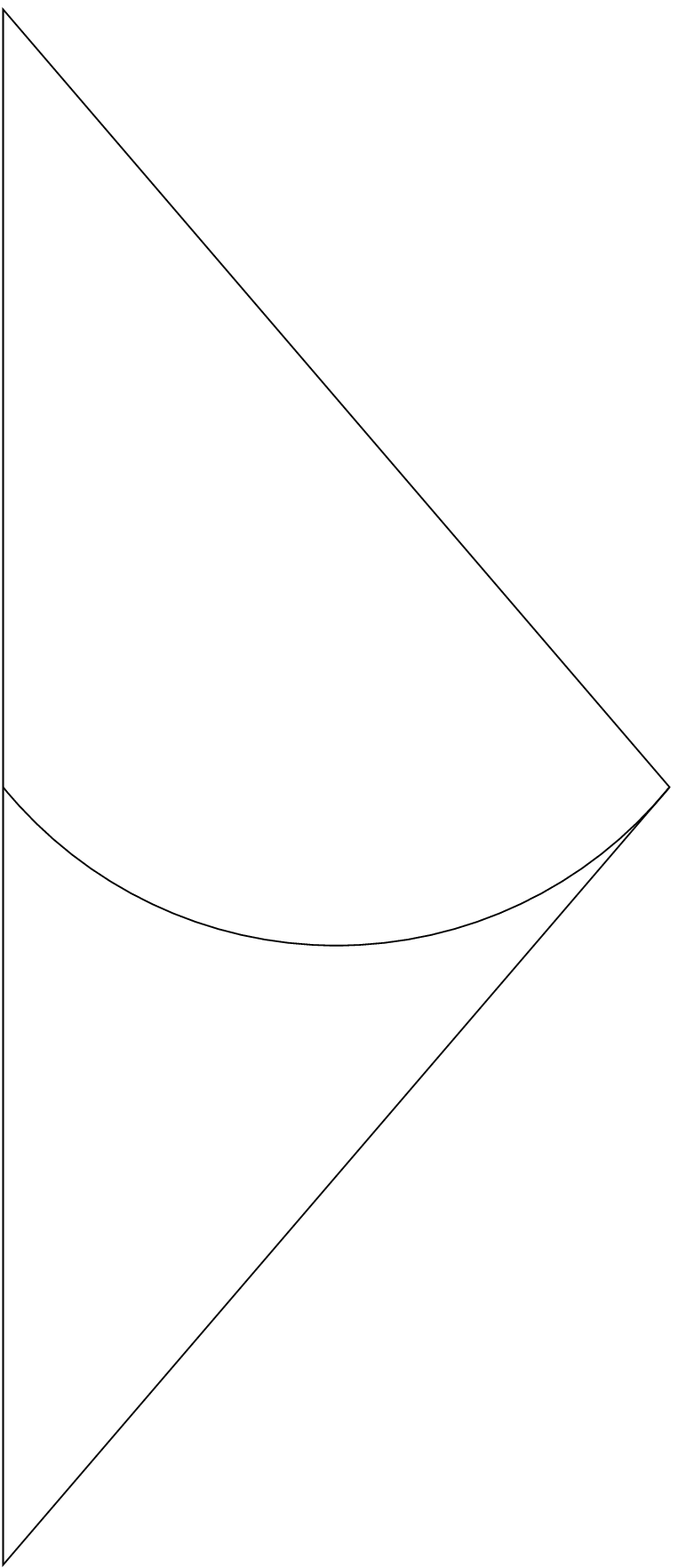,height=3truecm,width=2truecm}}
\vbox to 0pt{
\vspace{-4.0truecm}
\hspace{5.8truecm} $i^+$}
\vbox to 0pt{
\vspace{-3.7truecm}
\hspace{7.0truecm} $\ggg^+$}
\vbox to 0pt{
\vspace{-3.8truecm}
\hspace{6.0truecm} {\rm II}}
\vbox to 0pt{
\vspace{-3.8truecm}
\hspace{8.0truecm} $i^0$}
\vbox to 0pt{
\vspace{-3.5truecm}
\hspace{6.1truecm} {\rm I}}
\vbox to 0pt{
\vspace{-3.5truecm}
\hspace{7.0truecm} $\ggg^-$}
\vbox to 0pt{
\vspace{-3.4truecm}
\hspace{5.8truecm} $i^-$}
\vspace{-3.4truecm}
\caption{\footnotesize 
{}
}
\end{figure}

\begin{figure}[hb]
\vskip 1cm \hskip -1cm
\centerline{\epsfig{figure=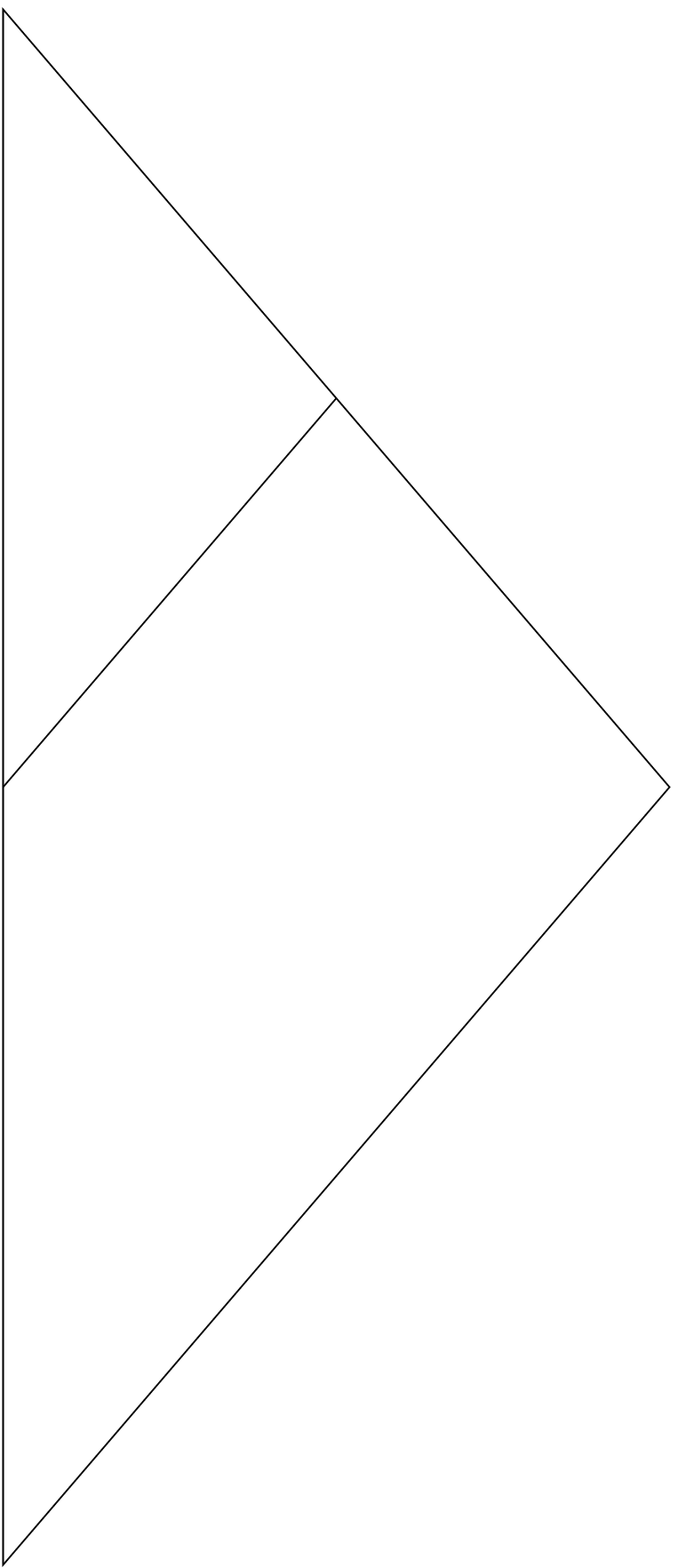,height=3truecm,width=2truecm}}
\vbox to 0pt{
\vspace{-4.0truecm}
\hspace{5.8truecm} $i^+$}
\vbox to 0pt{
\vspace{-3.7truecm}
\hspace{7.0truecm} $\ggg^+$}
\vbox to 0pt{
\vspace{-3.8truecm}
\hspace{6.0truecm} {\rm II}}
\vbox to 0pt{
\vspace{-3.8truecm}
\hspace{8.0truecm} $i^0$}
\vbox to 0pt{
\vspace{-3.5truecm}
\hspace{6.1truecm} {\rm I}}
\vbox to 0pt{
\vspace{-3.5truecm}
\hspace{7.0truecm} $\ggg^-$}
\vbox to 0pt{
\vspace{-3.4truecm}
\hspace{5.8truecm} $i^-$}
\vspace{-3.4truecm}
\caption{\footnotesize 
 {} 
}
\end{figure}

\begin{figure}[bth]
\vskip  1cm \hskip -1cm
\centerline{\epsfig{figure=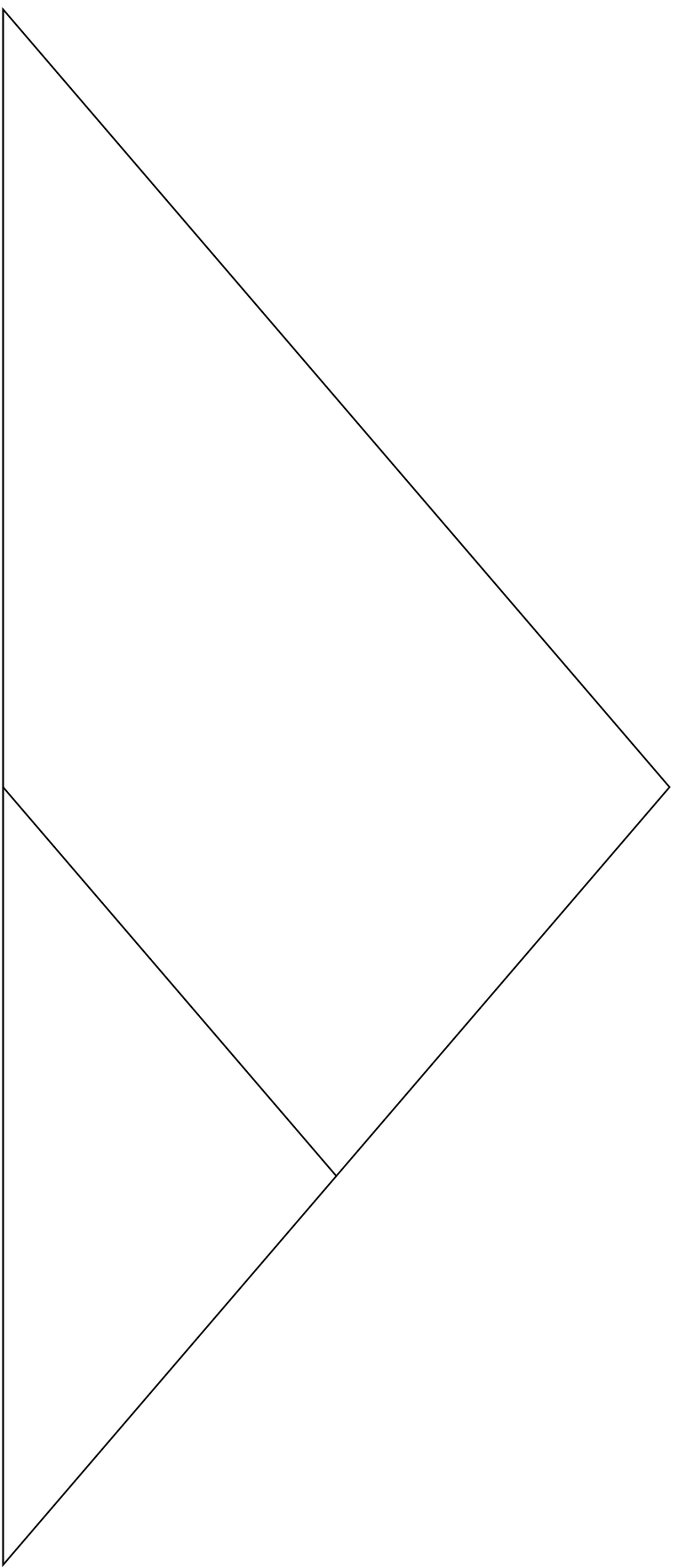,height=3truecm,width=2truecm}}
\vbox to 0pt{
\vspace{-4.0truecm}
\hspace{5.8truecm} $i^+$}
\vbox to 0pt{
\vspace{-3.7truecm}
\hspace{7.0truecm} $\ggg^+$}
\vbox to 0pt{
\vspace{-3.8truecm}
\hspace{6.0truecm} {\rm II}}
\vbox to 0pt{
\vspace{-3.8truecm}
\hspace{8.0truecm} $i^0$}
\vbox to 0pt{
\vspace{-3.5truecm}
\hspace{6.1truecm} {\rm I}}
\vbox to 0pt{
\vspace{-3.5truecm}
\hspace{7.0truecm} $\ggg^-$}
\vbox to 0pt{
\vspace{-3.4truecm}
\hspace{5.8truecm} $i^-$}
\vskip -3cm \hskip -1cm
\caption{\footnotesize 
{}
}
\end{figure}

In the special case $\cos\theta=- 1$, i.e.  $\psi=\varphi$,
the singularity
becomes the null line $\varphi=\psi$ (Fig. 3). In 
this case again the Penrose diagram
is  separated in two regions (II) and (I) by the singularity.

In the other special case $\cos\theta=1$, i.e.  $\theta=0$,
the singularity is a null line $\varphi=-\psi$ (Fig. 4). The Penrose diagram
is  separated in two regions (II) and (I) by the singularity.

In all the above cases, the conformal boundary is that
of Minkowski spacetime, i.e. it is represented by the null-lines
${\ggg}^+$ and ${\ggg}^-$ joining the points  $(0 ,\pi)$ with
$(\pi, 0)$ and $(\pi, 0)$ with $(0, -\pi)$, respectively.
 For a general angle $\theta$
(Figs. 1,2), a generic point
in the Penrose diagrams is a $S^{d-1}$ sphere. In the two special
cases (Figs. 3,4), the Penrose diagrams are two-dimensional.
In all these Penrose diagrams, there are special points $i^+=(0, \pi)$,
$i^-=(0, -\pi)$ and $i^0=(\pi, 0)$, representing the time-like future
infinity, the time-like past
infinity and the spatial infinity, respectively.
All the above diagrams are periodic in the angle $\varphi$ with period
$2\pi$.
For a general angle $\theta$, the singularity intersects the conformal boundary
at the point $i^0$. In the special case for which the singularity
is $\psi=\varphi$, the conformal boundary intersects the singularity at
$({\pi\over 2}, {\pi\over 2})$. In the other special case the intersection of
the singularity $\varphi=-\psi$ with the conformal boundary is at $({\pi\over 2}, -{\pi\over 2})$.

In the present plane wave spacetime, some time-like 
geodesics begin at past infinity $i^-$
and end at the singularity.
 Similarly,  some time-like geodesics originating
from the singularity end at future infinity $i^+$.
The rest of  time-like geodesics which
 go through the singularity will begin at $i^-$ and end at $i^+$
but they will not be $C^1$ differentiable
 at the singularity in conformal
coordinates.
 Some null geodesics,
those parallel to the plane wave, can reach ${\ggg}^+$ and ${\ggg}^-$
without passing through the singularity. Only null geodesics end at either
$\ggg^-$ or $\ggg^+$.

In the special case in which the singularity is $\psi=\varphi$,
there are  null geodesics in region (I), those parallel to the plane wave,
which can begin at   ${\ggg}^-$ and end in part of ${\ggg}^+$
without passing though the singularity.
Similarly,  in the other special case in which the singularity is
$\theta=-\pi$ in region (II) there are null geodesics which
begin in part of ${\ggg}^-$ and end in ${\ggg}^+$ without
passing through the singularity.

Incidentally, the equation for the
conformal boundary in the conformal compactification to the
Einstein static universe for which $q_1 q_2=1$ is
 \begin{equation}
\cos \varphi+\cos \psi = 2(2 \n-1) \bigl(\sin\varphi \ +\
\sin\psi \ \cos\theta\bigr) \ . \la{ople}
\end{equation}
The equation for the singularity is the same as above.



\end{document}